\documentclass[final]{siamltex}

\usepackage{graphicx}
\usepackage{amssymb,amsmath}
\usepackage{array}
\usepackage{hyperref}
\usepackage{pstricks}
\usepackage{pst-node}

\usepackage{vmargin}
\setpapersize{USletter}
\setmarginsrb{3.3cm}{1.5cm}{3.3cm}{1.5cm}{1cm}{0.5cm}{1cm}{0.5cm}

\newcommand{\myfig}[2]{\includegraphics[width=#1\textwidth]{#2}}

\newcommand{\leavethisout}[1]{}

\newcommand{\mass}{m}
\newcommand{\molmass}{M}
\newcommand{\radius}{R}
\newcommand{\nfiber}{N}
\newcommand{\height}{L}
\newcommand{\vol}{V}
\newcommand{\volmelt}{U}
\newcommand{\stefan}{\lambda}
\newcommand{\gasconst}{{\cal R}}
\newcommand{\Biot}{\mathrm{Bi}}
\newcommand{\units}[1]{\mbox{$\mathrm{#1}$}}
\newcommand{\Tvwleft}{T_{w1}^v}
\newcommand{\Tvwright}{T_{w2}^v}
\newcommand{\Rf}{\radius^f}
\newcommand{\Rv}{\radius^v}
\newcommand{\Kpit}{K}
\newcommand{\degK}{{}^\circ\hspace*{-0.1em}K}
\newcommand{\degC}{{}^\circ\hspace*{-0.1em}C}
\newcommand{\henry}{H}

\newcommand{\nondim}[1]{\bar{#1}}
\newcommand{\ndscale}[1]{\hat{#1}}
\newcommand{\dxone}{\partial_{\nondim{x}}}
\newcommand{\dxtwo}{\partial_{\nondim{x}\nondim{x}}}

\newcommand{\matlab}{MATLAB}

\title{A mathematical model of sap exudation in maple trees governed by
  ice melting, gas dissolution and osmosis%
  \thanks{This work was supported by grants from the Natural Science and
    Engineering Research Council of Canada, Mprime Network of Centres of
    Excellence, and North American Maple Syrup Council}.}

\author{Maurizio Ceseri \and John M. Stockie\thanks{Department of
    Mathematics, Simon Fraser University, 8888 University Drive,
    Burnaby, British Columbia, Canada, V5A 1S6 ({\tt mceseri@sfu.ca, 
      jstockie@sfu.ca}).}}

\newtheorem{remark}{Remark}[section]
\newcommand{\eqnref}[1]{(\ref{eq:#1})}

\begin{document}
\maketitle

\begin{abstract}
  We develop a mathematical model for sap exudation in a maple tree that
  is based on a purely physical mechanism for internal pressure
  generation in trees in the leafless state.  There has been a
  long-standing controversy in the tree physiology literature over
  precisely what mechanism drives sap exudation, and we aim to cast
  light on this issue.  Our model is based on the work of Milburn and
  O'Malley [\emph{Can.~J.~Bot.}, \underline{62}(10):2101--2106, 1984]
  who hypothesized that elevated sap pressures derive from compressed
  gas that is trapped within certain wood cells and subsequently
  released when frozen sap thaws in the spring.  We also incorporate the
  extension of Tyree [in \emph{Tree~Sap}, pp.~37--45, eds.\ M.~Terazawa
  et al., Hokkaido Univ.\ Press, 1995] who argued that gas bubbles are
  prevented from dissolving because of osmotic pressure that derives
  from differences in sap sugar concentrations and the selective
  permeability of cell walls.  We derive a system of
  differential-algebraic equations based on conservation principles that
  is used to test the validity of the Milburn--O'Malley hypothesis and
  also to determine the extent to which osmosis is required.  This work
  represents the first attempt to derive a detailed mathematical model
  of sap exudation at the micro-scale.
\end{abstract}



\pagestyle{myheadings}
\thispagestyle{plain}
\markboth{M. CESERI AND J. M. STOCKIE}{MODEL OF SAP EXUDATION IN MAPLE TREES}

\leavethisout{
  \printglossary 
}

\section{Introduction}
\label{sec:intro}

Sap flow in deciduous trees is driven during the growing season by the
process of transpiration, in which evaporation of water in the leaves
draws groundwater from the roots to the
crown~\cite{pallardy-2007,tyree-zimmermann-2002}.  As much as 90\%\ of
water taken up by the tree is transpired into the atmosphere to generate
the large pressure differential needed to overcome gravity; only 10\%\
of the water is actually consumed in photosynthesis reactions in the
leaves to produce the sugars needed for growth and other life processes.
These sugars are transported by the sap in dissolved form back to the
trunk and roots where they are either consumed immediately or else
stored as starch for later use.  Transpiration halts in late fall when
the leaves drop and the tree enters its winter dormant phase, which
lasts until the spring thaw.  At that time, the stored starch reserves
are released and provide the energy needed to initiate budding and leaf
growth before photosynthesis and transpiration can begin again.

In the sugar maple (\emph{Acer saccharum}), a process known as \emph{sap
  exudation} occurs in the time period between the dormant leafless
phase and the active transpiration phase, when certain processes are
triggered that build up positive sap pressure and convert stored
starches into sugars.  The stem pressure is large enough that when the
tree is tapped, sap exudes naturally from the tap-hole.  The mechanism
that generates this internal pressure is unique to the sugar maple and a
few other related species (black and red maple, birch, walnut, and
butternut) whose sap sugar content is also significant although not
typically as high as sugar maple.  Raw maple sap typically contains
2 to 3\%\ sugar by weight (primarily sucrose) and can be processed by
repeated boiling to produce sweet syrup and other edible maple products.
%

The scientific study of sap exudation has a long history that has seen
several competing hypotheses proposed for the mechanism that drives sap
flow in spring.  These hypotheses can be roughly divided into two
classes: ``vitalistic'' models that require some intervention by living
cells; and ``physical'' models that rely on passive, physical effects.
Early in the 20th century, Wiegand's experiments~\cite{wiegand-1906}
showed that even though up to one-quarter of the total volume of sapwood
or \emph{xylem} is filled with gas, the expansion and contraction of
this gas in response to temperature changes alone could still not
account for the observed pressure changes.  He concluded that some
active process initiated by living cells must be responsible, a claim
that was supported by the experiments of Johnson~\cite{johnson-1945}.

Later studies questioned this vitalistic hypothesis, for example Stevens
and Eggert~\cite{stevens-eggert-1945} who attributed the pressure
generation in leafless maple trees to volume changes arising from
freezing and thawing of sap.  This hypothesis was further explored in
subsequent years~\cite{sauter-etal-1973,tyree-1983}, culminating in the
ground-breaking paper of Milburn and
O'Malley~\cite{milburn-omalley-1984} who proposed that gas trapped in
the xylem is compressed by growth of ice crystals and subsequent uptake
of water when the tree freezes in fall; during the spring thaw, they
argued that this compressed gas generates the pressures that drive
exudation.  This freeze/thaw hypothesis was later supported by
experimental
evidence~\cite{johnson-tyree-dixon-1987,milburn-zimmermann-1986} showing
that gas expansion due to thawing sap is the primary cause of exudation
pressure; they also found no evidence to suggest that living cells play
a significant role.

However, the Milburn--O'Malley model was unable to account for all
observations, and so some
authors~\cite{sperry-etal-1988,tyree-1995,tyree-zimmermann-2002} cast
doubt on the ability of gas expansion alone to explain exudation,
arguing that heightened pressures in the xylem will dissolve any gas
bubbles within a few hours.  Several
studies~\cite{cirelli-etal-2008,cortes-sinclair-1985} suggested that
osmotic pressure, generated by differences in sugar concentration across
semi-permeable structures that separate various cells in the xylem,
could play a significant supporting role in sap exudation.  Most
notably, Tyree proposed~\cite{tyree-1995} that a combination of the
freeze/thaw mechanism and osmosis is the key to maintaining stable gas
bubbles in the xylem over observed exudation time scales.

This issue remains controversial and the current situation is perhaps
best summed up by Am\'eglio et~al.\ who concluded that ``no existing
single model explains all of the xylem winter pressure data,'' so that
both vitalistic and physical effects must play a
role~\cite{ameglio-etal-2001}.\ \
The lack of a clear consensus in the tree physiology literature
regarding the precise mechanisms driving sap exudation is compounded by
the fact that no detailed mathematical model has yet been developed for
the freeze/thaw mechanism, not to mention the other effects of osmosis
and gas dissolution.  

The aim of this paper is therefore to develop a mathematical model that
captures the essential processes driving sap exudation.  We begin in
Section~\ref{sec:tree-physiology} by providing details of tree
physiology, sap flow, and the Milburn--O'Malley and Tyree hypotheses for
pressure generation, at the core of which is the special role played by
two types of non-living xylem cells called vessels and fibers.  In
Section~\ref{sec:model} we derive a system of differential equations
governing the multiphase gas/liquid/ice system that incorporates the
effects of thawing sap in the fibers, dissolving gas bubbles in the
vessels, and the osmotic pressure pressure gradient between the two.
Numerical simulations in Section~\ref{sec:numerics} are used to validate
the model and conclusions are drawn regarding the various mechanism for
sap generation in Section~\ref{sec:conclusion}.

\section{Basic tree physiology and the Milburn--O'Malley hypothesis}
\label{sec:tree-physiology}

We first provide some background information on tree structure and the
hydraulics of sap flow that can be found in texts such as
\cite{pallardy-2007,tyree-zimmermann-2002}.  We also describe the
freeze/thaw model proposed by Milburn and O'Malley, and indicate how
other mechanisms such as gas dissolution and osmotic pressure enter the
picture. 

\begin{figure}[bthp]
  \footnotesize\em
  \centering
  \begin{tabular}{m{0.4\textwidth}cm{0.5\textwidth}}
    \mbox{\qquad\qquad}
    \includegraphics[height=0.33\textheight]{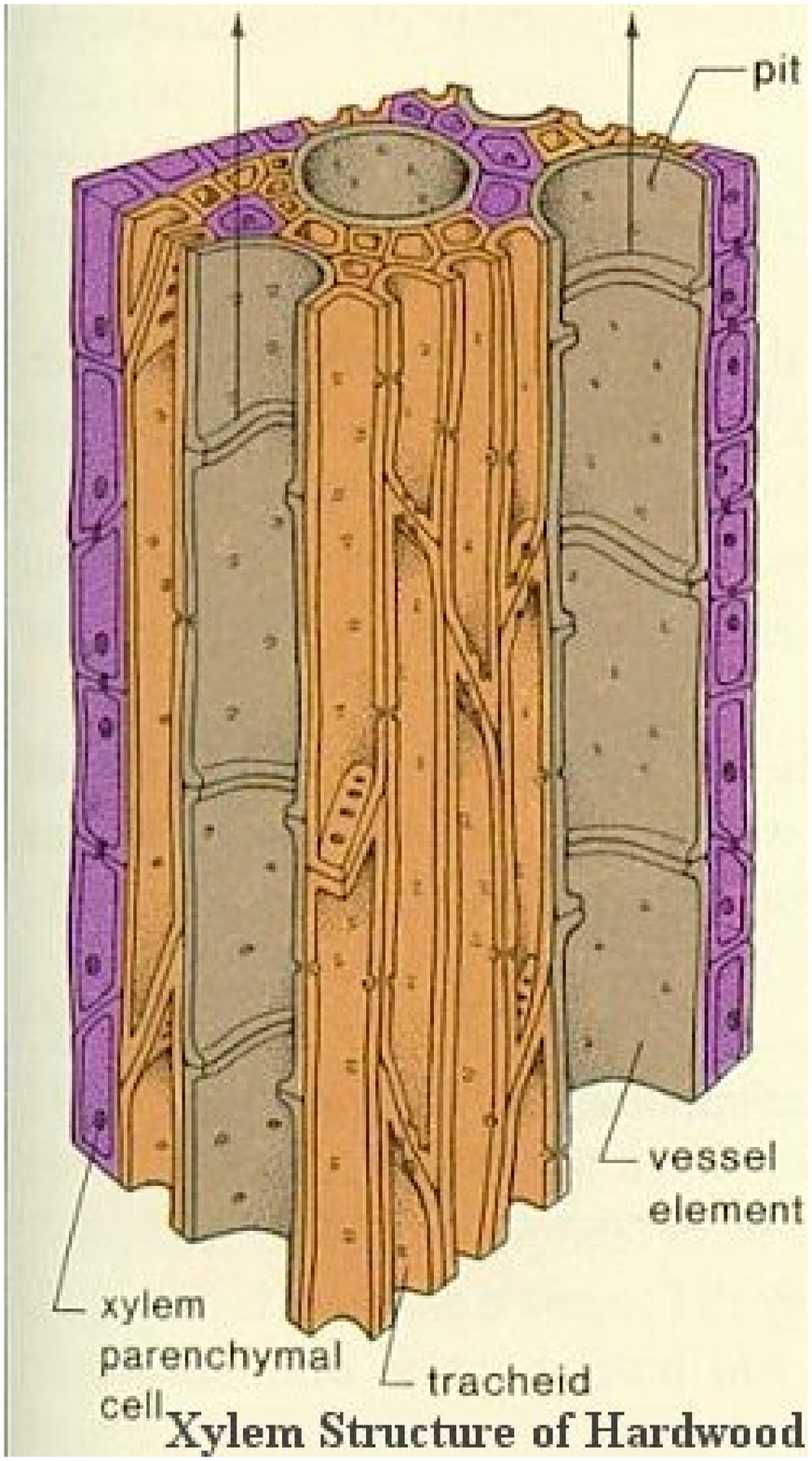}
    & & 
    \includegraphics[height=0.33\textheight]{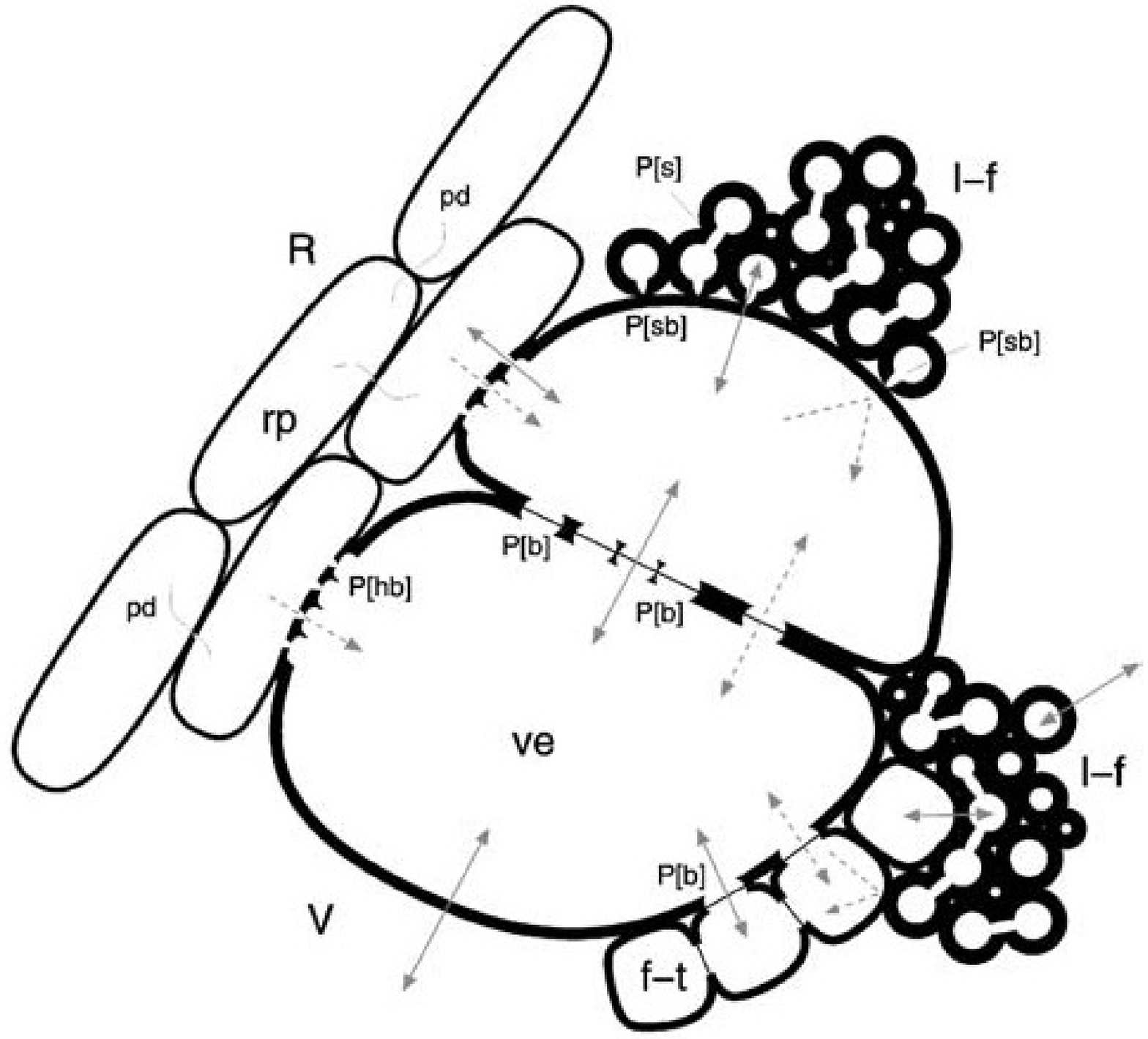}
    \\
    (a)~~3D view of the xylem, showing vessel elements surrounded by
    fibers and illustrating the difference in the cell diameters.   
    The pits forming the hydraulic connections between the cells are
    also shown.  Source: \cite{universe-review}.
    & & 
    (b)~~2D cross-section highlighting the vessel elements 
    (\,{\sffamily ve}), libriform fibers (\,{\sffamily l-f}), and
    fiber tracheids (\,{\sffamily f-t}).
    Source: Cirelli et
    al.~\cite[Fig.~6b]{cirelli-etal-2008} (reproduced with permission, 
    Oxford Univ.\ Press).
  \end{tabular}
  \caption{Two views of the layout of vessels and fibers in the xylem.}
  \label{fig:fibers-vessels}
\end{figure}

The Milburn--O'Malley model is a purely physical one that operates at
the cellular level and depends fundamentally on the structure of the
non-living cells making up the xylem.  As pictured in
Figure~\ref{fig:fibers-vessels}(a), the xylem consists of long, hollow,
roughly cylindrical cells of two types: \emph{fibers}, that are the main
structural members in the wood; and \emph{vessels}, that have a larger
radius and form the primary conduit for transmitting sap.  An individual
vessel can extend for up to tens of centimeters vertically through the
xylem, and is subdivided into \emph{elements} that are connected
end-to-end via perforated plates to form a continuous hydraulic
connection.  Vessels are connected to each other via numerous
\emph{pits} or cavities that perforate the vessel walls.  The fibers are
in turn subdivided into two sub-classes: \emph{libriform fibers} and
\emph{tracheids}.  According to Cirelli et al.~\cite{cirelli-etal-2008},
the tracheid walls also contain pits through which they exchange sap
with the vessels and hence integrating the tracheids into the water
transport system of the tree.  We will therefore ignore the tracheids,
treating them as a part of the vessel network, and focus instead on the
fibers.  

In comparison with tracheids, the libriform fibers (or simply
``fibers'') lack the pits that provide a direct hydraulic connection to
the vessels and so they are often considered to play a primarily
structural role.  However, the recent experiments of Cirelli
et~al.~\cite{cirelli-etal-2008} suggest that the fiber secondary wall
has a measurable permeability to water.  Other experiments show that
under normal conditions, maple trees (and other species that exude sap
in spring) are characterized by gas-filled fibers and water-filled
vessels~\cite{pallardy-2007,tyree-1995}.  This is in direct contrast
with most other tree species whose fibers are completely filled with
water.  Hence, it is reasonable to suppose that the presence of gas in
xylem fibers may be connected with the ability of maple trees to
generate exudation pressure in the leafless state.

The mechanism proposed by Milburn and
O'Malley~\cite{milburn-omalley-1984}, and discussed in complete detail
in~\cite{tyree-1995}, can be summarized as follows:
\begin{itemize}
\item During late fall or early winter when the tree freezes,
  ice crystals form on the inner wall of the fibers and the growing ice
  layer compresses the gas trapped inside.
\item In early spring when temperatures rise above freezing, ice
  crystals in the fiber melt and the pressure of the trapped gas drives
  the melted sap through the fiber walls into the vessel, hence leading
  to the elevated pressures observed during sap exudation.
  Tyree~\cite{tyree-1983} estimates that this effect is responsible for
  a pressure increase of roughly 30--60~\units{kPa}.
\item Exudation pressure is enhanced by the gravitational potential of
  sap that was drawn into the crown during the previous fall, and once
  thawed is then free to fall toward the roots.
\end{itemize}
As mentioned before, this model can account for the initiation of
exudation pressure but it cannot explain how pressures are sustained
over periods longer than about 12 hours because gas bubbles will
dissolve when pressurized.  Tyree~\cite{tyree-1995} explains how surface
tension effects require that any gas bubble is at higher pressure than
the surrounding fluid.  As a result, Henry's law necessitates that the
concentration of dissolved gas in the region adjacent to the bubble is
proportional to interfacial pressure.  Since the gas concentration is
highest there, dissolved gas diffuses away from the bubble and
eventually causes the bubble to disappear.

One clue to identifying a mechanism that can sustain gas bubbles over
longer time periods is provided by experiments that show maple trees
exude sap only when sugar is present~\cite{johnson-tyree-dixon-1987}.
With this in mind, Tyree~\cite{tyree-1995} proposed a modification of
the Milburn--O'Malley mechanism in which osmotic pressures arising from
differences in sucrose concentration could permit gas bubbles to remain
stable.  In particular, he suggested that the fiber secondary walls are
selectively permeable, permitting water to pass through but not larger
molecules like sucrose.  Therefore, the ice layer that forms on the
inner surface of the fibers in fall/winter is composed of pure water,
which leads to an osmotic potential difference between the fibers and
the vessels.

\begin{figure}[bthp]
  \centering
  \myfig{0.75}{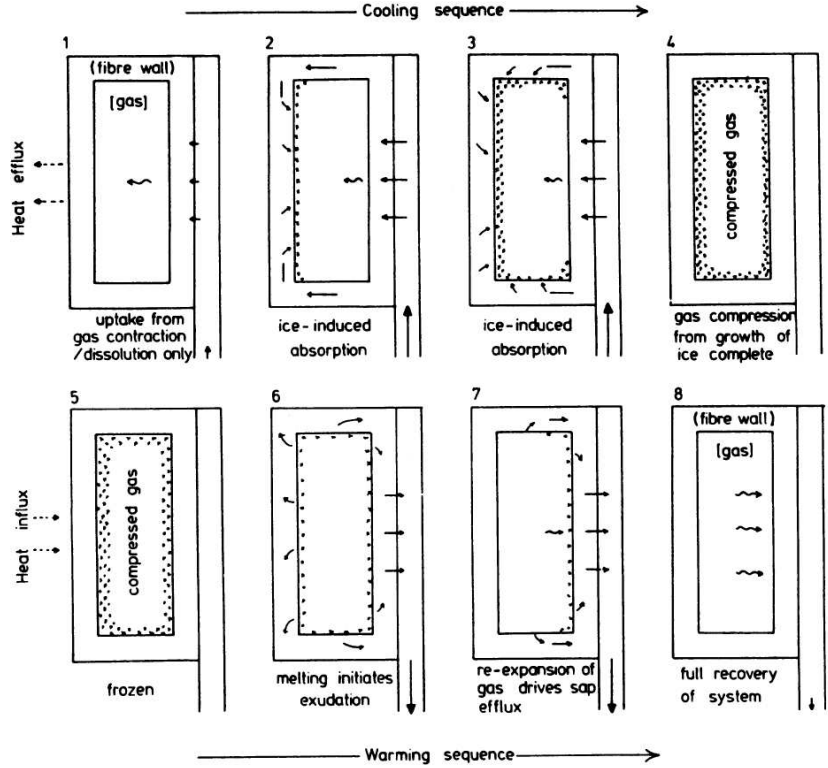}
  \caption{Picture of ice and compressed gas taken from Milburn and
    O'Malley~\cite[Fig.~7]{milburn-omalley-1984} (\copyright\ 2008
    Canadian Science Publishing, reproduced with permission).  We are
    concerned here only with the ``warming sequence'' in the bottom row,
    numbered 5--8.  The fiber is the rectangular structure on the left
    of each image, while the vessel is represented by the vertical
    channel on the right (not drawn to scale).}
  \label{fig:milburn-omalley}
\end{figure}

\section{Derivation of the model equations}
\label{sec:model}

\begin{figure}[bthp]
  \centering
  \myfig{0.7}{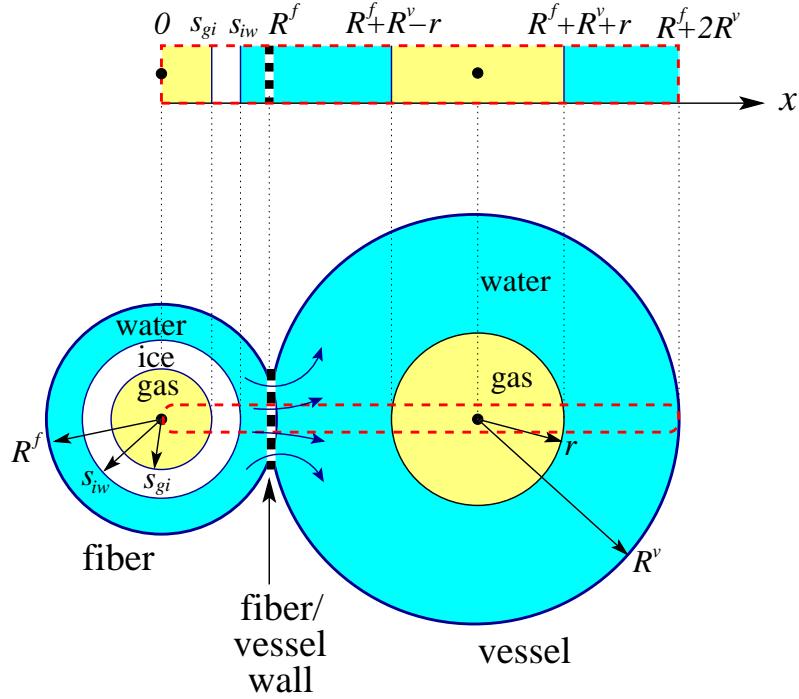}
  \caption{(Bottom) A 2D cross-section through a fiber--vessel pair
    showing the water, ice and gas regions, and the moving interfaces.
    (Top) The 1D region corresponding to our simplified model geometry
    is outlined with a red, dashed box in the bottom figure.  The
    locations of the various phase boundaries are indicated, with the
    origin $x=0$ at the center of the fiber.}
  \label{fig:fibves}  
\end{figure}

We now derive a compartment model describing the dynamics of gas, water
and ice in a single vessel that is in contact with the surrounding
fibers.  To begin with, we restrict ourselves to the time period before
the ice in the fiber melts completely.  The geometry is depicted in
Figure~\ref{fig:fibves} which is an idealized view of
Figure~\ref{fig:fibers-vessels}(b).  Before deriving the governing
equations, we make the following simplifying assumptions:
\begin{enumerate}
  \renewcommand{\theenumi}{A\arabic{enumi}}
\item We \label{assume:thaw} consider only the thawing phase, assuming
  that all water within the fibers is initially frozen.
\item Gravitational potential differences due to
  height \label{assume:gravity} can be neglected since we focus only
  locally on a small section of the xylem.
\item We \label{assume:element} consider a single vessel element
  and ignore interactions between adjacent elements.  This is a
  reasonable assumption if all elements experience similar conditions,
  and any gas contained within an element is isolated by the
  perforated plates from that of its neighbors.
\item The \label{assume:nfiber} vessel is surrounded by $\nfiber$
  identical fibers.  Although our model considers only a single fiber
  explicitly, the fluxes and other fiber quantities are scaled by a
  factor $\nfiber$ to determine the total influence of all $\nfiber$
  fibers on the vessel.
\item The \label{assume:cylinder} fiber and vessel are hollow and
  cylindrical in shape, with interior radii $\Rf$ and $\Rv$, and heights
  $\height^f$ and $\height^v$ respectively.  Both have large aspect
  ratio so that $\radius/\height \ll 1$.
\item We assume the problem is one-dimensional, with the domain
  extending horizontally from the center of the fiber to the rightmost
  outer edge of the vessel (pictured at the top of
  Figure~\ref{fig:fibves}).  The \label{assume:coord} coordinate system
  is chosen so that the horizontal ($x$) axis is directed along the line
  joining the centers of the fiber and vessel and with origin $x=0$ at
  the center of the fiber.  Referring to Figure~\ref{fig:fibves}, the
  fiber/vessel interface is located at $x=\Rf$ and the center of the
  vessel at $x=\Rf+\Rv$.
\item Gas is present in both \label{assume:vesselgas} fiber and vessel.
  Although the existence of gas in the fiber is a fundamental assumption
  of the Milburn--O'Malley model, the gas in the vessel is a novel
  feature introduced in our model.  Because water is incompressible and
  the xylem is a closed system when the tree is frozen, pressure cannot
  be transferred between the fiber and vessel compartments unless gas is
  also present in the vessel.  Further justification for this assumption
  is provided in the papers \cite{sperry-etal-1988,tyree-ewers-1991},
  which indicate that maple trees experience winter embolism (bubble
  formation) when gas in the vessels is forced out of solution upon
  freezing.
\item Gas \label{assume:bubble} in both fiber and vessel takes the form
  of a cylindrical bubble located at the center of the corresponding
  cell.  This seems reasonable in the vessel where the surface tension
  is of the order of $\sigma/\Rv\approx 4\times 10^3\;\units{Pa}$, which
  is several orders of magnitude smaller than the typical gas and liquid
  pressures.  In the fiber, the smaller radius gives rise to a much
  larger surface tension ($\sigma/\Rf\approx 2\times 10^4\;\units{Pa}$)
  which though still small relative to gas and liquid pressures could
  still potentially initiate a break-up into smaller bubbles owing to
  the Plateau-Rayleigh instability~\cite{eggers-1997}.  However,
  regardless of the precise configuration of the gas in the fiber, we
  assume that the net effect on gas pressure is equivalent to that of a
  single large bubble.
\item Heat \label{assume:ice} from outside the tree enters from the
  right in Figure~\ref{fig:fibves}.  Consequently, the sap in the vessel
  is taken to be initially in liquid form and the ice in the fiber
  begins melting on the inner surface of the fiber wall.
\item Gas and ice temperatures in the fiber can be taken as constant and
  equal to the freezing point.  This is justified by the fiber length
  scale \label{assume:Tfiber} being so much smaller than that of the
  vessel ($\Rf\ll \Rv$).
\item Time \label{assume:Tvessel} scales for heat and gas diffusion are
  much shorter than those corresponding to ice melting and subsequent
  phase interface motion (which our simulations show are on the order of
  minutes to hours).  This is justified by considering the
  various diffusion coefficients ($D$) and a typical length scale
  ($x\approx 10^{-5}\;\units{m}$), and in each case calculating the
  corresponding diffusion time scale $t \sim x^2/4D$:
  \begin{itemize}
  \item[--] \emph{diffusion in gas bubbles:} using the 
    air self-diffusion coefficient $D\approx 2\times
    10^{-5}\;\units{m^2/s}$, the time scale is $t\approx 1.3\times
    10^{-6}\;\units{s}$;  
  \item[--] \emph{diffusion of dissolved gas:} diffusivity
    $D\approx 1\times 10^{-9}$ of air in water gives
    $t\approx 0.025\;\units{s}$;
  \item[--] \emph{diffusion of heat:} thermal diffusivities $D\approx
    1.9\times 10^{-5}$ (for air) and $1.4\times 10^{-7}$ (for water)
    yield $t\approx 1.3\times 10^{-6}$ and $1.8\times 10^{-4}$
    respectively.
  \end{itemize}
  Therefore, heat transport is essentially quasi-steady and the 
  temperature equilibrates rapidly to any change in local
  conditions.  Also, gas concentration and density (in both gaseous
  and dissolved forms) can be taken as constants in space.
\item A significant osmotic \label{assume:osmosis} pressure arises owing
  to differences in sugar content between liquid in the vessel and
  fiber.  This is motivated by recent experiments which show that the
  fiber cell wall is permeable to water but not to larger molecules such
  as sucrose~\cite{cirelli-etal-2008}, suggesting that ice contained in
  the fiber is composed of pure water.  A simple order of magnitude
  estimate based on a sap sugar concentration of 2\%\ then implies that
  the osmotic pressure $\Pi\sim O(10^5)\;\units{Pa}$, which is of the
  same order as typical gas and liquid pressures.
\item There \label{assume:post-dissolve} is no need to consider
  modelling the system beyond the time when either fiber or vessel
  bubble is completely dissolved, since then no further exchange of
  pressure is possible.
\end{enumerate}

Model equations are derived for the fiber and vessel compartments in the
next two sections.  For easy reference, all parameters and variables are
listed in Table~\ref{tab:params} along with physical values where
appropriate.  There are some similarities that can be found with other
models developed for related problems such as ice lensing in porous
soils~\cite{talamucci-2003} and freeze/thaw processes in the food
industry~\cite{lewis-seetharamu-1995}; however, neither problem contains
all of the physical mechanisms encountered during sap exudation.

\begin{table}[!htb]
  \begin{center}
    \caption{A list of nomenclature for all physical parameters and
      variables, including values in SI units.  References are given for
      each numerical value, either as a citation to the literature (in
      square brackets) or else an equation number (in parentheses) for
      those values calculated from other parameters.}
  \label{tab:params}
  \footnotesize
  \begin{tabular}{|c|p{0.35\textwidth}|c|c|c|}\hline
    & {\bfseries Description} & {\bfseries Units}
    & {\bfseries Value} & {\bfseries Ref.}
    \\\hline
    \multicolumn{3}{l}{\emph{Constant parameters:}} \\\hline
    $A$         & Area of vessel wall that is permeable 
    & \units{m^2} & $6.28\times 10^{-8}$ & \eqnref{wall-area} \\
    $c_s$       & Sucrose concentration in vessel sap 
    & \units{mol/m^3} & 58.4 & \\
    $g$         & Gravitational acceleration 
    & \units{m/s^2} & 9.81 & \cite{crc-1994} \\ 
    $h$         & Heat transfer coefficient
    & \units{W/m^2\,\degK} & 10.0 & \cite{potter-andresen-2002} \\ 
    $\henry$    & Henry's constant 
    & -- & 0.0274 & \cite{Sander1999} \\
    $k_g$       & Thermal conductivity of air 
    & \units{W/m\,\degK} & 0.0243 & \cite{EngineeringTool} \\ 
    $k_w$       & Thermal conductivity of water 
    & \units{W/m\,\degK} & 0.580 & \cite{EngineeringTool} \\
    $\Kpit$     & Cell wall hydraulic conductivity 
    & \units{m/s} & $1.98\times 10^{-14}$ & \cite{petty-palin-1983} \\ 
    $\height^f$ & Length of fiber
    & \units{m} & $1.0\times 10^{-3}$ & \cite{panshin-dezeeuw-1980} \\ 
    $\height^v$ & Length of vessel element
    & \units{m} & $0.5\times 10^{-3}$ & \cite{panshin-dezeeuw-1980} \\
    $\molmass_g$& Molar mass of air
    & \units{kg/mol} & 0.0290 & \cite{crc-1994} \\ 
    $\molmass_s$& Molar mass of sucrose
    & \units{kg/mol} & 0.3423 & \cite{crc-1994} \\ 
    $\nfiber$   & Number of fibers per vessel
    & -- & 16 & \eqnref{nfiber} \\ 
    $\ndscale{p}$   & Pressure scale 
    & \units{N/m^2} & $1.01\times 10^{5}$ & \eqnref{rho-p-bar} \\ 
    $\Rf$ & Interior radius of fiber
    & \units{m} & $3.5\times 10^{-6}$ & \cite{yang-tyree-1992} \\
    $\Rv$ & Interior radius of vessel
    & \units{m} & $2.0\times 10^{-5}$ & \cite{yang-tyree-1992} \\
    $\gasconst$ & Universal gas constant
    & \units{J/mol\,\degK} & 8.314 & \cite{crc-1994} \\
    $\ndscale{t}$   & Time scale 
    & \units{s} & 46.1 & \eqnref{tbar} \\
    $T_c$       & Freezing temperature of ice 
    & \units{\degK} & 273.15 & \\
    $T_a$       & Ambient temperature of gas
    & \units{\degK} & $0.005 + T_c$ & \\
    $\vol^f$    & Volume of fiber 
    & \units{m^3} & $3.85\times 10^{-14}$ & \eqnref{vbarf} \\
    $\vol^v$    & Volume of vessel 
    & \units{m^3} & $6.28\times 10^{-13}$ & \eqnref{vbarv} \\
    $W$         & Thickness of fiber + vessel wall 
    & \units{m} & $3.64\times 10^{-6}$ & \cite{watson-etal-2003} \\
    $\stefan$   & Latent heat of fusion
    & \units{J/kg} & $3.34\times 10^5$ & \cite{EngineeringTool} \\
    $\rho_i$    & Ice density
    & \units{kg/m^3} & 917 & \cite{EngineeringTool} \\
    $\rho_w$    & Water density
    & \units{kg/m^3} & 1000 & \cite{crc-1994} \\
    $\ndscale{\rho}$& Density scale 
    & \units{kg/m^3} & 1.29 & \cite{tyree-1995} \\
    $\sigma$    & Air-water surface tension 
    & \units{N/m} & 0.0756 & \cite{crc-1994} \\
    \hline
    \multicolumn{3}{l}{\emph{Dependent and independent variables:}} \\\cline{1-3}
    $c_g$       & Dissolved gas concentration 
    & \units{mol/m^3} \\
    $\mass$    & Mass 
    & \units{m} \\
    $p$         & Pressure 
    & \units{N/m^2} \\
    $r$         & Radius of gas bubble in vessel 
    & \units{m} \\
    $s$         & Fiber interface location
    & \units{m} \\
    $t$         & Time 
    & \units{s} \\
    $T$         & Temperature 
    & \units{\degK} \\
    $\volmelt$  & Volume of melted ice 
    & \units{m^3} \\
    $\vol$      & Volume 
    & \units{m^3} \\
    $x$         & Spatial location 
    & \units{m} \\
    $\rho$      & Density 
    & \units{kg/m^3} \\
    \cline{1-3}
    \multicolumn{3}{l}{\emph{Subscript (denotes phase):}} \\\cline{1-2}
    $g$         & \multicolumn{1}{l|}{Gas phase} \\
    $i$         & \multicolumn{1}{l|}{Ice phase} \\
    $w$         & \multicolumn{1}{l|}{Water phase} \\
    \cline{1-2}
    \multicolumn{3}{l}{\emph{Superscript (denotes cell component):}} \\\cline{1-2}
    $f$         & \multicolumn{1}{l|}{Fiber} \\
    $v$         & \multicolumn{1}{l|}{Vessel} \\\cline{1-2}
  \end{tabular}
  \end{center}
\end{table}

\subsection{Fiber model}

We now develop a set of governing equations for the fiber that is
restricted to the time period when ice is present, so that an ice layer
separates the gas from the water adjacent to the wall.  The
modifications that are required to the model once the ice layer has
disappeared and the fiber bubble is in direct contact with the sap are
described in Section~\ref{sec:ice-melted}.

The inner fiber wall is initially coated by an ice layer of uniform
thickness, inside of which is a cylindrical volume of compressed gas
with radius $s_{gi}(t)$.  As the ice layer melts, an ice/water interface
$x=s_{iw}(t)$ appears at the inner edge of the fiber wall
($s_{iw}(0)=R^f$) and propagates inward.  The speed of the interface
depends not only on the volume difference between ice and water but also
on the seepage of liquid through the porous fiber wall.  The gas
pressure that drives water through the wall into the vessel is
transmitted through the ice layer to the adjacent water.

If we denote by $\vol^f_g(t)$, $\vol^f_w(t)$ and $\vol^f_i(t)$
the volumes of gas, water and ice within the fiber, then owing to
the cylindrical symmetry 
\begin{align}
  \vol^f_g(t) &= \pi \height^f \, s_{gi}(t)^2, \label{eq:volfg}\\
  \vol^f_w(t) &= \pi \height^f \left( (\Rf)^2 -
    s_{iw}(t)^2 \right), \label{eq:volfw}\\
  \vol^f_i(t) &= 
  \pi \height^f \left(s_{iw}(t)^2 - s_{gi}(t)^2\right),
  \label{eq:volfi}
\end{align}
which also satisfy
\begin{align}
  \vol^f_g(t) + \vol^f_w(t) + \vol^f_i(t) = \pi \height^f \left(
    \Rf \right)^2 =: \vol^f. \label{eq:vbarf}
\end{align}
Recalling Assumption~\ref{assume:Tfiber}, which takes the gas
temperature $T^f_g$ to be constant and equal to the freezing point
$T_c=273.15\,\units{\degK}$, the gas density varies only due to changes
in volume so that
\begin{gather}
  \rho^f_g(t) = \rho^f_g(0) \, \frac{\vol^f_g(0)}{\vol^f_g(t)}.  
  \label{eq:rhofg}
\end{gather}
The corresponding gas pressure is given by the ideal gas law
\begin{gather*}
  p^f_g(t) = \frac{\rho^f_g(t) \gasconst T_c}{\molmass_g},
\end{gather*}
where $\molmass_g$ is the molar mass of air.  This can be combined
with~\eqnref{volfg} and \eqnref{rhofg} to obtain
\begin{gather}
  p^f_g(t) = p^f_g(0) \, \left( \frac{s_{gi}(0)}{s_{gi}(t)} \right)^2.
  \label{eq:pfg2}
\end{gather}

As ice thaws, some volume of melt-water (denoted $\volmelt(t)$) is
forced through the fiber wall into the vessel, while the remainder stays
in the fiber sandwiched between the ice layer and the wall.  Let the
(constant) mass of gas in the fiber be $\mass^f_{g0}$ and the mass of
water and ice be
\begin{gather}
  \mass^f_i(t) = \rho_i \vol^f_i(t), \label{eq:massfi} \\
  \mass^f_w(t) = \rho_w \vol^f_w(t), \label{eq:massfw} 
\end{gather}
where $\rho_w$ and $\rho_i$ are the densities of water and ice
respectively.  Conservation of mass in the fiber implies that
\begin{gather}
  \mass^f_{g0} + \mass^f_i(t) + \mass^f_w(t) + \rho_w \volmelt(t) =
  \mass^f_0, 
  \label{eq:mf0}
\end{gather}
where initially $\volmelt(0)=0$ and $\mass^f_0 = \mass^f_{g0} +
\mass^f_i(0)+\mass^f_w(0)$.  Differentiating \eqnref{mf0} yields
\begin{gather*}
  \dot{\mass}^f_i(t) + \dot{\mass}^f_w(t) + \rho_w\dot{\volmelt}(t) = 0, 
\end{gather*}
where the ``dot'' denotes a time derivative.  Substituting
\eqnref{vbarf} and \eqnref{massfi}--\eqnref{massfw} into the
last expression yields
\begin{gather*}
  -\rho_i\dot{\vol}^f_g + (\rho_w-\rho_i) \dot{\vol}^f_w + \rho_w
  \dot{\volmelt} = 0,
\end{gather*}
and the volume terms can be replaced using \eqnref{volfg}
and \eqnref{volfw} to obtain
\begin{gather}
  2\rho_i s_{gi} \dot{s}_{gi} +
  2(\rho_w-\rho_i) s_{iw} \dot{s}_{iw} - 
  \frac{\rho_w}{\pi\height^f}\, \dot{\volmelt} = 0.
  \label{eq:sgi-siw}
\end{gather}

Recall that the temperature in both gas and ice is assumed constant and
equal to the freezing point $T_c$.  The water layer, on the other hand,
is heated from the vessel side so that the water temperature
$T^f_w(x,t)$ is not constant but instead obeys the steady-state heat
equation and boundary condition
\begin{subequations}\label{eq:heat-Tfw}
  \begin{align}
    \partial_{xx} T^f_w &= 0 \qquad \text{for all} \; 
    x \in \left(s_{iw}, \Rf\right),
    \label{eq:heat-Tfw-pde}\\
    T^f_w &= T_c \qquad \text{at} \; x=s_{iw}.
    \label{eq:heat-Tfw-bc2}
  \end{align}
\end{subequations}
The speed $\dot{s}_{iw}$ of the ice/water interface is determined by the
Stefan condition
\begin{gather}
  \stefan\rho_w \dot{s}_{iw} = - k_w\partial_x T^f_w
  \qquad \text{at}\; x=s_{iw},
  \label{eq:stefan}
\end{gather}
where $\stefan$ is the latent heat of fusion and $k_w$ is the thermal 
conductivity of water.

In summary, \eqnref{pfg2}, \eqnref{sgi-siw}, \eqnref{heat-Tfw-pde} and
\eqnref{stefan} represent a coupled system of equations for the fiber
unknowns $s_{iw}(t)$, $s_{gi}(t)$, $p^f_g(t)$, and $T^f_w(x,t)$.  In the
next section, we will derive an equation for $\dot{\volmelt}$ and also
determine the missing boundary condition for $T^f_w$ as a matching
condition with the vessel temperature at the fiber/vessel wall $x=\Rf$.

\subsection{Vessel model and matching conditions}

Let $\vol^v_w(t)$ and $\vol^v_g(t)$ denote the volume of the water and
gas compartments.  The time derivative of the total volume (gas +
water) must be zero so that
\begin{gather*}
  \frac{d}{dt} \left( \vol^v_w(0) + \nfiber \volmelt(t) + \vol^v_g(t)
  \right) = 0, 
\end{gather*}
or simply
\begin{gather}
  \dot{\vol}^v_g = -\nfiber\dot{\volmelt}.
  \label{eq:Vvg0}
\end{gather}
This equation embodies the assumption that there is no flow between
vessel elements so that the portion of the volume consisting of water
equals the initial value $\vol^v_w(0)$ plus whatever water enters from
the $\nfiber$ surrounding fibers.  Using the fact that the bubble radius
and volume are connected by
\begin{gather}
  \vol^v_g(t) = \pi \height^v \, r(t)^2,
  \label{eq:Vvg}
\end{gather}
equation \eqnref{Vvg0} can then be rewritten as
\begin{gather}
  r \dot{r} = -\frac{\nfiber\dot{\volmelt}}{2\pi\height^v}. 
  \label{eq:rg}  
\end{gather}
Assuming that fibers are packed tightly around the vessel and that both
vessel and fiber have wall thickness $W$, a simple geometric argument
provides an estimate for the constant $N$:
\begin{gather}
  \nfiber \approx \frac{2\pi(\Rf+\Rv + W)}{2\Rf+W} . 
  \label{eq:nfiber}
\end{gather}

We next focus on porous transport through the fiber
wall which is governed by Darcy's law
\begin{gather}
  \nfiber\dot{\volmelt}(t) = -\frac{\Kpit A}{\rho_w g W} (p^v_w(t) -
  p^f_g(t)),  
  \label{eq:volmelt}
\end{gather}
where $p^v_w(t)$ is the vessel water pressure, $\Kpit$ is the hydraulic
conductivity, $W$ is the wall thickness, and $A$ is the area of the wall
that is permeable to water.  In the absence of experimental measurements
for hydraulic conductivity in maple, we use the value $K=1.98\times
10^{-14}\;\units{m/s}$ obtained for birch trees~\cite{petty-palin-1983}.
Finally, $A$ is taken equal to the internal surface area of the
cylindrical vessel
\begin{gather}
  A = 2 \pi \Rv \height^v. 
  \label{eq:wall-area}
\end{gather}

We next describe the temperature evolution within the various vessel
compartments, recalling that the diffusive time scale is short in
relation to either melting or gas dissolution.  We therefore employ a
quasi-steady approximation in which temperatures obey a steady-state
heat equation.  Denote by $T^v_g(x,t)$ the temperature within the gas
bubble, and $\Tvwleft(x,t)$ and $\Tvwright(x,t)$ the water temperatures
to the left and right of the bubble respectively.  Then, in the water
region on the left (adjacent to the vessel wall) 
$\Tvwleft(x,t)$ obeys
\begin{subequations}\label{eq:heat-Tvw1}
  \begin{alignat}{3}
    \partial_{xx} \Tvwleft &= 0 \qquad &\text{for all} \; 
    x &\in \left(\Rf, \Rf+\Rv-r\right).
    \label{eq:heat-Tvw1-pde}\\
    \intertext{Both temperature and heat flux are continuous at 
      the vessel wall}
    \Tvwleft &= T^f_w \qquad &\text{at} \; x &= \Rf, 
    \label{eq:heat-Tvw1-bc1}\\
    k_w \partial_x \Tvwleft &= k_w \partial_x T^f_w 
    \qquad &\text{at} \; x &= \Rf, 
    \label{eq:heat-Tvw1-bc2}\\
    \intertext{and at the water/gas interface}
    \Tvwleft &= T^v_g \qquad &\text{at} \; 
    x &= \Rf+\Rv - r, 
    \label{eq:heat-Tvw1-bc3}\\
    k_w \partial_x \Tvwleft &= k_g \partial_x T^v_g  \qquad &\text{at} \;
    x &= \Rf+\Rv - r.
    \label{eq:heat-Tvw1-bc4}
  \end{alignat}
\end{subequations}
The gas temperature $T^v_g(x,t)$ inside the bubble obeys
\begin{subequations}\label{eq:heat-Tvg}
  \begin{alignat}{3}
    \partial_{xx} T^v_g &= 0 \qquad &\text{for all}\; 
    x &\in \left(\Rf+\Rv-r,
      \Rf+\Rv+r\right), 
    \label{eq:heat-Tvg-pde}\\ 
    \intertext{with temperature and flux continuity conditions}
    T^v_g &= \Tvwright
    \qquad &\text{at} \; x &= \Rf+\Rv+r,
    \label{eq:heat-Tvg-bc1}\\
    k_g \partial_x T^v_g &= k_w \partial_x \Tvwright
    \qquad &\text{at} \; x &= \Rf+\Rv+r.
    \label{eq:heat-Tvg-bc2}
  \end{alignat}
\end{subequations}
Finally, the water temperature $\Tvwright(x,t)$ to the right of the 
gas obeys
\begin{subequations}\label{eq:heat-Tvw2}
  \begin{alignat}{3}
    \partial_{xx} \Tvwright &=0 \qquad &\text{for all}\; 
    x &\in \left(\Rf+\Rv+r,
      \Rf+2\Rv\right),
    \label{eq:heat-Tvw2-pde}\\
    \intertext{with a Robin condition on the right-most boundary} 
    k_w \partial_x \Tvwright &= h(T_a-\Tvwright) \qquad
    &\text{at} \; x &= \Rf+2\Rv,
    \label{eq:heat-Tvw2-bc1}
  \end{alignat}
\end{subequations}
where $h$ is a convective heat transfer coefficient.  The ambient
temperature $T_a$ refers to the temperature in the neighboring layer of
xylem cells; using a temperature difference of $10\units{\degK}$ through
the sapwood over a tap depth of 0.05~\units{m}, we obtain a rough
estimate of $T_a\approx 0.005\units{\degK}$ for the temperature
difference on the cell scale.

The final component of the model is a description of the process whereby
elevated pressures in the gas bubble cause the gas to dissolve at the
gas/water interface.  Because of the small bubble size, it is essential
to take into account the effect of surface tension.  Gas and liquid
pressures in the vessel are connected by the Young--Laplace
equation\footnote{The Young--Laplace equation states that the pressure
  difference is proportional to the mean curvature, $p_g-p_w=\sigma
  \left(\frac{1}{R_1}+\frac{1}{R_2}\right)$, where the proportionality
  constant is the interfacial surface tension $\sigma$.  For a
  cylindrical gas bubble with radius $r$, the principal radii of
  curvature are $R_1=r$ and $R_2=\infty$.},
\begin{gather}
  \label{eq:surftens}
  p^v_w(t) = p^v_g(x,t) - \frac{\sigma}{r(t)} \qquad
  \text{at $x=\Rf+\Rv-r$},
\end{gather}
where $\sigma$ is the air-water surface tension and $r(t)$ is the bubble
radius.  In this formula, the gas pressure can be written using the
ideal gas law
\begin{gather}
  p^v_g(x,t) = \frac{\rho^v_g(t) \gasconst T^v_g(x,t)}{\molmass_g},
  \label{eq:pvg}
\end{gather}
where $\rho^v_g(t)$ is the gas density which we have taken uniform in
space according to Assumption~\ref{assume:Tvessel}.  The concentration
$c^v_g(t)$ of dissolved gas in the sap is then related to the gas
density by Henry's law
\begin{gather}
  c^v_g(t) = \frac{\henry}{\molmass_g}\rho^v_g(t),
  \label{eq:henry}
\end{gather}
where $\henry$ is a dimensionless constant.  The gas density in the
preceding equations decreases over time owing to dissolution and can be
written 
\begin{align}
  \rho^v_g(t) = \frac{\rho^v_g(0) \vol^v_g(0) -
    \molmass_g\int_{\Omega^v_w}c^v_g(t)\,dV}{\vol^v_g(t)}
  = \frac{\rho^v_g(0) \vol^v_g(0) - \molmass_g
    c^v_g(t)(\vol^v-\vol^v_g(t))}{\vol^v_g(t)}, 
  \label{eq:rhovg}
\end{align} 
where $\vol^v_g(t)$ is given by \eqnref{Vvg} and the integral in the
first expression is taken over $\Omega_w$, the annular portion of the
vessel filled with liquid.

In summary, equations \eqnref{rg}, \eqnref{volmelt} and
\eqnref{heat-Tvw1}--\eqnref{rhovg} represent a nonlinear system of
differential-algebraic equations for the nine unknown functions
$r(t)$, $p^v_w(t)$, $p^v_g(x,t)$, $\Tvwleft(x,t)$, $\Tvwright(x,t)$,
$T^v_g(x,t)$, $\rho^v_g(t)$, $c^v_g(t)$ and $\volmelt(t)$.  The vessel
and fiber solutions are coupled via matching conditions
\eqnref{volmelt}, \eqnref{heat-Tvw1-bc1} and
\eqnref{heat-Tvw1-bc2}.

\subsection{Osmotic effects}
\label{sec:osmosis}

Tyree and co-workers have argued that a pressurized gas bubble in the
xylem will dissolve entirely over a period of 12 hours or less, so that
some other pressure-generating mechanism must operate to sustain the
bubbles actually observed in xylem
sap~\cite{cirelli-etal-2008,tyree-1995,tyree-zimmermann-2002}.  Recent
experiments by Cirelli et al.~\cite{cirelli-etal-2008} suggest that the
fiber secondary wall is selectively-permeable, allowing water to pass through
but preventing the passage of larger molecules such as sucrose and
thereby generating a significant osmotic potential difference between
fibers and vessels.  This leads naturally to the hypothesis that osmotic
pressure might provide the extra mechanism needed to enhance flow of sap
from fiber to vessel and hence prevent bubbles in the fiber from totally
dissolving once the ice layer has completely melted.

Osmosis can be described mathematically by means of the Morse equation
that relates osmotic pressure $\Pi$ within a solution to the dissolved
solute concentration $c_s$ via $\Pi = c_s \gasconst T$.\ \ 
Following the approach in~\cite{Radu20101}, we apply this equation to
the sap solutions in fiber and vessel and hence obtain a modified
version of the melt volume equation \eqnref{volmelt}
\begin{gather*}
  \nfiber \dot{\volmelt} = 
  - \frac{\Kpit A}{\rho_w g W}\, \left[(p^v_w - p^f_g) - \gasconst
    T^f_w(R^f,t) \, (c^v_s-c^f_s)\right],
\end{gather*}
where $c^f_s$ and $c^v_s$ are the sucrose concentrations in the fiber
and vessel respectively.  According to Tyree and
Zimmermann~\cite{tyree-zimmermann-2002}, sucrose is present only in the
vessels, so that $c^f_s\equiv 0$ and
\begin{gather}
  \nfiber \dot{\volmelt} = 
  - \frac{\Kpit A}{\rho_w g W}\, \left[(p^v_w - p^f_g) - \gasconst
    T^f_w(R^f,t) c^v_s \right].
  \label{eq:volmelt-osmotic}
\end{gather}
This equation replaces \eqnref{volmelt}, but otherwise the governing
equations remain unchanged.  For the sake of simplicity, we assume
that the vessel sucrose concentration $c^v_s$ is a constant
which for a 2\%\ sucrose solution corresponds to taking $c^v_s = 0.02
\rho_w/M_s \approx 58.4\;\units{mol/m^3}$, where the molar mass of
sucrose is $M_s=0.3423\;\units{kg/mol}$.  This assumption can be justified
by arguing that ray parenchyma cells (which provide the bulk of the
sucrose to the tree vascular system) are so numerous in maples that they
should be capable of maintaining the sucrose concentration at a
relatively constant level.

\subsection{Non-dimensional variables}
\label{sec:nondim}

To simplify the governing equations, we reduce the number of parameters
in the problem by introducing the following change of variables:
\begin{gather}
  \begin{array}{ccc}
    x = \Rv \nondim{x}, & 
    s = \Rv \nondim{s}, &
    r = \Rv \nondim{r}, \\[0.2cm]
    t = \ndscale{t} \, \nondim{t}, & 
    \volmelt = \vol^v \nondim{\volmelt}, & 
    p = \ndscale{p} \, \nondim{p}, \\[0.2cm]
    \rho = \ndscale{\rho} \, \nondim{\rho}, &
    c = \frac{\ndscale{\rho}}{\molmass_g} \, \nondim{c}, & 
    T = T_c + (T_a-T_c) \nondim{T}, 
  \end{array}
  \label{eq:change-vars} 
\end{gather}
where an overbar denotes a dimensionless quantity.  The
melt volume is non-dimensionalized using 
\begin{gather}
  \vol^v = \pi \height^v (\Rv)^2, 
  \label{eq:vbarv}
\end{gather}
density and pressure scales are chosen based on initial
values as 
\begin{gather}
  \ndscale{\rho} = \rho^v_g(0) 
  \qquad \text{and} \qquad
  \ndscale{p} = p^v_g(0) = \frac{\ndscale{\rho} \gasconst T_c}{\molmass_g}, 
  \label{eq:rho-p-bar}
\end{gather}
and the time scale $\ndscale{t}$ will be specified shortly.

The above expressions are then substituted into the dimensional
equations from Section~\ref{sec:model}, after which we immediately
simplify notation by dropping asterisks.  The differential equations and
boundary conditions \eqnref{heat-Tfw} and
\eqnref{heat-Tvw1}--\eqnref{heat-Tvw2} governing the temperatures
$\nondim{T}^f_w(x,t)$, $\nondim{T}^v_{w1}(x,t)$,
$\nondim{T}^v_{w2}(x,t)$ and $\nondim{T}^v_g(x,t)$ become:
\begin{subequations}\label{eq:adim-Tfw1}
  \begin{alignat}{6}
    & \text{\emph{Fiber water:}} \qquad&
    \dxtwo \nondim{T}_w^f &= 0 \qquad\qquad\quad\;\;\;
    & \text{for all} \; \nondim{x} &\in \left(s_{iw}, \delta\right),
    & \qquad\qquad\qquad\;\; \label{eq:adim-Tfw1-pde}\\
    & & \nondim{T}_w^f &= 0 
    \qquad & \text{at} \; \nondim{x} &= s_{iw}, & 
    \label{eq:adim-Tfw1-bc}
  \end{alignat}
\end{subequations}
\vspace*{-0.5cm}
\begin{subequations}\label{eq:adim-Tvw1}
  \begin{alignat}{6}
    & \text{\emph{Vessel water (left):}} \qquad &
    \dxtwo \nondim{T}^v_{w1} &= 0 \qquad\qquad\quad\;\;\; &\text{for all} \; 
    \nondim{x} &\in \left(\delta, \delta+1-\nondim{r}\right), & \qquad\quad\;\;\;
    \label{eq:adim-Tvw1-pde}\\
    & & \nondim{T}^v_{w1} &= \nondim{T}^f_w \qquad &
    \text{at} \; \nondim{x} &= \delta, & \label{eq:adim-Tvw1-bc1}\\
    & & \dxone \nondim{T}^v_{w1} &= \dxone \nondim{T}^f_w 
    \qquad\quad &\text{at} \; \nondim{x} &= \delta, & 
    \label{eq:adim-Tvw1-bc2}\\
    & & \nondim{T}^v_{w1} &= \nondim{T}^v_g \qquad &\text{at} \; 
    \nondim{x} &= \delta+1-\nondim{r}, & 
    \label{eq:adim-Tvw1-bc3}\\
    & & \dxone \nondim{T}^v_{w1} &= \eta \dxone \nondim{T}^v_g  \qquad &\text{at} \;
    \nondim{x} &= \delta+1-\nondim{r}, & 
    \label{eq:adim-Tvw1-bc4}
  \end{alignat}
\end{subequations}
\vspace*{-0.5cm}
\begin{subequations}\label{eq:adim-Tvg}
  \begin{alignat}{5}
    & \text{\emph{Vessel gas bubble:}} \qquad &
    \dxtwo \nondim{T}^v_g &= 0 \qquad\qquad\quad\;\; &\text{for all}\; 
    \nondim{x} &\in \left(\delta+1-\nondim{r}, \delta+1+\nondim{r}\right), 
    \label{eq:adim-Tvg-pde}\\ 
    & & \nondim{T}^v_g &= \nondim{T}^v_{w2}
    \qquad &\text{at} \; \nondim{x} &= \delta+1+\nondim{r}, \label{eq:adim-Tvg-bc1}\\
    & & \eta \dxone \nondim{T}^v_g &= \dxone \nondim{T}^v_{w2}
    \qquad &\text{at} \; \nondim{x} &= \delta+1+\nondim{r},
    \label{eq:adim-Tvg-bc2}
  \end{alignat}
\end{subequations}
\vspace*{-0.5cm}
\begin{subequations}\label{eq:adim-Tvw2}
  \begin{alignat}{5}
    & \text{\emph{Vessel water (right):}} \qquad &
    \dxtwo \nondim{T}^v_{w2} &=0 &\text{for all}\; 
    \nondim{x} &\in \left(\delta+1+\nondim{r}, \delta+2\right), \qquad\;\; 
    \label{eq:adim-Tvw2-pde}\\
    & & \dxone \nondim{T}^v_{w2} &= \Biot(1-\nondim{T}^v_{w2})
    \quad &\text{at} \; \nondim{x} &= \delta+2.
    \label{eq:adim-Tvw2-bc1}
  \end{alignat}
\end{subequations}
The dimensionless parameter $\Biot=h \Rv/k_w$ is the Biot number.

Next are five algebraic equations for $\nondim{p}^f_g(\nondim{t})$,
$\nondim{p}^v_g(\nondim{x},\nondim{t})$, $\nondim{p}^v_w(\nondim{t})$,
$\nondim{c}^v_g(\nondim{t})$ and $\nondim{\rho}^v_g(\nondim{t})$:
\begin{gather}
  \nondim{p}^f_g = \nondim{p}^f_g(0) \, \left(
    \frac{\nondim{s}_{gi}(0)}{\nondim{s}_{gi}}
  \right)^2,  \label{eq:pfg2-nondim}\\ 
  \nondim{p}^v_g = \nondim{\rho}^v_g (1+\theta
  \nondim{T}^v_g), \label{eq:pvg-nondim}\\ 
  \nondim{p}^v_w = \nondim{p}^v_g -
  \frac{\omega}{\nondim{r}}, \label{eq:surftens-nondim}\\  
  \nondim{c}^v_g = \henry \nondim{\rho}^v_g,\label{eq:henry-nondim}\\
  \nondim{\rho}^v_g = \frac{(\nondim{r}(0))^2 - \nondim{c}^v_g (1 -
    \nondim{r}^2)}{\nondim{r}^2}.\label{eq:rhovg-nondim} 
\end{gather}
The non-dimensionalized Stefan condition \eqnref{stefan} becomes
\begin{gather}
  \dot{\nondim{s}}_{iw} = - \dxone \nondim{T}^f_w 
  \qquad \text{at} \; \nondim{x} = \nondim{s}_{iw},
  \label{eq:stefan-nondim} 
\end{gather}
where we have chosen the time scale 
\begin{gather}
  \ndscale{t} = \frac{\stefan\rho_w (\Rv)^2}{k_w (T_a-T_c)}, 
  \label{eq:tbar}
\end{gather}
so as to eliminate the coefficient in the Stefan condition.  This is
clearly an appropriate time scale for our problem because the melting
process governs the dynamics of the ice layer and hence also the water
transport from fiber to vessel.  Using parameter values from
Table~\ref{tab:params}, we find that a typical value of the time scale
is $\ndscale{t}\approx 46.1\;\units{s}$.  Finally, the remaining
differential equations \eqnref{sgi-siw}, \eqnref{rg} and
\eqnref{volmelt} for the quantities $\nondim{s}_{gi}(\nondim{t})$,
$\nondim{r}(\nondim{t})$ and $\nondim{\volmelt}(\nondim{t})$ become
\begin{gather}
  2\beta \nondim{s}_{gi}\dot{\nondim{s}}_{gi} + 2(1-\beta)
  \nondim{s}_{iw}\dot{\nondim{s}}_{iw} - \gamma \dot{\nondim{\volmelt}}
  = 0, \label{eq:sgi-siw-nondim} \\ 
  2 \nondim{r} \dot{\nondim{r}} = -\nfiber
  \dot{\nondim{\volmelt}}, \label{eq:rg-nondim} \\ 
  \nfiber \dot{\nondim{\volmelt}} = -\alpha \left[ \nondim{p}^v_w -
    \nondim{p}^f_g - (1+\theta \nondim{T}^f_w(\delta,\nondim{t}))
    \nondim{c}^v_s \right].  \label{eq:n-udot-nondim}
\end{gather}

\leavethisout{
  \begin{remark}
    Note that the time scale $\ndscale{t}$ may be written as
    \begin{gather*}
      \ndscale{t} = \frac{\stefan}{C_w(T_a-T_c)} \, 
      \frac{C_w\rho_w (\Rv)^2}{k_w} 
      = \frac{1}{St} \, \frac{(R^v)^2}{\alpha_w}
    \end{gather*}
    where $C_p$ is the specific heat of water, $St=k_w(T_a-T_c)/\stefan$ is
    the Stefan number, and $\alpha_w=k_w/(C_w\rho_w )$ is the thermal
    diffusivity.  This clearly indicates that  $\ndscale{t}$ is a time scale
    integrally associated with heat transport in the water and phase change
    at the interface between water and ice.
  \end{remark}
}

\begin{table}
  \centering
  \footnotesize
  \caption{A list of dimensionless parameters and their numerical values.}
  \label{tab:dimless-params}
  \renewcommand{\extrarowheight}{5pt}
  \begin{tabular}{|ccc|}\hline
    Symbol & Definition & Value \\\hline
    $\Biot$      & $\frac{h \Rv}{k_w}$            & $3.45\times 10^{-4}$ \\
    $\alpha$     & $\frac{\Kpit A \ndscale{p} \ndscale{t}}{\rho_w g W \vol^v}$ & 0.258 \\
    $\beta$      & $\frac{\rho_i}{\rho_w}$        & 0.917 \\
    $\gamma$     & $\frac{\height^v}{\height^f}$  & 0.500 \\
    $\delta$     & $\frac{\Rf}{\Rv}$              & 0.175 \\
    $\eta$       & $\frac{k_g}{k_w}$              & 0.0419 \\
    $\theta$     & $\frac{T_a-T_c}{T_c}$          & $1.83\times 10^{-5}$ \\
    $\omega$     & $\frac{\sigma}{\ndscale{p}\Rv}$& 0.0374 \\
    \hline
  \end{tabular}
\end{table}

\subsection{Onset of gas dissolution in fiber after ice melts
  completely} 
\label{sec:ice-melted}

As the ice in the fiber melts, we eventually reach a critical time when
the ice layer disappears and the two interfaces $\nondim{s}_{gi}$ and
$\nondim{s}_{iw}$ merge into a single gas--water interface whose
position we denote by $\nondim{x} = \nondim{s}(\nondim{t})$.  Similar to
the gas/water interface in the vessel, the dynamics of $\nondim{s}$ are
governed by dissolution of gas within the fiber water and changes in
pressure/volume of the gas bubble.  Assuming as before that the gas in
the fiber diffuses much faster than it dissolves, we again employ a
quasi-steady approximation in which the dissolved gas concentration is
assumed constant at any given time.  Therefore, we replace
\eqnref{pfg2-nondim} with
\begin{gather}
  \nondim{p}^f_g = \nondim{\rho}^f_g (1 + \theta \nondim{T}^f_g) =
  \nondim{\rho}^f_g, 
  \label{eq:pfg-nondim-noice}
\end{gather}
and eliminate \eqnref{stefan-nondim} and \eqnref{sgi-siw-nondim} in
favor of the single equation
\begin{gather}
  2\nondim{s}\dot{\nondim{s}} = \gamma
  \dot{\nondim{\volmelt}}. \label{eq:s-nondim-noice} 
\end{gather}
These last two equations are already dimensionless and are analogous to
\eqnref{pvg-nondim} and \eqnref{rg-nondim} for the vessel.  The modified
system has as additional unknowns the water pressure $\nondim{p}^f_w$,
dissolved gas concentration $\nondim{c}^f_g$, and gas density
$\nondim{\rho}^f_g$, which are governed by the following analogues of
the vessel equations \eqnref{rhovg-nondim} and \eqnref{henry-nondim}:
\begin{gather}
  \nondim{p}^f_w = \nondim{p}^f_g - \frac{\omega}{\nondim{s}}
  \label{eq:surftens-nondim-noice}\\
  \nondim{c}^f_g = H \nondim{\rho}^f_g, 
  \label{eq:henry-nondim-noice}\\
  \nondim{\rho}^f_g = \frac{\xi (\nondim{s}_{gi}(0))^2 -
    \nondim{c}^f_g(1-\nondim{s}^2)}{\nondim{s}^2}, 
  \label{eq:rhofg-nondim-noice}
\end{gather}
where $\xi=\rho^f_g(0)/\rho^v_g(0)$ is a dimensionless parameter
corresponding to the ratio of initial gas densities.  Finally, without
the ice layer to transmit the gas pressure directly to the fiber water
located on the other side, Darcy's law \eqnref{n-udot-nondim} must be
modified so as to replace the fiber gas pressure by $\nondim{p}^f_w$:
\begin{gather}
  \nfiber \dot{\nondim{\volmelt}} = -\alpha \left[ \nondim{p}^v_w -
    \nondim{p}^f_w - (1 + \theta \nondim{T}^f_w(\delta,\nondim{t}))
    \nondim{c}^v_s \right]. 
  \label{eq:n-udot-nondim-noice}
\end{gather}

\subsection{Analytical solution for temperatures}
\label{sec:Texact}

Because the quasi-steady temperature equations
\eqnref{adim-Tfw1}--\eqnref{adim-Tvw2} have such a simple form, they can
be solved analytically.  In fact, the temperature is a linear function
of $\nondim{x}$ in each sub-domain, and applying the corresponding
boundary and matching conditions leads to the following explicit
solutions in terms of $\nondim{x}$, $\nondim{t}$ and the non-dimensional
parameters:
\begin{align}
  \nondim{T}^f_w(\nondim{x},\nondim{t}) \equiv
  \nondim{T}^v_{w1}(\nondim{x},\nondim{t}) &= \frac{\eta (\nondim{x}-
    \nondim{s}_{iw}(\nondim{t}))}{\eta (\delta+2+\Biot^{-1}) +
    2(1-\eta) \nondim{r}(\nondim{t}) - \eta
    \nondim{s}_{iw}(\nondim{t})},
  \label{eq:Tfw-Tvw1-exact}\\ 
  \nondim{T}^v_g(\nondim{x},\nondim{t}) &= \frac{\nondim{x}-(1-\eta)
    \left( \delta + 1 - \nondim{r}(\nondim{t})\right) - \eta
    \nondim{s}_{iw}(\nondim{t})}{\eta(\delta+2+\Biot^{-1}) +
    2(1-\eta) \nondim{r}(\nondim{t}) - \eta
    \nondim{s}_{iw}(\nondim{t})},
  \label{eq:Tvg-exact} \\
  \nondim{T}^v_{w2}(\nondim{x},\nondim{t}) &= \frac{\eta
    \nondim{x}+2(1-\eta)\nondim{r}(\nondim{t}) - \eta
    \nondim{s}_{iw}(\nondim{t})}{\eta(\delta+2+\Biot^{-1}) +
    2(1-\eta) \nondim{r}(\nondim{t}) - \eta
    \nondim{s}_{iw}(\nondim{t})}. 
  \label{eq:Tvw2-exact}
\end{align}
The temperature solution on the whole domain may then be summarized as
\begin{gather}
  \nondim{T}(\nondim{x},\nondim{t}) = 
  \begin{cases}    
    0,           & \text{if $\nondim{x} \in [0, \, \nondim{s}_{iw}]$,}\\
    \nondim{T}^f_w(\nondim{x}, \nondim{t}),    & \text{if $\nondim{x} \in
      (\nondim{s}_{iw}, \, \delta+1-\nondim{r}]$,}\\
    \nondim{T}^v_g(\nondim{x},\nondim{t}),    & \text{if $\nondim{x} \in
      (\delta+1-\nondim{r}, \, \delta+1+\nondim{r}]$,}\\ 
    \nondim{T}^v_{w2}(\nondim{x},\nondim{t}), & \text{if $\nondim{x} \in
      (\delta+1+\nondim{r}, \, \delta+2]$.} 
  \end{cases}
  \label{eq:T-summary}
\end{gather}
Note that the matching conditions on temperature and heat flux at the
vessel wall imply that the functions $\nondim{T}^f_w$ and
$\nondim{T}^v_{w1}$ are identical.

\subsection{Initial conditions}
\label{sec:ic}

We choose a ``base case'' for our simulations in which the initial
conditions are taken consistent with the typical scenario described by
Tyree~\cite{tyree-1995}:
\begin{itemize}
\item $p^v_g(0) = 100\;\units{kPa}$: gas in
  the vessel is initially at atmospheric pressure.
\item $p^f_g(0) = 200\;\units{kPa}$: gas in the fiber is at twice the
  pressure in the vessel owing to compression by the frozen sap.  Based
  on Tyree's data, we expect that this value is an upper bound, so that
  $p^f_g(0)$ actually lies in the range $[100,200]\;\units{kPa}$.
\item $\nondim{s}_{iw}(0) = \delta$, $\nondim{s}_{gi}(0) =
  \delta/\sqrt{2}\approx 0.7\delta$ and $\nondim{\volmelt}(0) = 0$:
  corresponding to a fiber gas bubble that is initially half the total
  fiber volume, with the other half filled by ice.  This initial state
  is established during the previous fall/winter freeze, when gas-filled
  fibers are at atmospheric pressure in equilibrium with the vessels.
  As the temperature falls below freezing, ice forms on the inside of
  the fiber wall and the gas pressure increases by a factor inversely
  proportional to the ice thickness according to equation
  \eqnref{pfg2-nondim}; hence, the value
  $\nondim{s}_{gi}(0)=0.7\delta$ is consistent with the initial gas
  pressures chosen above.
\item $\nondim{r}(0)$: is the quantity in this study with the greatest
  uncertainty since we haven't yet found data on gas concentration or
  bubble size in the vessels; hence, this is a natural parameter to vary
  in our simulations.  We expect that the initial vessel volume fraction
  taken up by gas is significantly less than in the fiber, otherwise a
  large embolus is likely to remain in the vessel and hinder sap
  transport once transpiration recommences.
\end{itemize}
Initial values are not required for the temperatures, although
they can be obtained by substituting the above values into
\eqnref{Tfw-Tvw1-exact}--\eqnref{T-summary} and the algebraic relations
in Section~\ref{sec:nondim}.

\section{Numerical simulations}
\label{sec:numerics}

Owing to the nonlinear coupling in the governing equations derived in
the previous section, numerical simulations are required in order to
obtain solutions of the full system.  We employ a straightforward
numerical approach in which the four ordinary differential
equations~\eqnref{stefan-nondim} and
\eqnref{sgi-siw-nondim}--\eqnref{n-udot-nondim} are solved for
$\nondim{s}_{gi}$, $\nondim{s}_{iw}$, $\nondim{r}$ and
$\nondim{\volmelt}$ using the stiff solver \texttt{ode15s} in \matlab,
with error tolerances $\text{RelTol}=10^{-10}$ and
$\text{AbsTol}=10^{-8}$.  Temperatures are calculated using the
analytical expressions in \eqnref{Tfw-Tvw1-exact}--\eqnref{Tvw2-exact},
and the system is closed using the algebraic equations
\eqnref{pfg2-nondim}--\eqnref{rhovg-nondim}.

\begin{figure}[bthp]
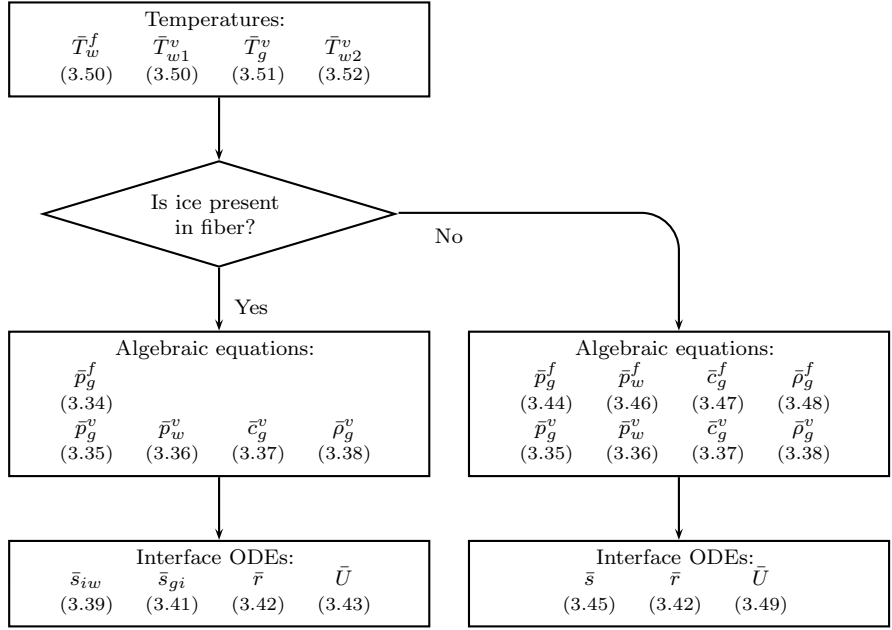

  \centering
  \footnotesize
  \begin{psmatrix}[rowsep=0.8,colsep=0.5]
    \psframebox{%
      \begin{minipage}{0.35\linewidth}
        \begin{center}
          Temperatures:\\
          \begin{tabular}{cccc}
            $\nondim{T}^f_w$ & $\nondim{T}^v_{w1}$ & $\nondim{T}^v_g$
            & $\nondim{T}^v_{w2}$ \\
            \scriptsize \eqnref{Tfw-Tvw1-exact}
            & \scriptsize \eqnref{Tfw-Tvw1-exact}
            & \scriptsize \eqnref{Tvg-exact}
            & \scriptsize \eqnref{Tvw2-exact}
          \end{tabular}
        \end{center}
      \end{minipage}
    } & \\
    \psdiabox{%
      \begin{minipage}{0.14\linewidth}
        \begin{center}
          Is ice present\\
          in fiber?
        \end{center}
      \end{minipage}
    }\\
    \psframebox{%
      \begin{minipage}{0.35\linewidth}
        \begin{center}
          Algebraic equations:\\
          \begin{tabular}{cccc}
            $\nondim{p}^f_g$ \\
            \scriptsize \eqnref{pfg2-nondim} \\
            $\nondim{p}^v_g$ & $\nondim{p}^v_w$
            & $\nondim{c}^v_g$ & $\nondim{\rho}^v_g$ \\
            \scriptsize \eqnref{pvg-nondim}
            & \scriptsize \eqnref{surftens-nondim}
            & \scriptsize \eqnref{henry-nondim}
            & \scriptsize \eqnref{rhovg-nondim}
          \end{tabular}
        \end{center}
      \end{minipage}
    } &
    \psframebox{%
      \begin{minipage}{0.35\linewidth}
        \begin{center}
          Algebraic equations:\\
          \begin{tabular}{cccc}
            $\nondim{p}^f_g$ & $\nondim{p}^f_w$ & 
            $\nondim{c}^f_g$ & $\nondim{\rho}^f_g$ \\
            \scriptsize \eqnref{pfg-nondim-noice}
            & \scriptsize \eqnref{surftens-nondim-noice}
            & \scriptsize \eqnref{henry-nondim-noice}
            & \scriptsize \eqnref{rhofg-nondim-noice}\\
            $\nondim{p}^v_g$ & $\nondim{p}^v_w$ & 
            $\nondim{c}^v_g$ & $\nondim{\rho}^v_g$ \\
            \scriptsize \eqnref{pvg-nondim}
            & \scriptsize \eqnref{surftens-nondim}
            & \scriptsize \eqnref{henry-nondim}
            & \scriptsize \eqnref{rhovg-nondim}
          \end{tabular}
        \end{center}
      \end{minipage}
    } \\
    \psframebox{%
      \begin{minipage}{0.35\linewidth}
        \begin{center}
          Interface ODEs:\\
          \begin{tabular}{cccc}
            $\nondim{s}_{iw}$ & $\nondim{s}_{gi}$ & 
            $\nondim{r}$  & $\nondim{U}$ \\
            \scriptsize \eqnref{stefan-nondim}
            & \scriptsize \eqnref{sgi-siw-nondim}
            & \scriptsize \eqnref{rg-nondim}
            & \scriptsize \eqnref{n-udot-nondim}\\
          \end{tabular}
        \end{center}
      \end{minipage}
    } 
    & 
    \psframebox{%
      \begin{minipage}{0.35\linewidth}
        \begin{center}
          Interface ODEs:\\
          \begin{tabular}{ccc}
            $\nondim{s}$ & $\nondim{r}$  & $\nondim{U}$ \\
            \scriptsize \eqnref{s-nondim-noice}
            & \scriptsize \eqnref{rg-nondim}
            & \scriptsize \eqnref{n-udot-nondim-noice} 
          \end{tabular}
        \end{center}
      \end{minipage}
    } 
    \ncline{->}{1,1}{2,1}
    \ncline{->}{2,1}{3,1}>{Yes}
    \ncangle[angleB=90,armB=0,linearc=0.5]{->}{2,1}{3,2}_{No}
    \ncline{->}{3,1}{4,1}
    \ncline{->}{3,2}{4,2}
  \end{psmatrix}
  \caption{List of variables and governing equations depending
    on whether or not ice is present in the fiber.}
  \label{fig:algo}
\end{figure}

The above procedure applies when ice is present in the fiber.  In order
to permit \matlab\ to modify the equations automatically once the ice
disappears, the event detection feature in {\tt ode15s} is used to
determine the instant when $\nondim{s}_{gi} = \nondim{s}_{iw}$ and
$\vol^f_i\rightarrow 0$; at this time, the simulation is halted and the
calculation is restarted with the equations modified as described in
Section~\ref{sec:ice-melted}.  To assist in understanding which
equations are solved under which circumstances, we have provided in
Figure~\ref{fig:algo} a graphical summary of the equations and variables
in the two cases.

In the remainder of this section, we perform a series of simulations
with increasing model complexity, first studying the dynamics of gas
dissolution and flow through the fiber wall when the pressurized fibers
contain no ice, then introducing osmotic effects, and finally
incorporating the ice layer and phase change.  Even though the code is
written in dimensionless variables, all parameters and results from this
point on are converted to dimensional variables using equations
\eqnref{change-vars}.

\subsection{Case~1: Dissolution and porous transport only (no ice, no osmosis)}
\label{sec:sims-case1}

We begin by considering the simple situation in which there is no
sucrose in the vessel and no ice in the fiber (and hence, no phase
change).  As a result, the sap dynamics are governed primarily by two
effects: gas dissolution and the flow of sap driven by pressure exchange
between gas bubbles in the fiber and vessel.  We note that temperature
variations are relatively small and so they have a negligible effect on
the solution.

Gas bubble dissolution has been studied extensively in the context of
sap flow by several
authors~\cite{konrad-rothnebelsick-2003,shen-etal-2003,yang-tyree-1992},\  
who concluded based on experimental evidence and simple physical
arguments that pressurized gas bubbles in xylem sap should dissolve
completely over a time period ranging anywhere from 20 minutes to
several hours.  These results are partially supported by the
mathematical study of bubble dissolution by Keller~\cite{keller-1964},
who proved that for an under-saturated solution (such as xylem sap) that
fills an \emph{unbounded domain}, the only stable equilibrium solution
is one in which the gas bubble shrinks and eventually disappears.  An
analogous situation was studied by Yang and
Tyree~\cite{yang-tyree-1992}, who performed experiments on excised maple
branches \emph{open to the air} and developed a corresponding
mathematical model in which the pressure is maintained at atmospheric
conditions at the open boundary.  Their experimental and numerical
results are consistent with the complete dissolution of gas bubbles in
finite time.\ \

In our study of sap exudation however, we are interested in the behavior
of xylem sap within an intact maple tree stem that is not exposed to the
open air; consequently, the open boundary conditions treated by
Yang-Tyree above are not applicable here.  Fortunately, Keller also
considered a pressurized gas bubble suspended in a \emph{closed fluid
  container}~\cite{keller-1964} for which he proved that the bubble will
still shrink in size as it dissolves, but in this case will reach an
equilibrium state having non-zero radius.  This behavior is contrary to
that observed above in experiments on excised maple branches, and so
bears careful consideration.

\begin{figure}[bthp]
  \begin{center}
    \footnotesize
    \begin{tabular}{ccc}
      (a) Water pressures & (b) Gas pressures & (c) Gas bubble radii\\
      \myfig{0.31}{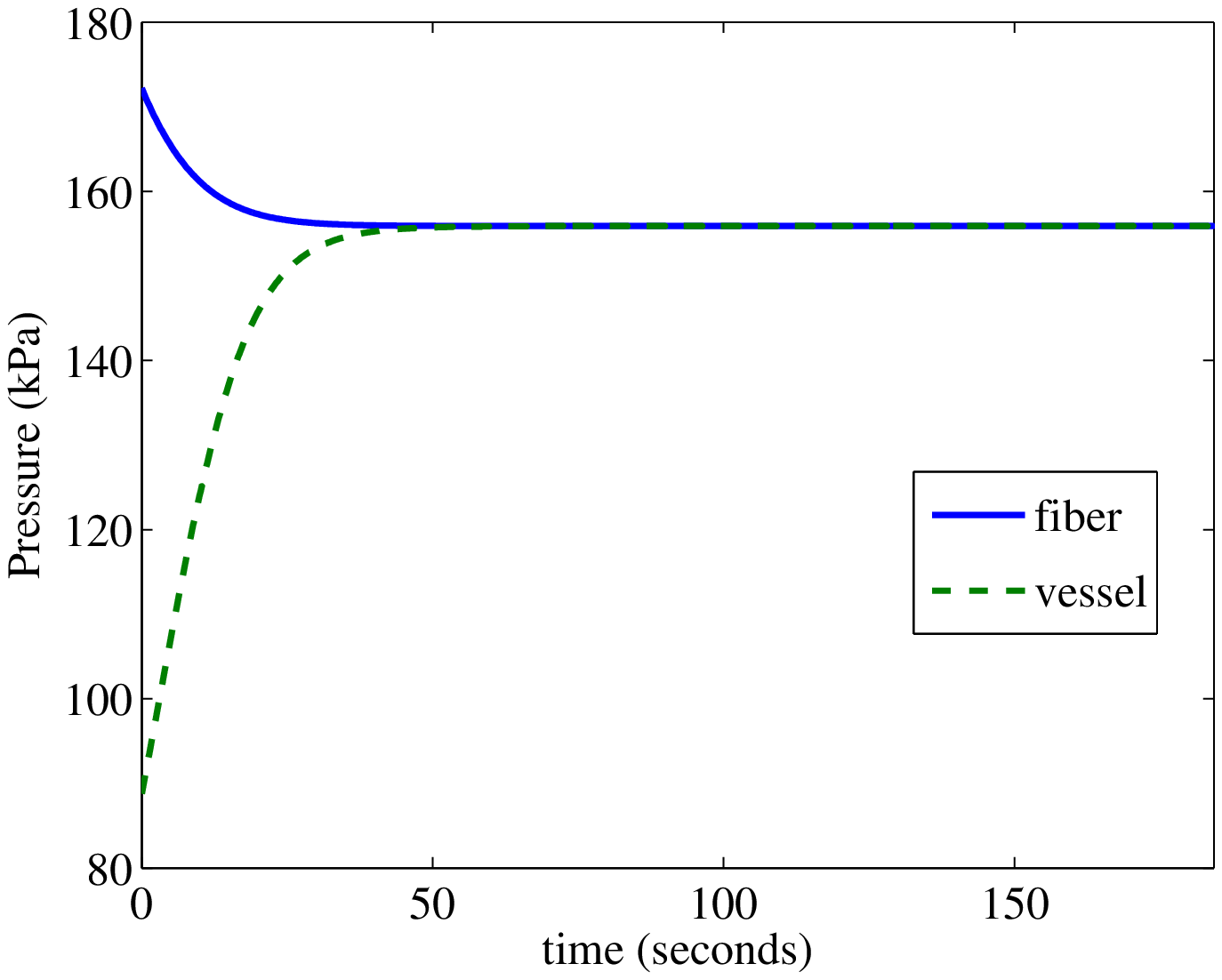}
      & 
      \myfig{0.31}{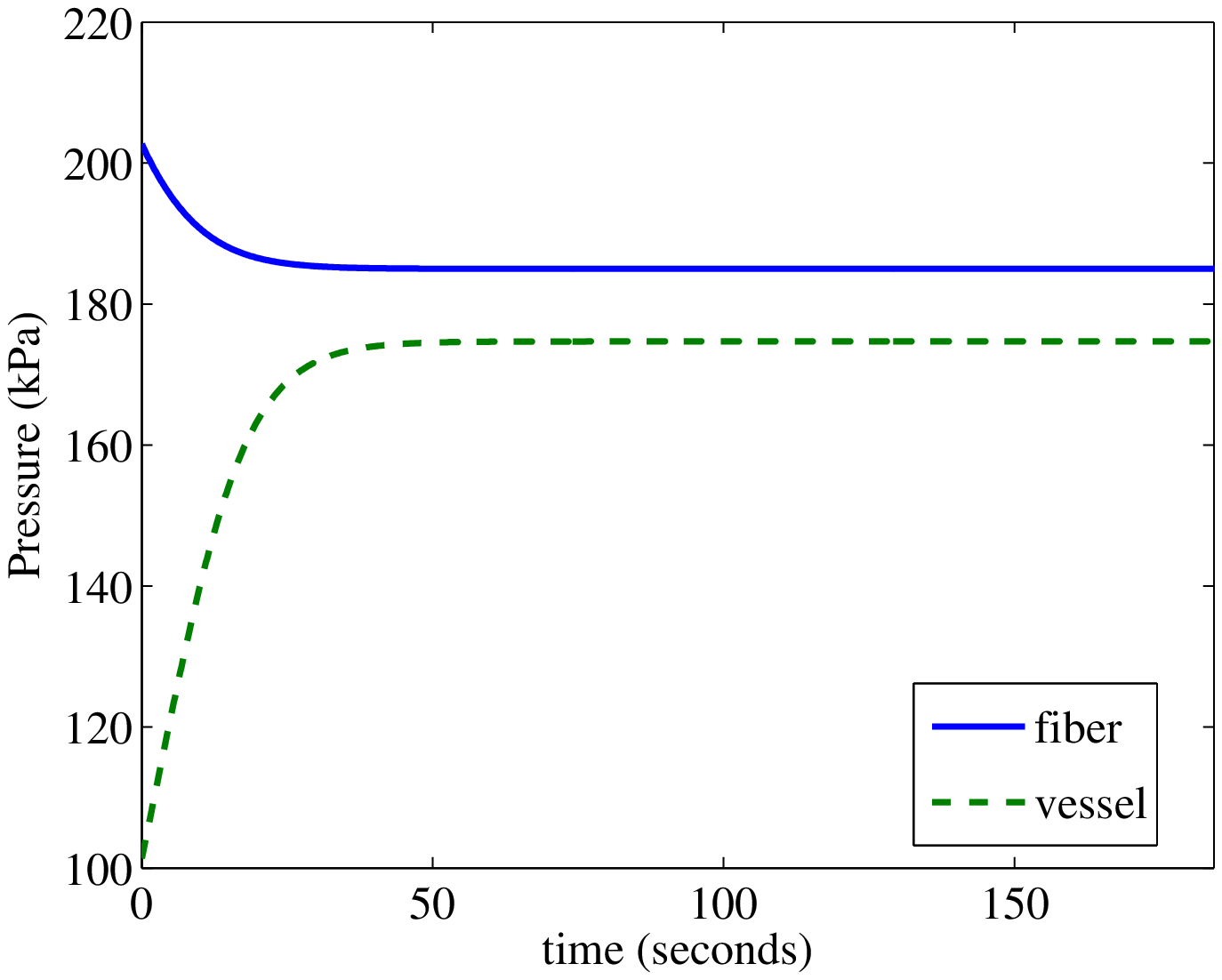}
      & 
      \myfig{0.31}{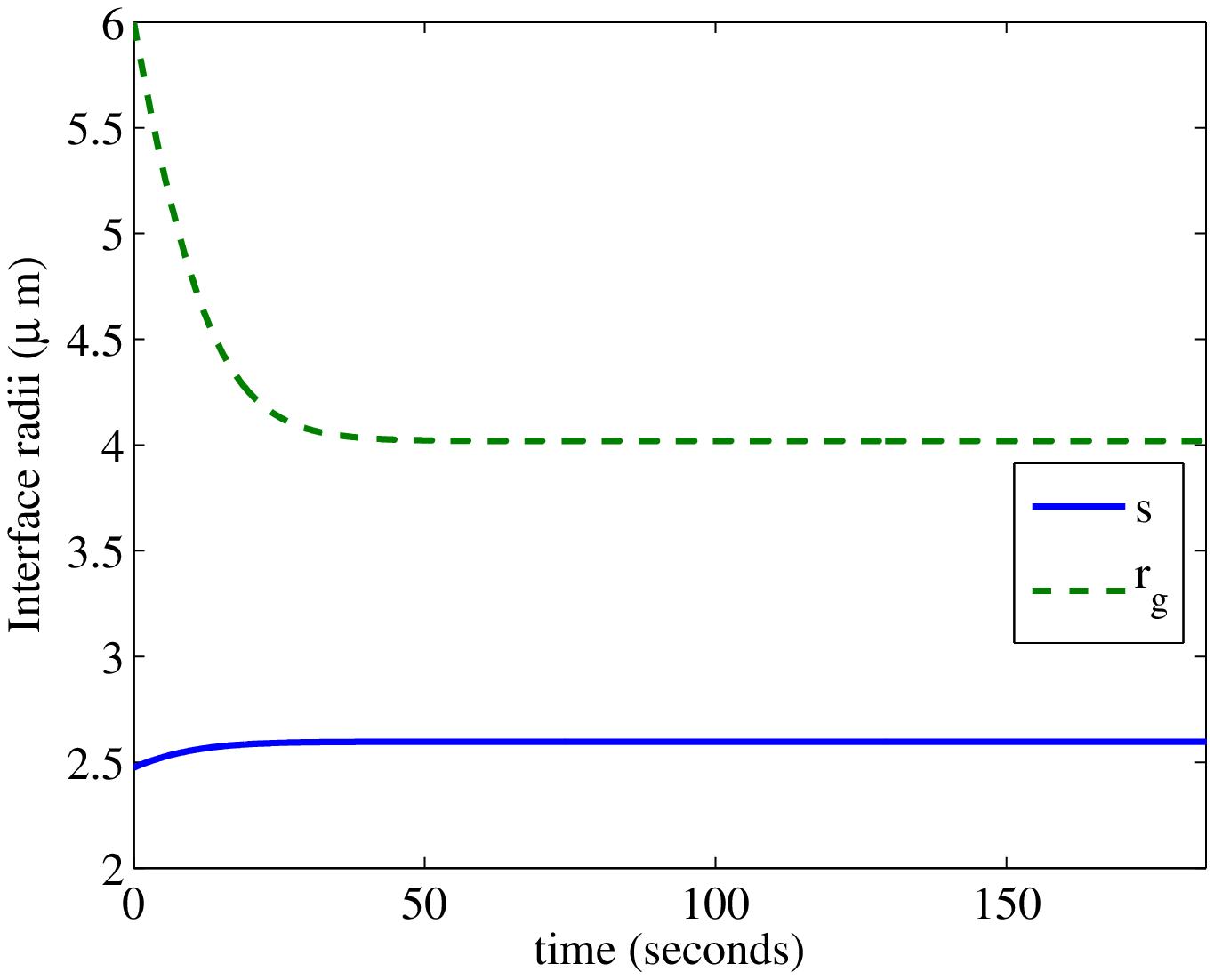}
      \\
      & (d) Temperatures & \\
      & 
      \myfig{0.31}{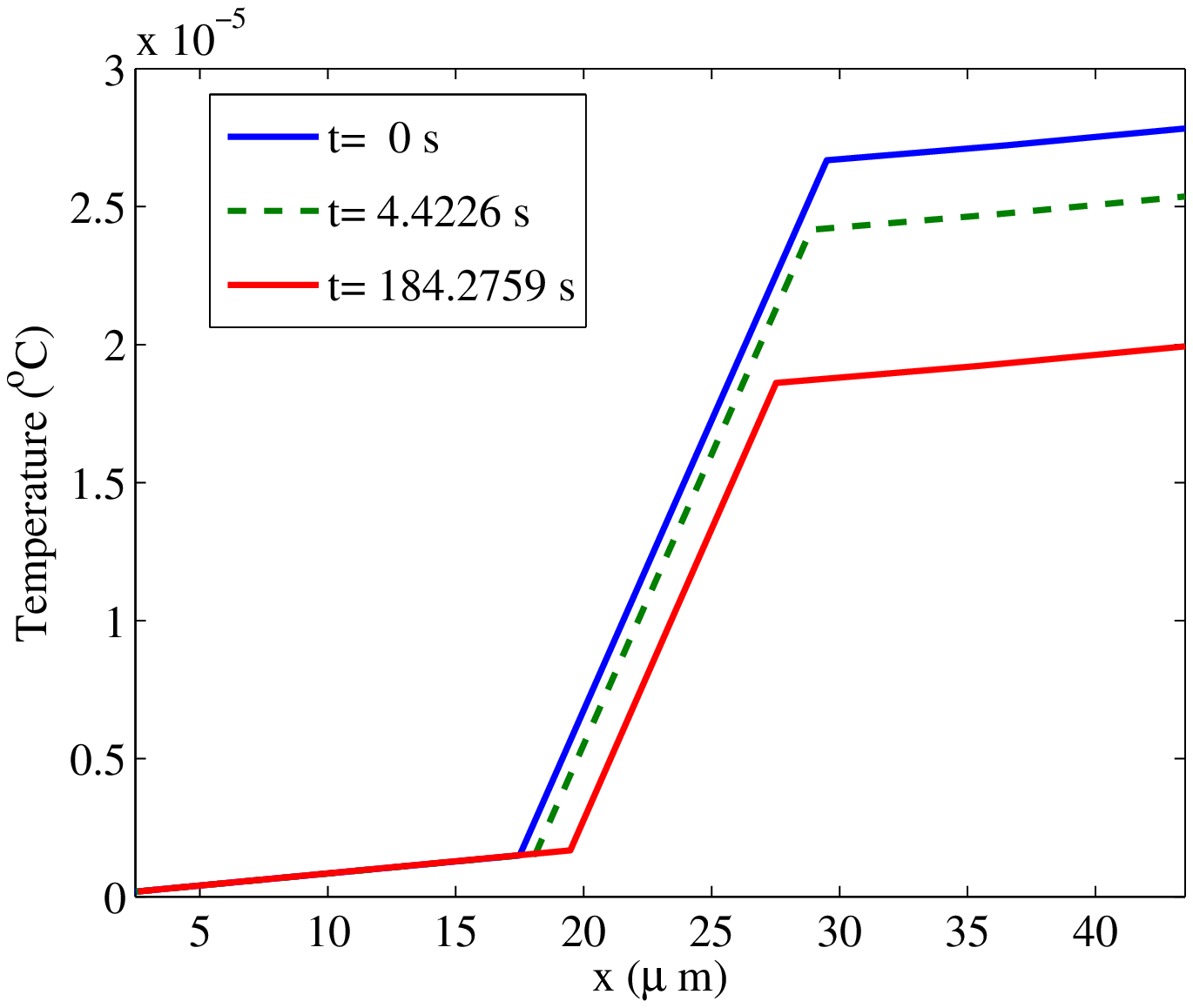}
      & 
    \end{tabular}
    \caption{Case~1 -- base case: No ice, initial radii $s(0) = 0.7\Rf$,
      $r(0)=0.3 \Rv$, and pressure $p^f_g(0)=200\;\units{kPa}$.}
    \label{fig:case1a}
  \end{center}
\end{figure}

To investigate the ability of our model to capture Keller's results for
a closed container, we performed several simulations with bubbles of
various sizes in both fiber and vessel.  We start with a scenario in
which the fiber bubble takes up 50\%\ of the fiber volume and the vessel
bubble only 10\%, corresponding to $s_{gi}(0) = \Rf/\sqrt{2} \approx 0.7
\Rf$ and $r(0)=\Rv/\sqrt{10}\approx 0.3\Rv$.  Based on the discussion of
initial conditions in Section~\ref{sec:ic}, the gas in the vessel is
initially at atmospheric pressure (100~\units{kPa}) while that in the
fiber is twice as large (200~\units{kPa}).  The resulting dynamics for
the pressures and bubble radii are shown in Figure~\ref{fig:case1a},
from which we observe that the fiber bubble grows in size while the
vessel bubble shrinks, until both reach an equilibrium state at a time
of roughly $t=40$ seconds.  Figure~\ref{fig:case1a}(a) shows shows that
the system approaches an equilibrium state in which the water pressure
in the two compartments is equal (at approximately 155~\units{kPa}).
This is clearly consistent with \eqnref{n-udot-nondim} since there can
be no flux across the porous cell wall at the equilibrium state, which
requires that $\nondim{p}^f_w = \nondim{p}^g_w$ (remembering also that
$\nondim{c}^v_s=0$).  The plots show that the gas and water pressures
vary inversely with the radii as expected, and the piecewise linear
dependence of temperature on position in equations
\eqnref{Tfw-Tvw1-exact}--\eqnref{T-summary} is evident in
Figure~\ref{fig:case1a}(d).

The plot of gas pressure shows a decrease in $p^f_g$ over time as the
fiber bubble expands and a corresponding decrease in $p^v_g$ as the
vessel bubble is compressed.  This behavior is consistent with Keller's
results for bubbles in a closed system, and also captures the transfer
of pressure from fiber to vessel that we would expect in sap exudation.
It is important to note that the increase in vessel water pressure from
roughly 90 to 155\;\units{kPa} (see Figure~\ref{fig:case1a}(a)) is close
to the upper limit of the range of 30--60~\units{kPa} that is to be
expected for positive pressures in maple according to~\cite{tyree-1983}.
Finally, we note that the magnitude of the pressure change in the fiber
arising from dissolution of gas is only 5~\units{kPa}, which is small
relative to the other pressure differences between fiber and vessel;
consequently, dissolution has only a small effect on sap flow over this
time scale.

\begin{figure}[bthp]
  \centering
    \footnotesize
    \begin{tabular}{ccc}
      (a) Vessel bubble radius & & (b) Vessel water pressure \\
      \myfig{0.42}{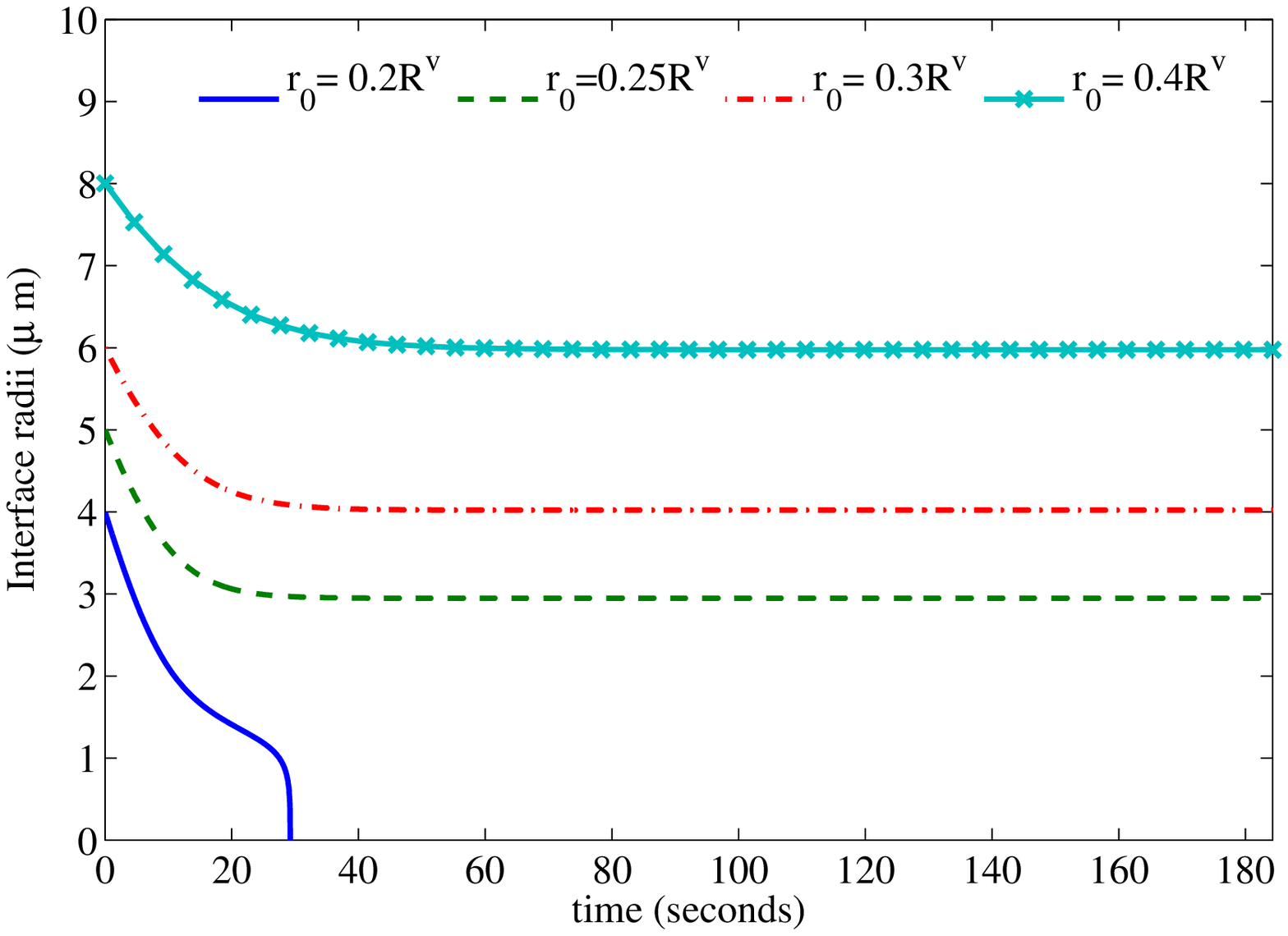}
      & & 
      \myfig{0.42}{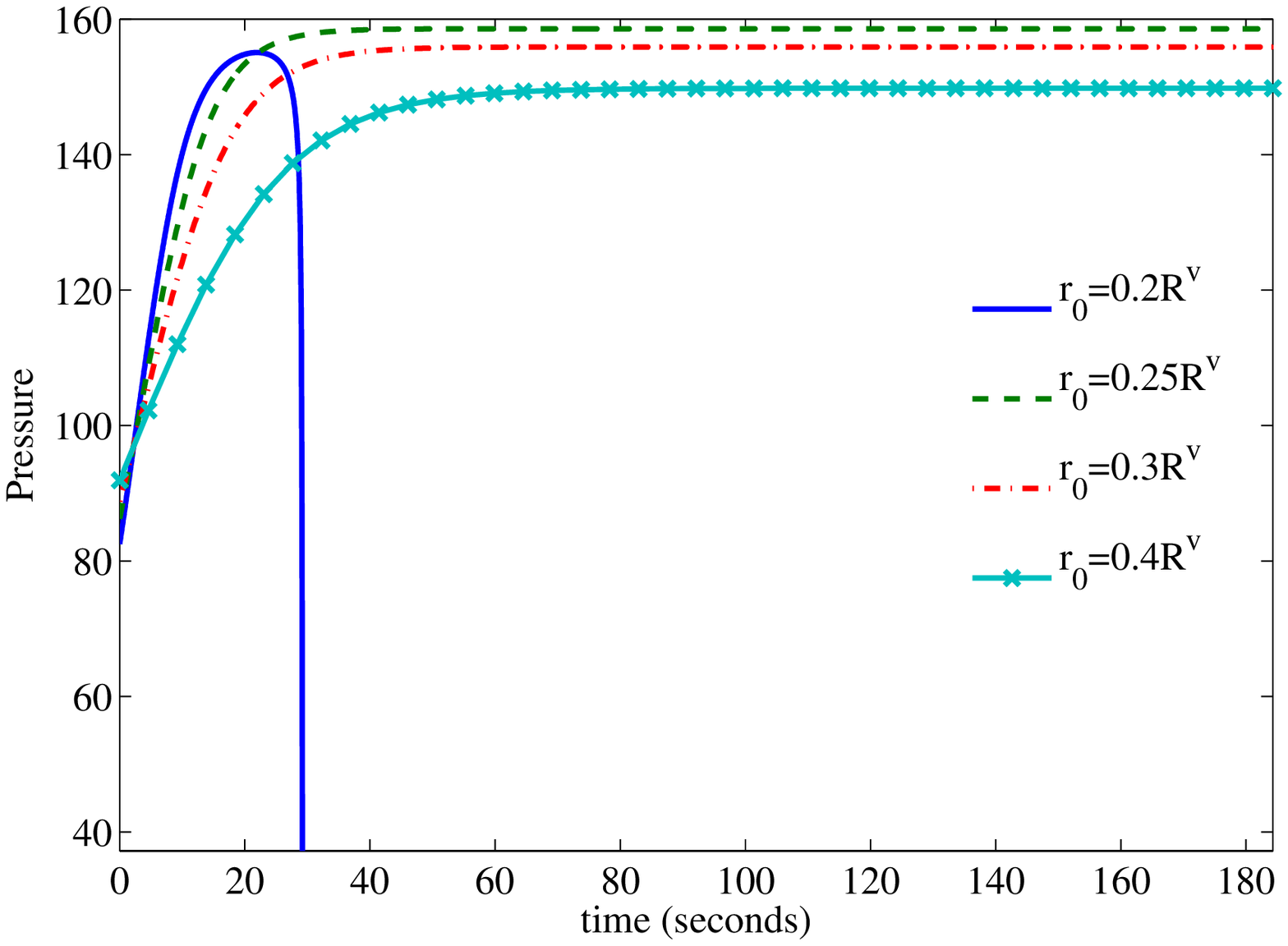}
    \end{tabular}
    \caption{Case~1 -- varying $r(0)$: No ice, initial radius
      $s(0)=0.7\Rf$, and pressure $p^f_g(0)=200\;\units{kPa}$.}
  \label{fig:case1-summary-r}
\end{figure}

\begin{figure}[bthp]
  \begin{center}
    \footnotesize
    \begin{tabular}{ccc}
      (a) Water pressures & (b) Gas pressures & (c) Gas bubble radii\\
      \myfig{0.31}{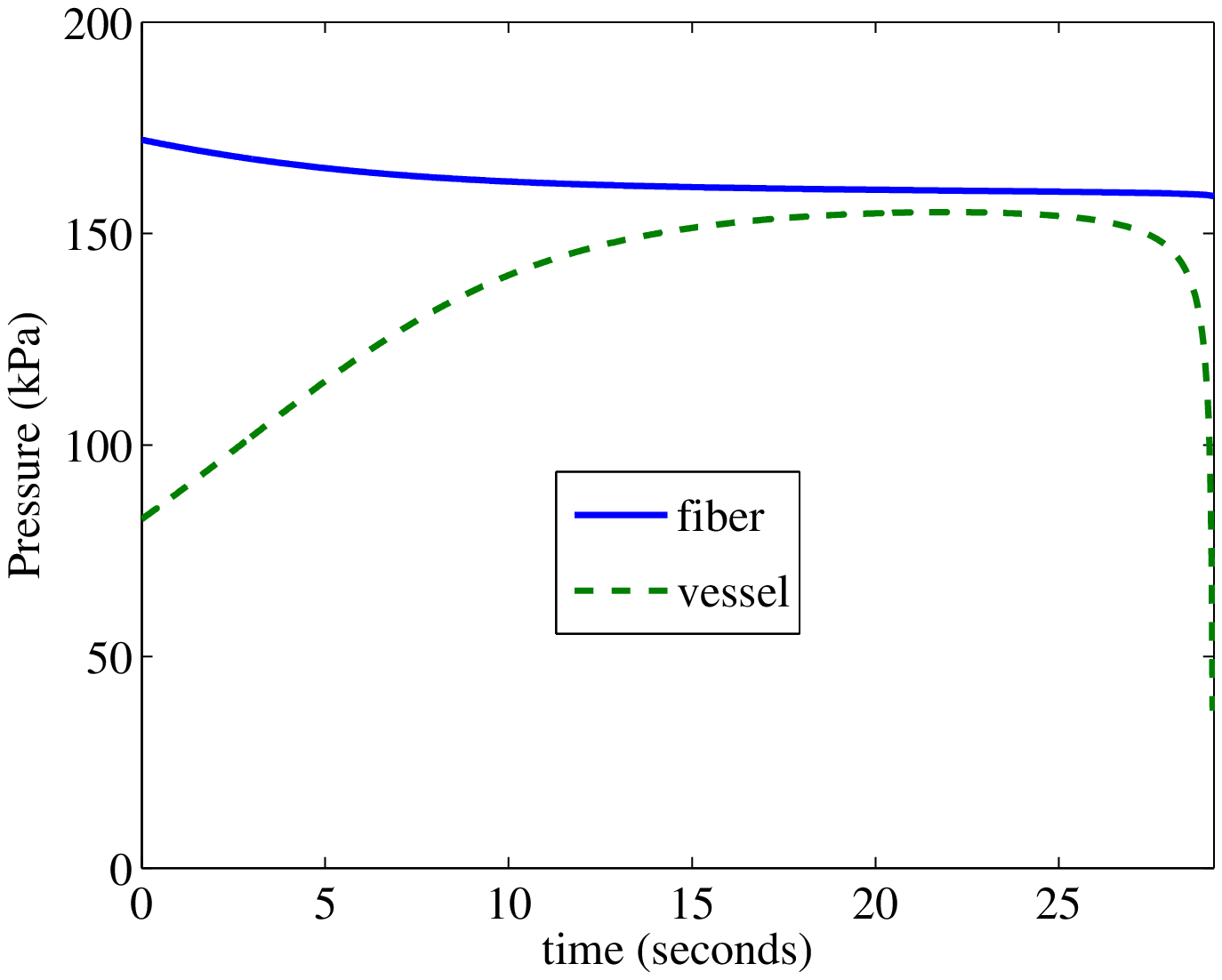}
      &
      \myfig{0.31}{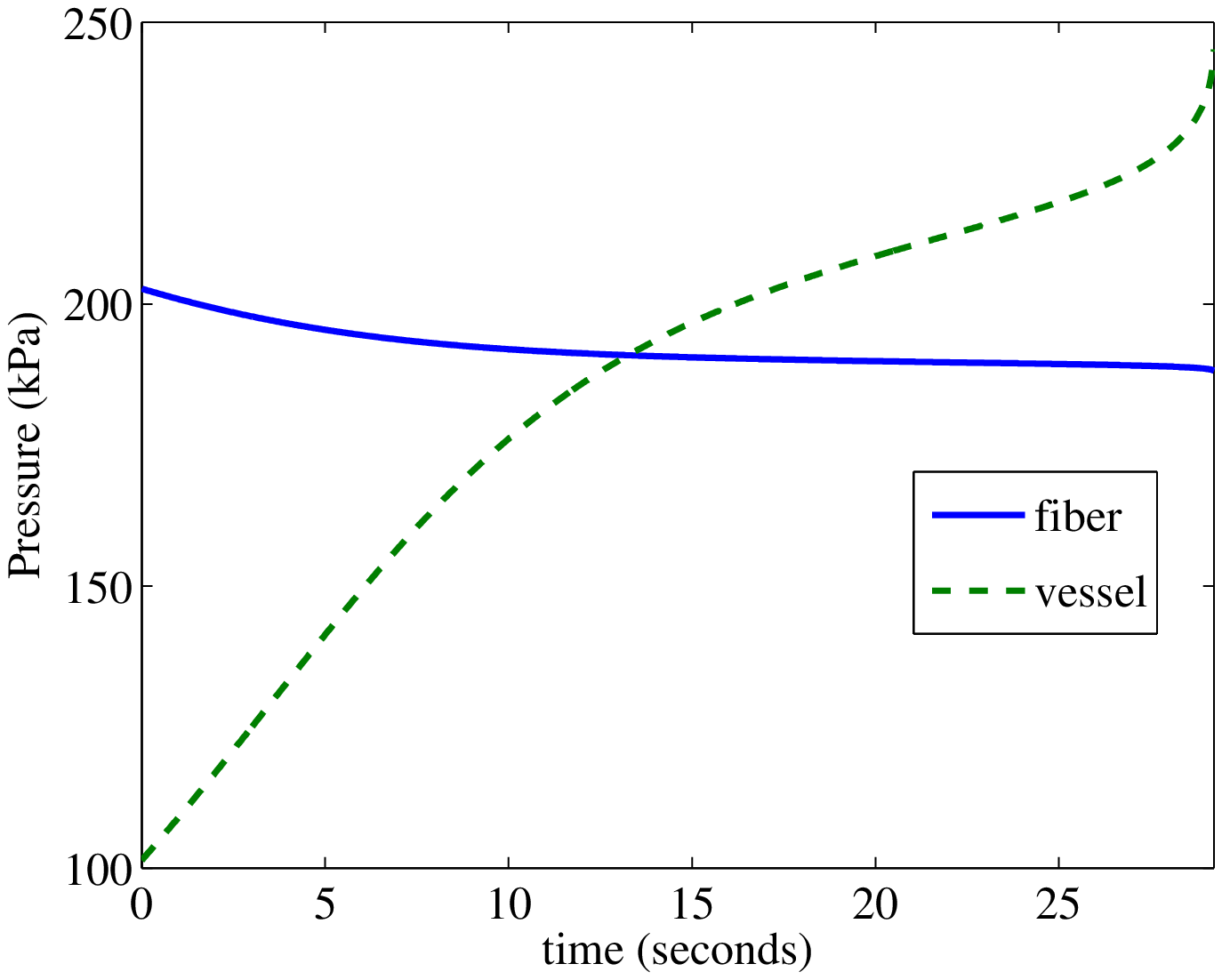}
      & 
      \myfig{0.31}{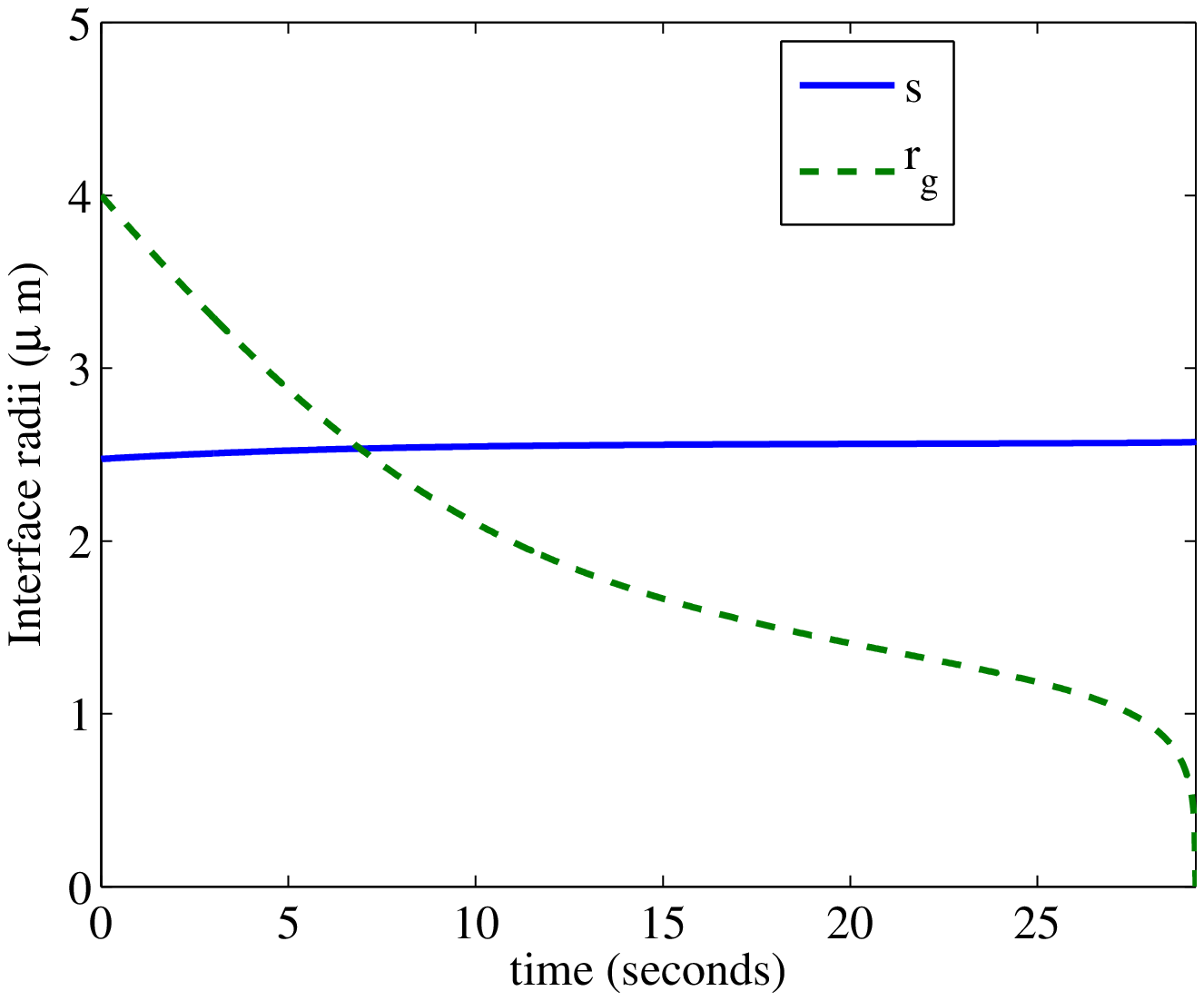}
    \end{tabular}
    \caption{Case~1 -- collapsing vessel bubble: No ice, initial radii
      $s(0)=0.7\Rf$, $r(0)=0.2 \Rv$, and pressure
      $p^f_g(0)=200\;\units{kPa}$.  Here, the initial vessel bubble is
      so small that it dissolves entirely.}
    \label{fig:case1i}
  \end{center}
\end{figure}

The parameter value with the most uncertainty is the initial value of
vessel bubble radius, and so we focus next on the effect of reducing
$r(0)$ in Figure~\ref{fig:case1-summary-r}.  As long as $r(0)$ is
greater than approximately $0.2\Rv$, we observe behavior similar to the
earlier simulations in Figure~\ref{fig:case1a} in that the fiber gas
bubble expands, driving water through the cell wall and hence
compressing the gas in the vessel.  However, when $r(0)=0.2\Rv$ the
vessel gas bubble dissolves entirely, causing the gas pressure to become
unbounded as $r\rightarrow 0$ owing to the air/water surface tension
term in \eqnref{surftens-nondim}.  At the precise moment when the vessel
bubble disappears, the model breaks down because there is no further
transfer of pressure between fiber and vessel, nor is there any flow of
sap through the fiber/vessel wall.  This point actually corresponds to a
terminal steady state of the system, and in our computations we have
found it necessary to specify a very small threshold value of the bubble
radius below which we halt the simulation.  The collapsing vessel bubble
case $r(0)=0.2\Rv$ is illustrated in more detail in the plots in
Figure~\ref{fig:case1i}.

The dissolution of gas in the vessel has an intriguing connection with
the phenomenon of embolism repair~\cite{yang-tyree-1992}, in which gas
bubbles within the vessel that could potentially hinder sap transport
are known to dissolve during the spring thaw.  The physical mechanism
underlying embolism repair is still not fully understood, and has
implications for a wide range of other tree species; hence, a more
detailed study of bubble dissolution as relates to embolism would form a
fascinating avenue for future research.

\begin{figure}[bthp]
  \centering
    \footnotesize
    \begin{tabular}{ccc}
      (a) Vessel bubble radius & & (b) Vessel water pressure \\
      \myfig{0.42}{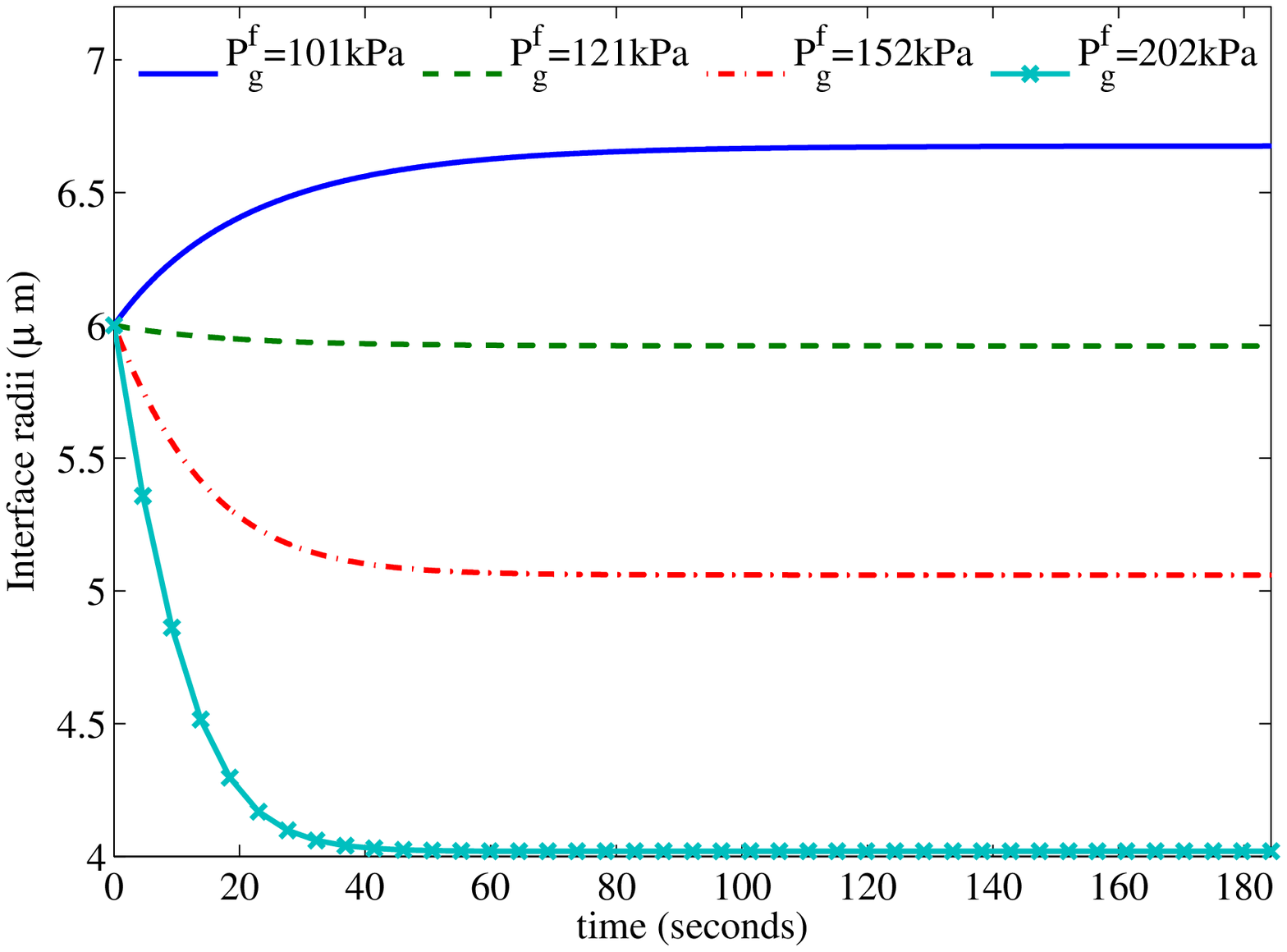}
      & & 
      \myfig{0.42}{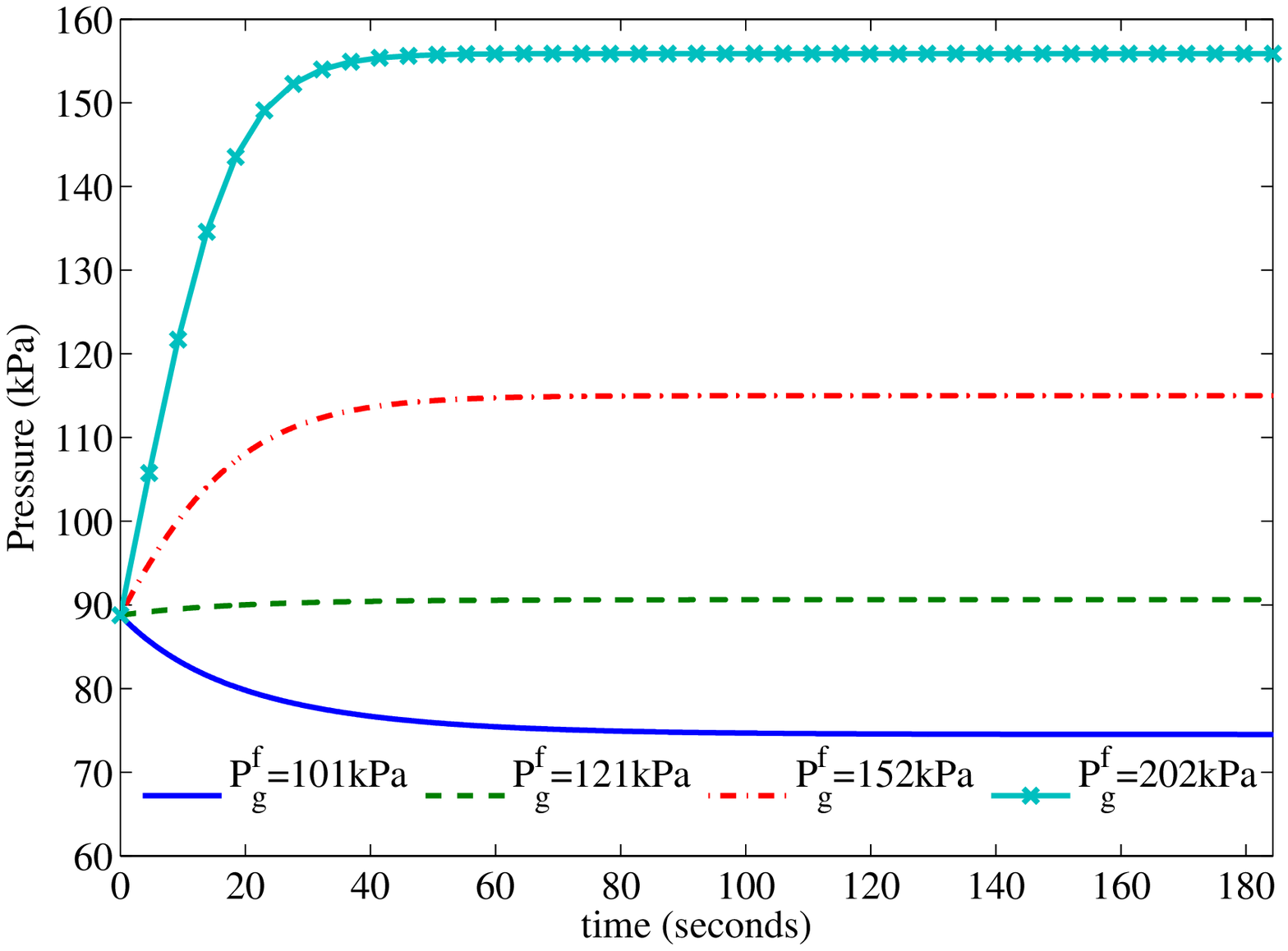}
    \end{tabular}
    \caption{Case~1 -- varying $p^f_g(0)$: No ice, initial radii
      $s(0)=0.7\Rf$, $r(0)=0.3\Rv$.}
    \label{fig:case1-summary-s-and-p}
\end{figure}

\begin{figure}[bthp]
  \begin{center}
    \footnotesize
    \begin{tabular}{ccc}
      (a) Water pressures & (b) Gas pressures & (c) Gas bubble radii\\
      \myfig{0.31}{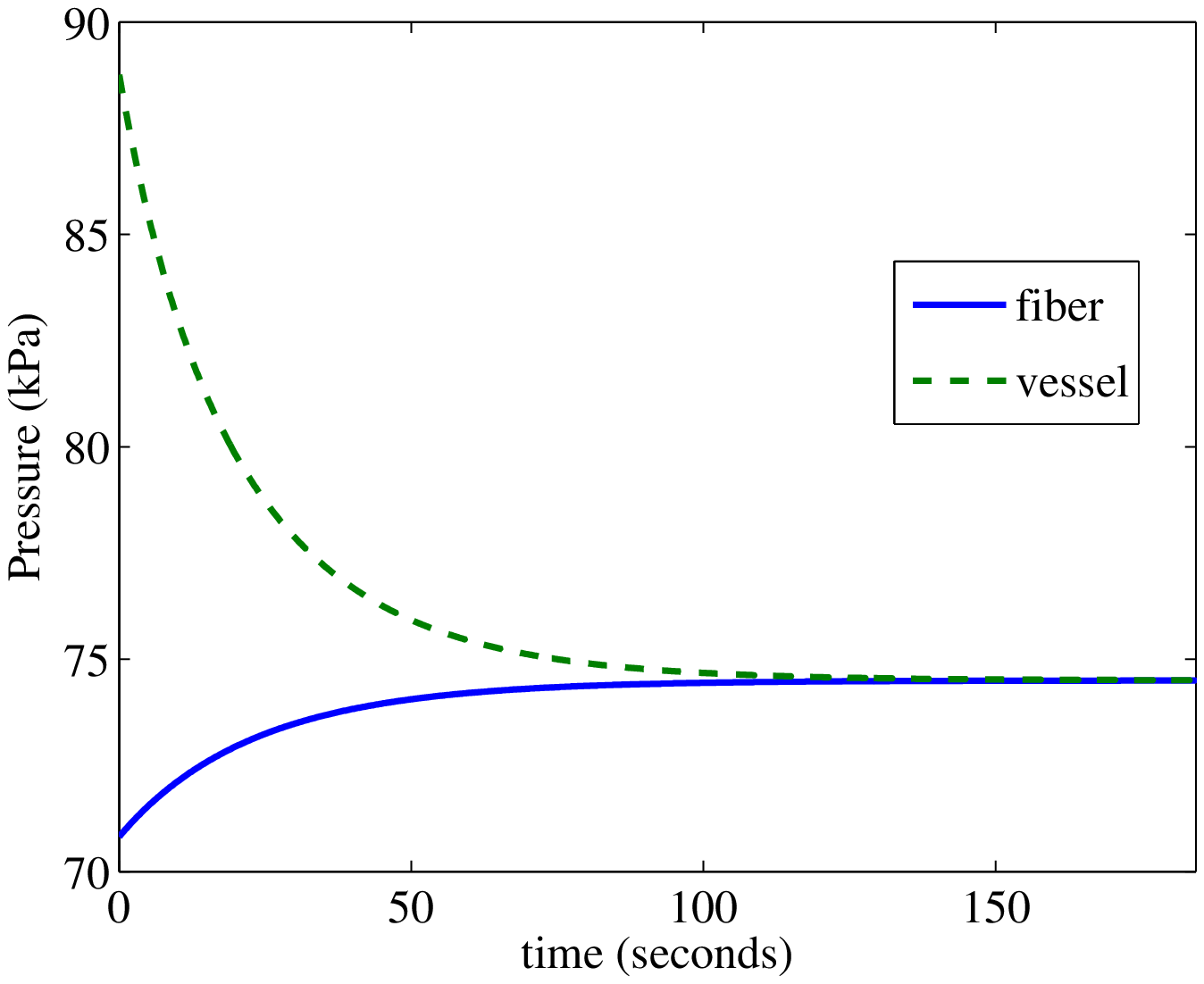}
      &
      \myfig{0.31}{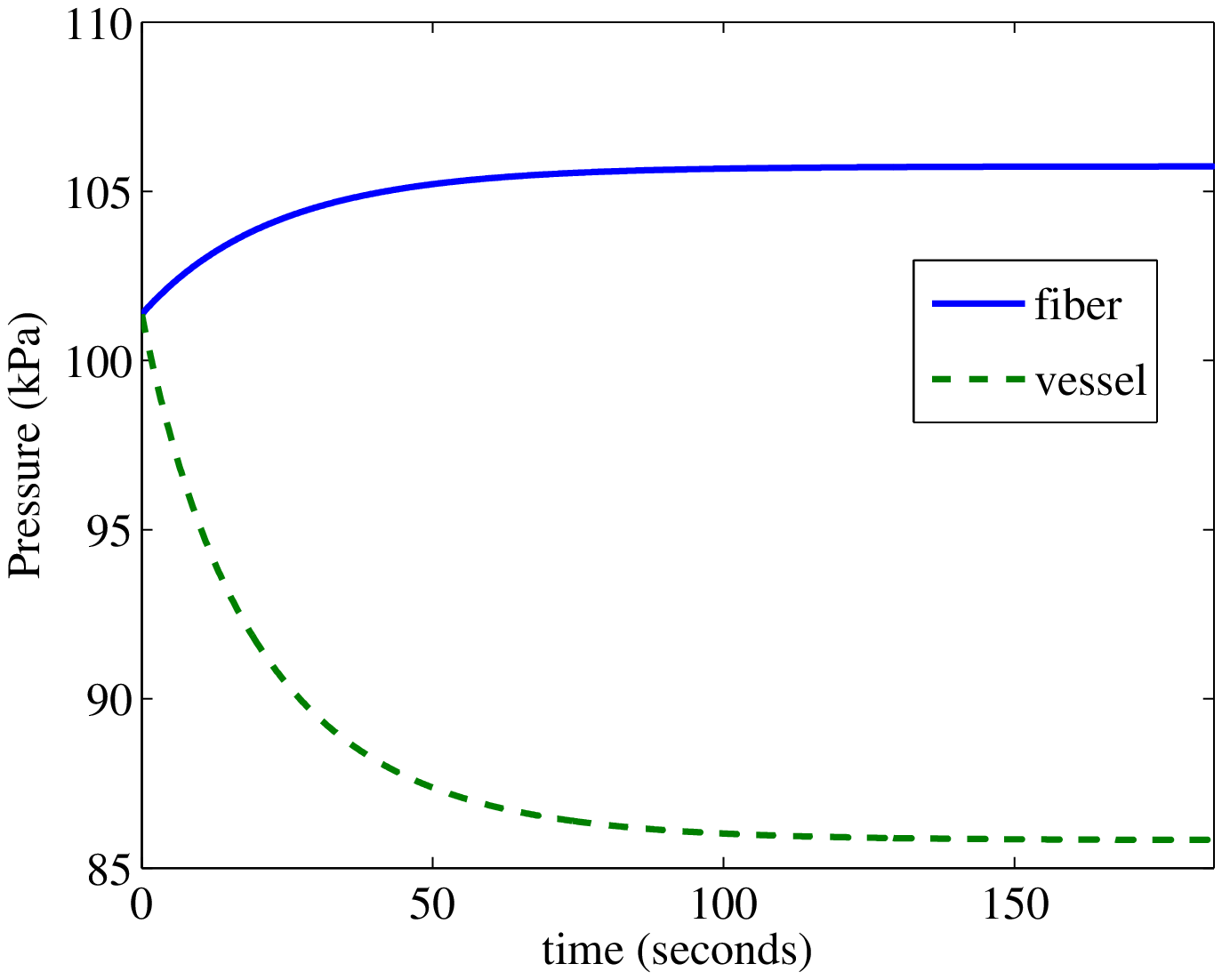}
      & 
      \myfig{0.31}{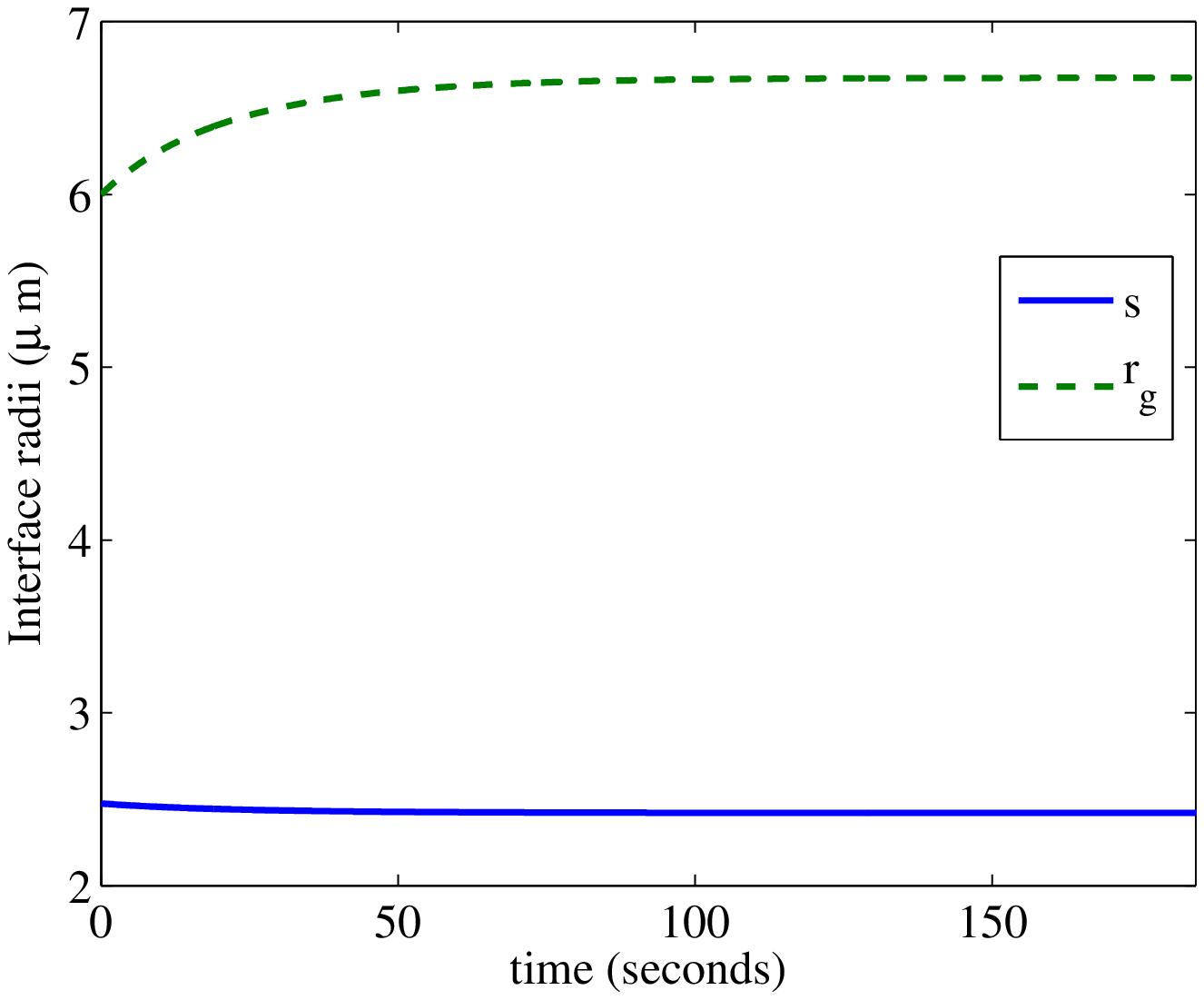}
    \end{tabular}
    \caption{Case~1 -- reduced fiber pressure: No ice, initial radii
      $s(0) = 0.7\Rf$ and $r(0)=0.3 \Rv$, and pressure
      $p^f_g(0)=100\;\units{kPa}$.}
    \label{fig:case1c}
  \end{center}
\end{figure}

Since $p^f_g(0)=200\;\units{kPa}$ is at the upper end of the range of
fiber gas pressures we would expect in actual trees, we next consider
the effect of varying $p^f_g(0)$ over the range
$[100,200]\;\units{kPa}$.  As seen in
Figure~\ref{fig:case1-summary-s-and-p}, the qualitative behavior is
similar until the initial fiber gas pressure approaches the atmospheric
level in the vessel; for $p^f_g(0)$ low enough, the dynamics are
reversed in the sense that the fiber bubble shrinks and the vessel
bubble expands.  As seen in the detailed plots for
$p^f_g(0)=100\;\units{kPa}$ in Figure~\ref{fig:case1c}, the fiber and
vessel again equilibrate at a state where the water pressures are equal,
although in this case the level of $p_w\approx 75\;\units{kPa}$ is much
lower than that observed in the base case.  Clearly, when the fiber
begin at lower pressure, then the tree is less able to pressurize the
vessel sap; furthermore, the fiber compartment must be above some
threshold pressure for there to be any pressure transfer from fiber to
vessel that is consistent with exudation.
In fact, if the initial fiber gas pressure and radius are both small
enough, then it is the gas in the fiber that dissolves entirely.
Although this situation is not likely to occur in practice, we have
nonetheless included a simulation with $p^f_g(0)=100\;\units{kPa}$ and
$s(0)=0.25\Rf$ in Figure~\ref{fig:case1i} to illustrate this point.

\begin{figure}[bthp]
  \begin{center}
    \footnotesize
    \begin{tabular}{ccc}
      (a) Water pressures & (b) Gas pressures & (c) Gas bubble radii\\
      \myfig{0.31}{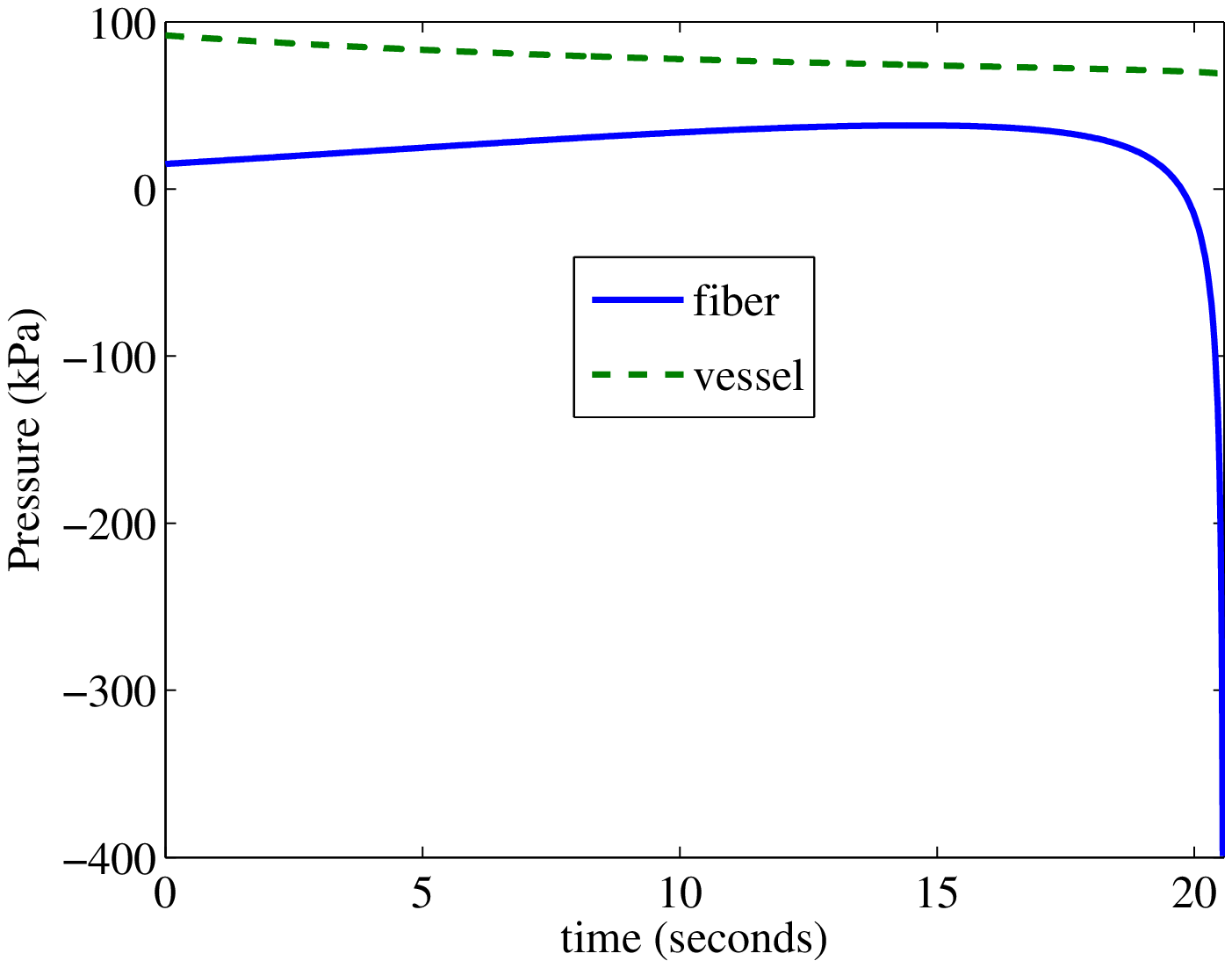}
      &
      \myfig{0.31}{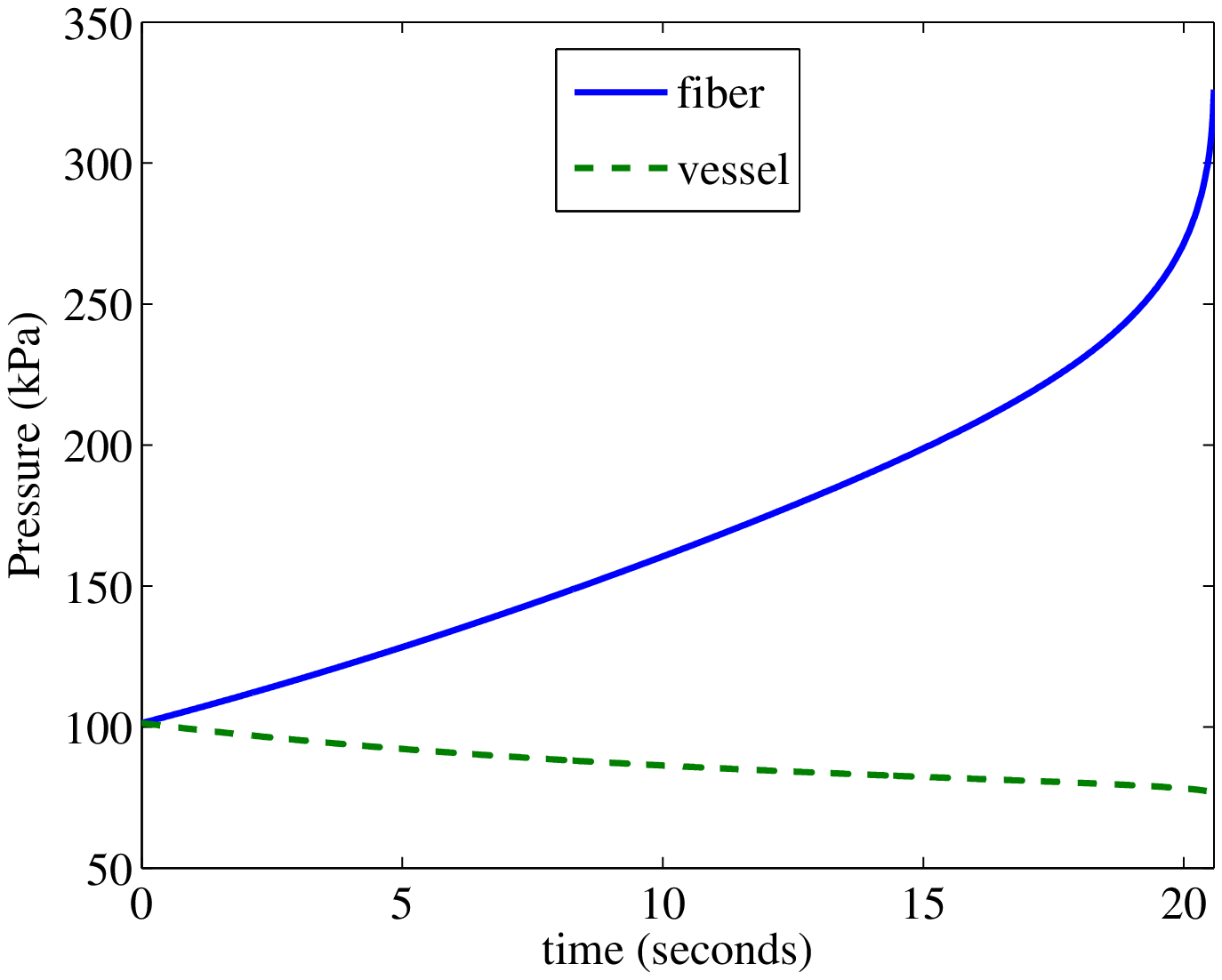}
      & 
      \myfig{0.31}{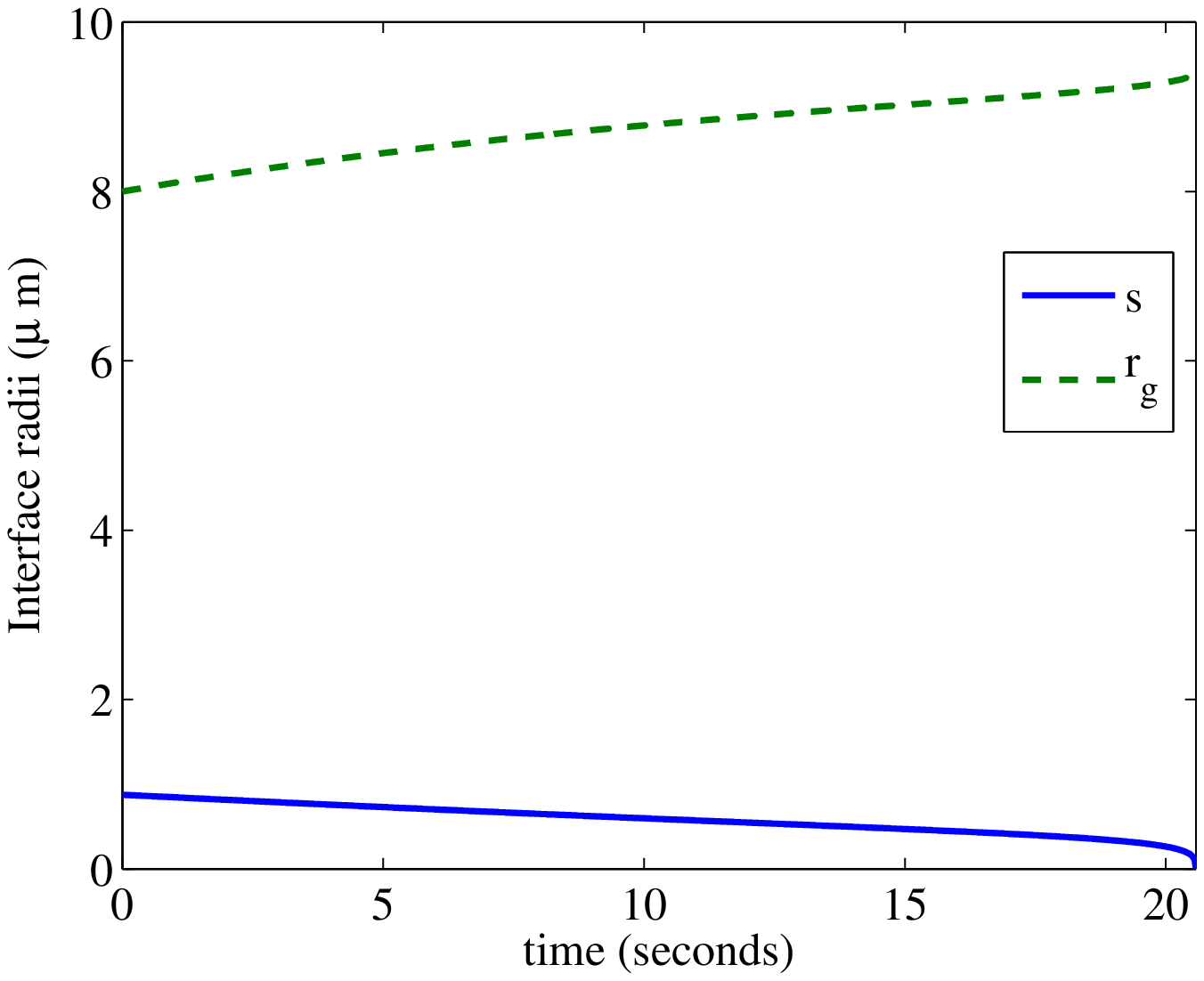}
    \end{tabular}
    \caption{Case~1 -- collapsing fiber bubble: No ice, initial radii
      $s(0)=0.25 \Rf$, $r(0)=0.4 \Rv$, and pressure
      $p^f_g(0)=100\;\units{kPa}$.  The fiber bubble radius and pressure
      are initially so small in this case that the fiber bubble
      dissolves entirely.}
    \label{fig:case1e}
  \end{center}
\end{figure}

\subsection{Case~2: No ice, with osmosis}
\label{sec:sims-case2}

For the next round of simulations, we introduce osmotic effects
arising from the presence of sucrose in the vessel, and still maintain our
focus on the effect of porous flow and gas dissolution by assuming all
sap is in the liquid phase.  The vessel sap is taken to contain 2\%\
sucrose by mass (with 0\%\ in the fiber), but otherwise the initial
conditions are similar to Case~1.

\begin{figure}[bthp]
  \begin{center}
    \footnotesize
    \begin{tabular}{ccc}
      (a) Water pressures & (b) Gas pressures & (c) Gas bubble radii\\
      \myfig{0.31}{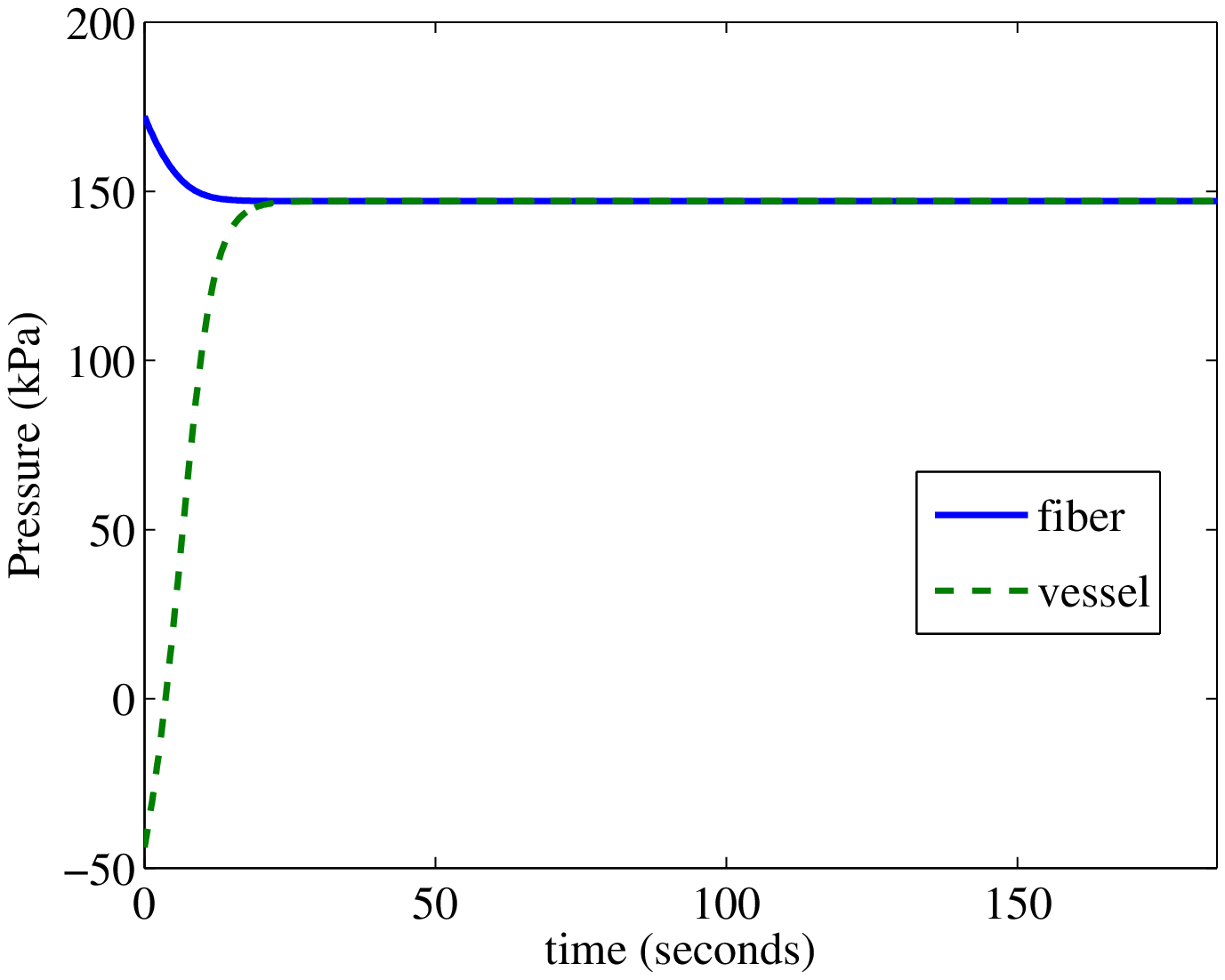}
      &
      \myfig{0.31}{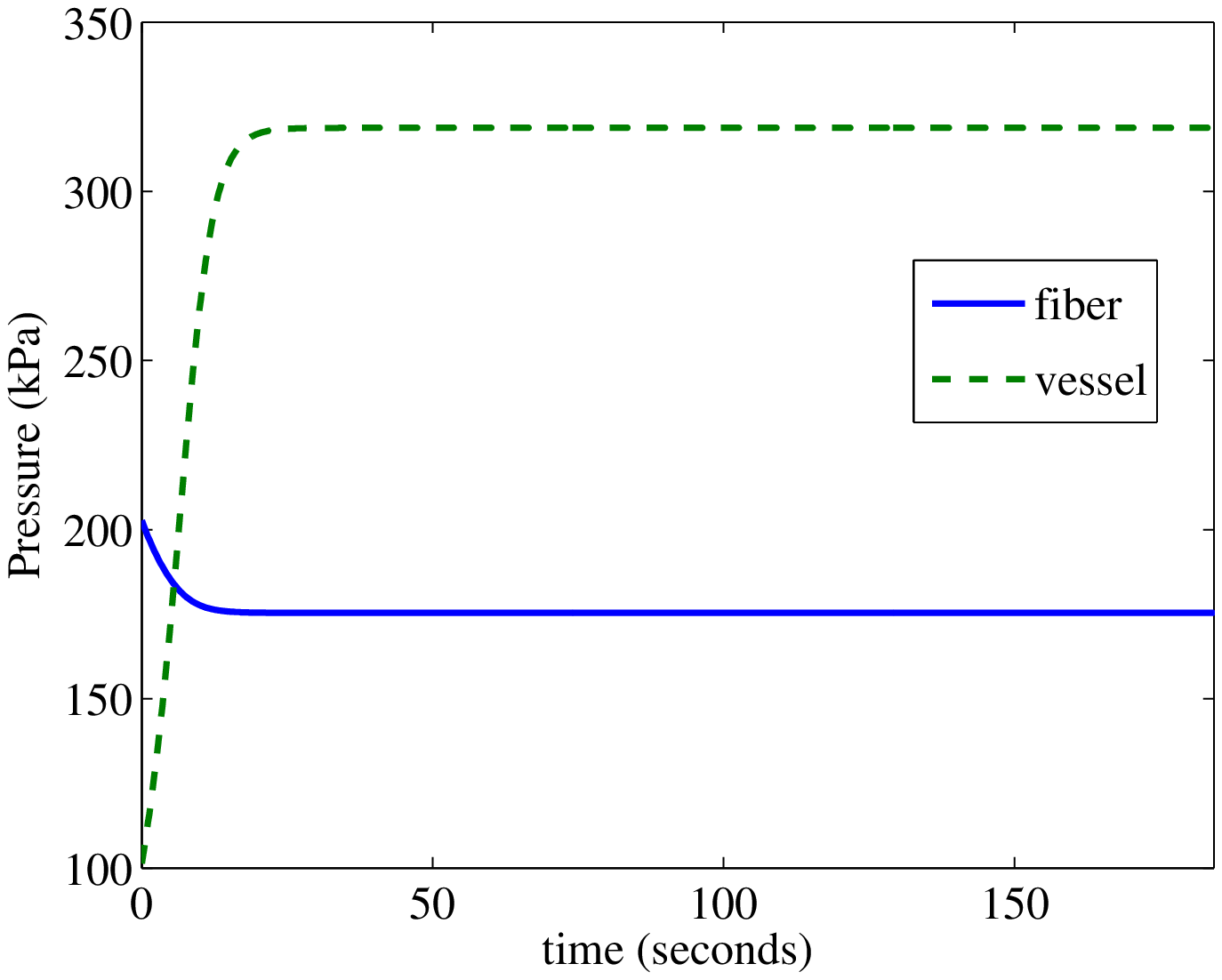}
      & 
      \myfig{0.31}{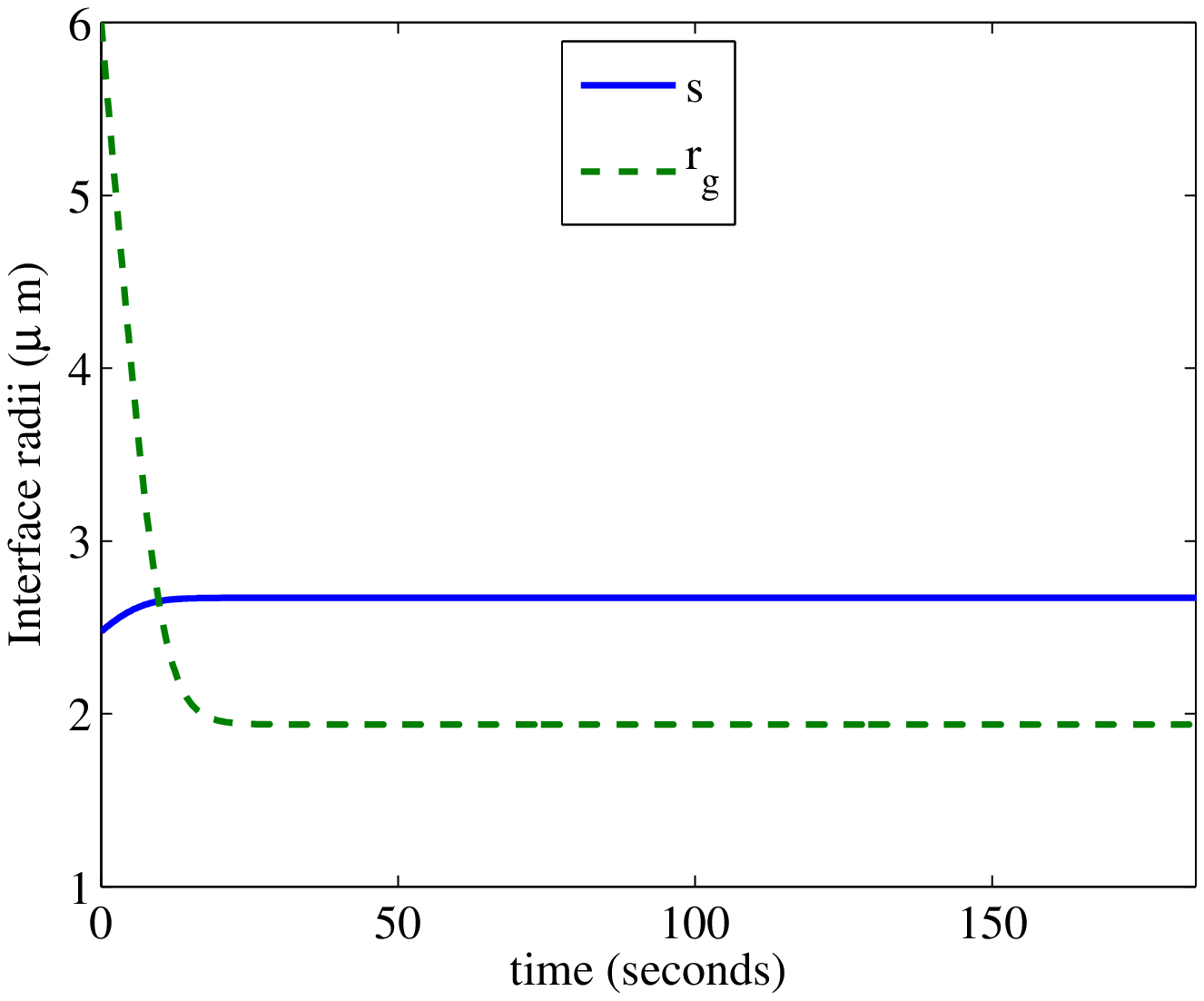}
    \end{tabular}
    \caption{Case~2 -- base case with osmosis: No ice, with osmosis,
      initial radii $s(0) = 0.7\Rf$, $r(0)=0.3 \Rv$, and pressure
      $p^f_g(0)=200\;\units{kPa}$.}
    \label{fig:case2a}
  \end{center}
\end{figure}

The first simulation with $r(0)=0.3\Rv$, $p^v_g(0)=100\;\units{kPa}$ and
$p^f_g(0)=200\;\units{kPa}$ is displayed in Figure~\ref{fig:case2a} and
should be compared with Figure~\ref{fig:case1a} (base case, without
osmosis).  Osmotic pressure has a relatively small impact on the fiber
bubble dynamics, with the fiber radius being only slightly smaller and
the equilibrium water pressure dropping from 155~\units{kPa} to just
under 150~\units{kPa}.  In contrast, there is a major change in the
vessel bubble evolution, with the equilibrium radius dropping to roughly
half that of the base case.  The source of the difference is evident from 
the pressure plots: the initial vessel water pressure is
nearly 140~\units{kPa} lower than the base case in
Figure~\ref{fig:case1a}(a) owing to the osmotic pressure, and the
equilibrium vessel gas pressure increases by roughly the same amount.

\begin{figure}[bthp]
  \centering
    \footnotesize
    \begin{tabular}{ccc}
      (a) Vessel bubble radius & & (b) Vessel water pressure \\
      \myfig{0.42}{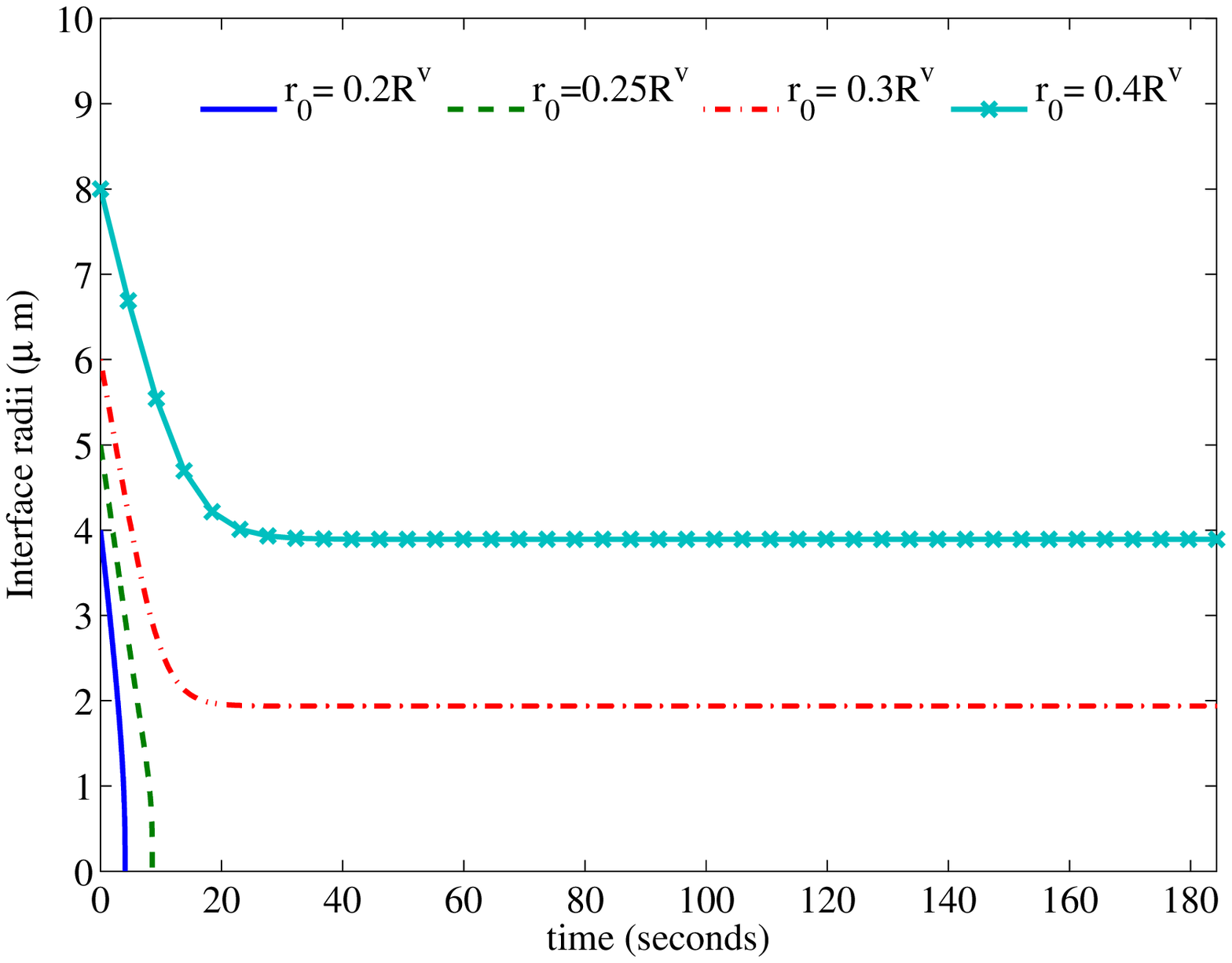}
      & & 
      \myfig{0.42}{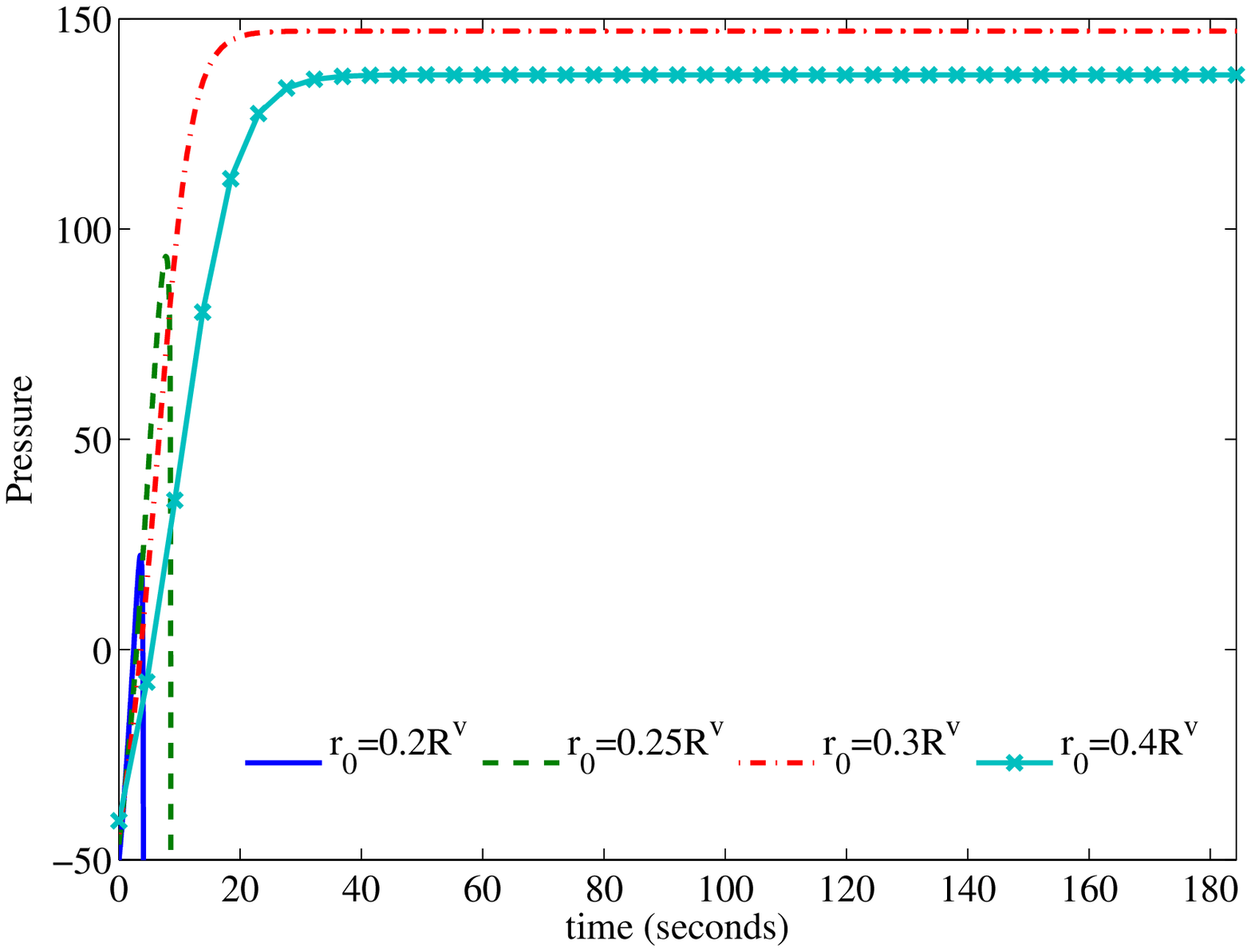}
    \end{tabular}
    \caption{Case~2 -- varying $r(0)$: No ice, with osmosis, initial
      radius $s(0)=0.7\Rf$, and pressure $p^f_g(0)=200\;\units{kPa}$.}
  \label{fig:case2-summary}
\end{figure}

Maintaining the fiber bubble at pressure $p^f_g(0)=100\;\units{kPa}$, we
next draw comparisons with Figure~\ref{fig:case1-summary-r} which varied
the vessel bubble radius over values $r(0)=\{0.2, 0.25, 0.3, 0.4\}\Rv$.
From the summary of results shown in Figure~\ref{fig:case2-summary}, we
see that the presence of osmosis enhances the flow from fiber to vessel
and leads to a much larger pressure increase in the vessel sap.  The
smallest vessel gas bubble with $r(0)=0.2$ disappears sooner in the
presence of osmosis, which is to be expected.

\begin{figure}[bthp]
  \begin{center}
    \footnotesize
    \begin{tabular}{ccc}
      (a) Water pressures & (b) Gas pressures & (c) Gas bubble radii\\
      \myfig{0.31}{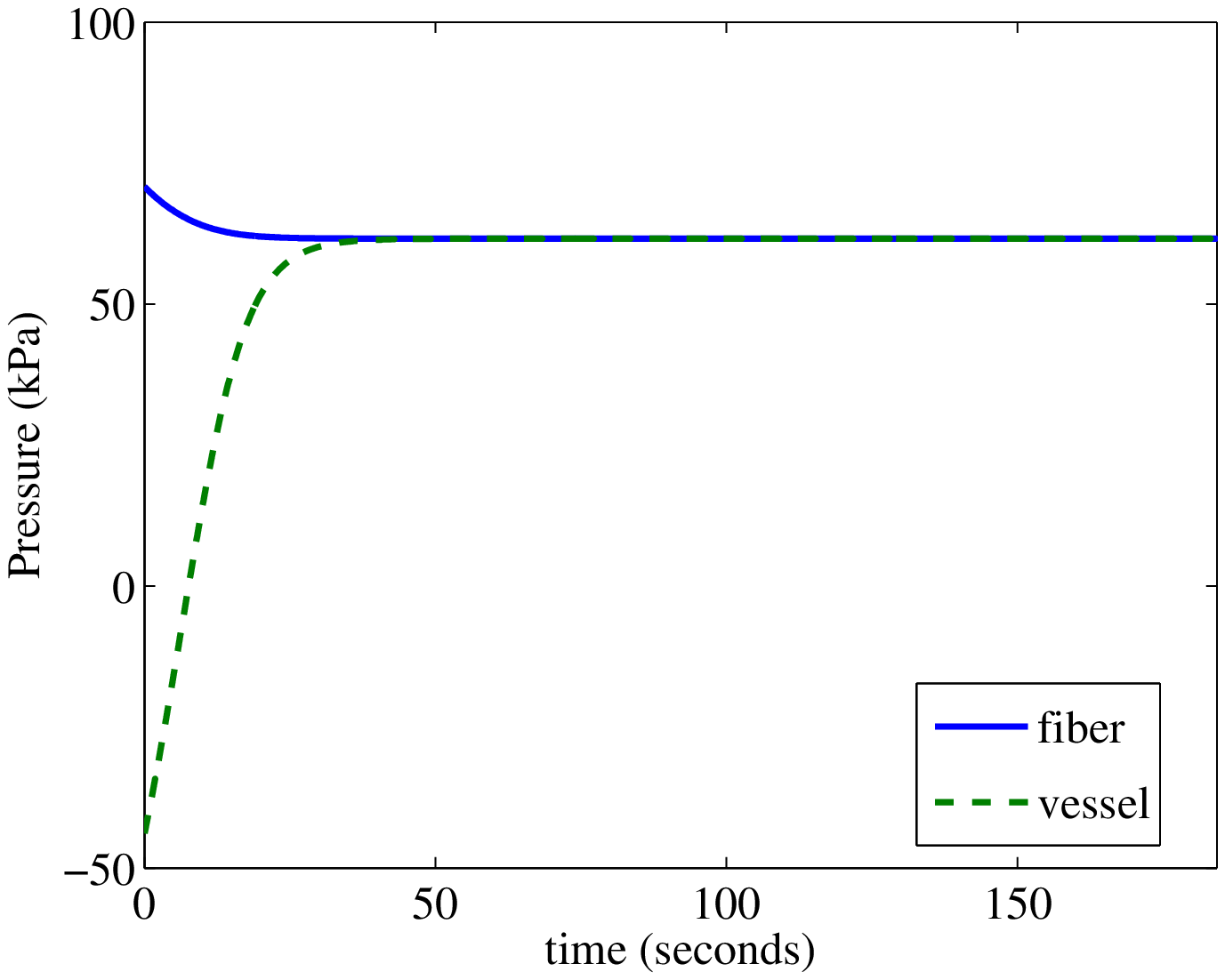}
      &
      \myfig{0.31}{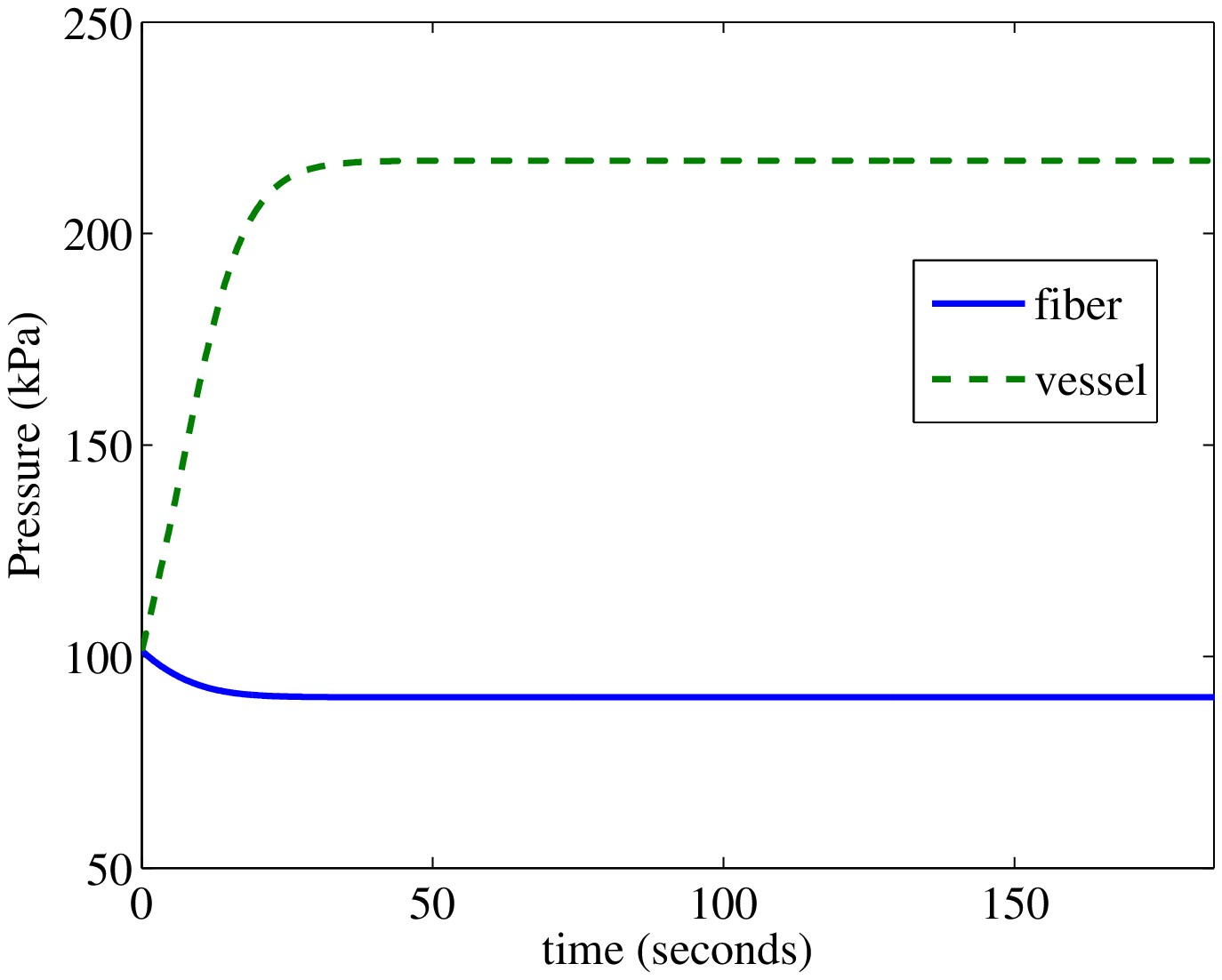}
      & 
      \myfig{0.31}{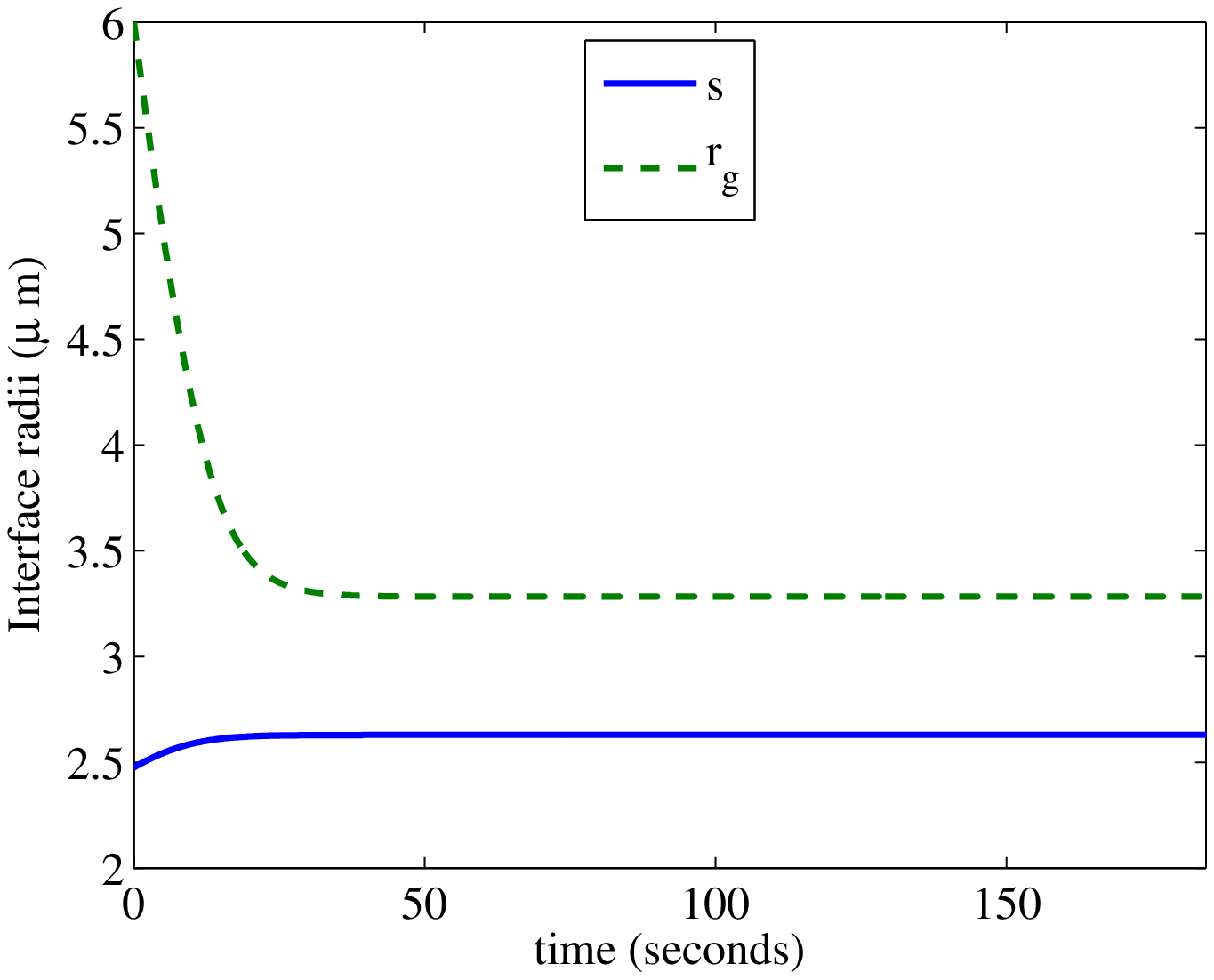}
    \end{tabular}
    \caption{Case~2 -- reduced fiber pressure: No ice, with osmosis,
      initial radii $s(0) = 0.7\Rf$, $r(0)=0.3 \Rv$, and pressure
      $p^f_g(0)=100\;\units{kPa}$.}
    \label{fig:case2c}
  \end{center}
\end{figure}

A more insightful comparison is afforded by reducing the fiber gas
pressure to $p^f_g(0)=100\;\units{kPa}$, and comparing
Figure~\ref{fig:case2c} to the corresponding case without osmosis in
Figure~\ref{fig:case1c} (where we observed a reversed sap flow, from
vessel to fiber).  Clearly, osmotic effects act to return the sap
dynamics to ``normal,'' with water now being driven from fiber to
vessel.  Except for the shift in equilibrium pressures, the qualitative
solution behavior is now consistent with the base case simulation in
Figure~\ref{fig:case1a}.  Consequently, our model demonstrates that
osmotic pressures owing to sucrose in the vessel can provide a mechanism
for enhancing sap exudation, even when the gas pressure in the fiber is
as low as that in the vessel.  In this respect, our model is consistent
with the hypothesis of Tyree~\cite{tyree-1995} that claims osmosis
prevents gas bubble dissolution in the fiber and sustains sap exudation
pressures over longer time periods.

\subsection{Case~3: Milburn--O'Malley model (with ice, no osmosis)}
\label{sec:sims-case3}

We next test Milburn and O'Malley's hypothesis for sap exudation that
includes the effects of melting ice and gas dissolution, but ignores
osmosis.  We restrict ourselves to the time period over which ice is
present, so that gas in the fiber is isolated from the water phase by
the frozen ice layer and hence gas dissolution occurs only in the vessel
(the long-time dynamics after the ice layer melts are considered in
Case~5).  Furthermore, gas--water interfacial tension plays no role in
the fiber in this case, so that the pressure of the melted liquid
adjacent to the fiber wall is simply equal to the gas pressure.

\begin{figure}[bthp]
  \begin{center}
    \footnotesize
    \begin{tabular}{ccc}
      (a) Water pressures & (b) Gas pressures & (c) Fiber phase interfaces\\
      \myfig{0.31}{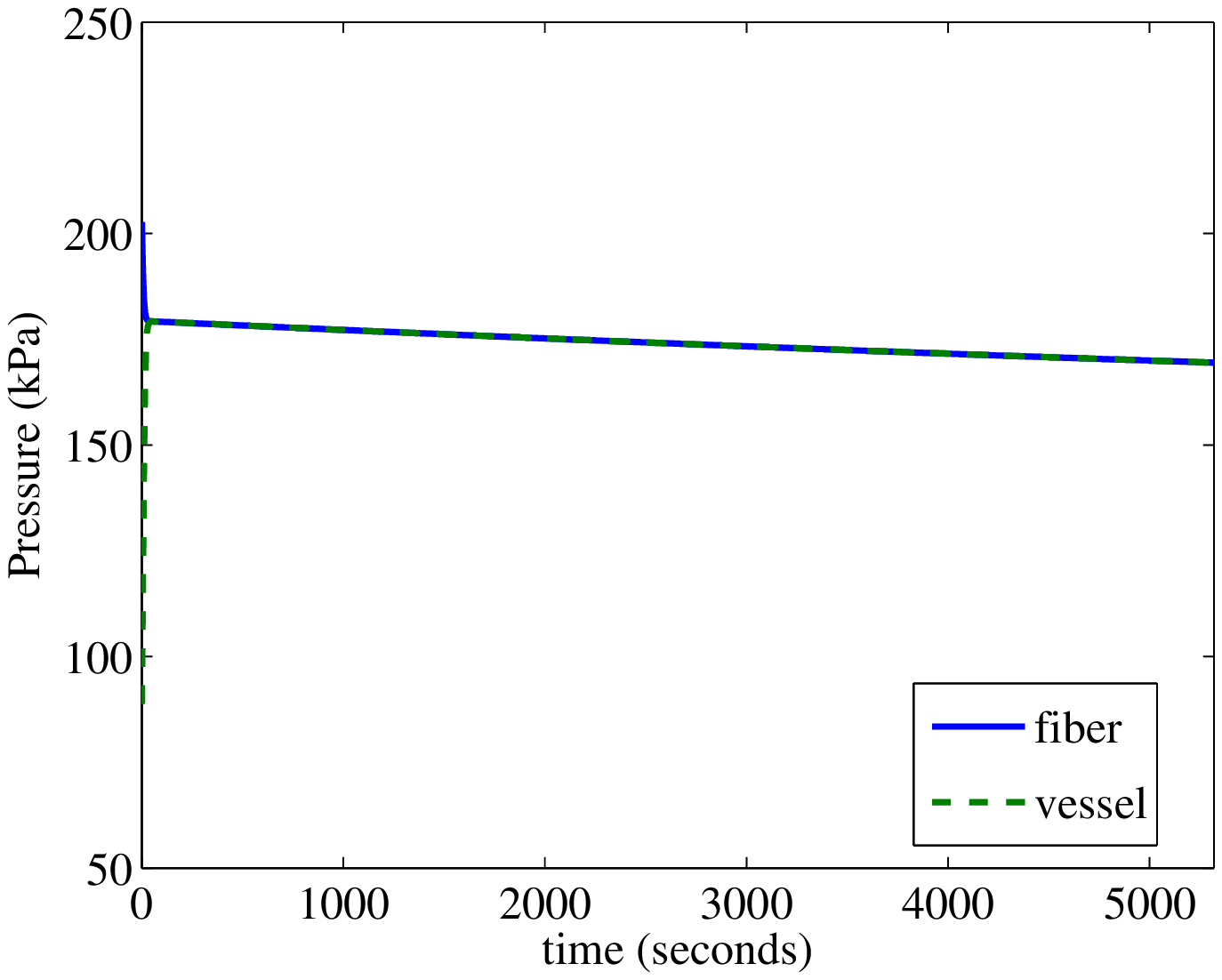}
      &
      \myfig{0.31}{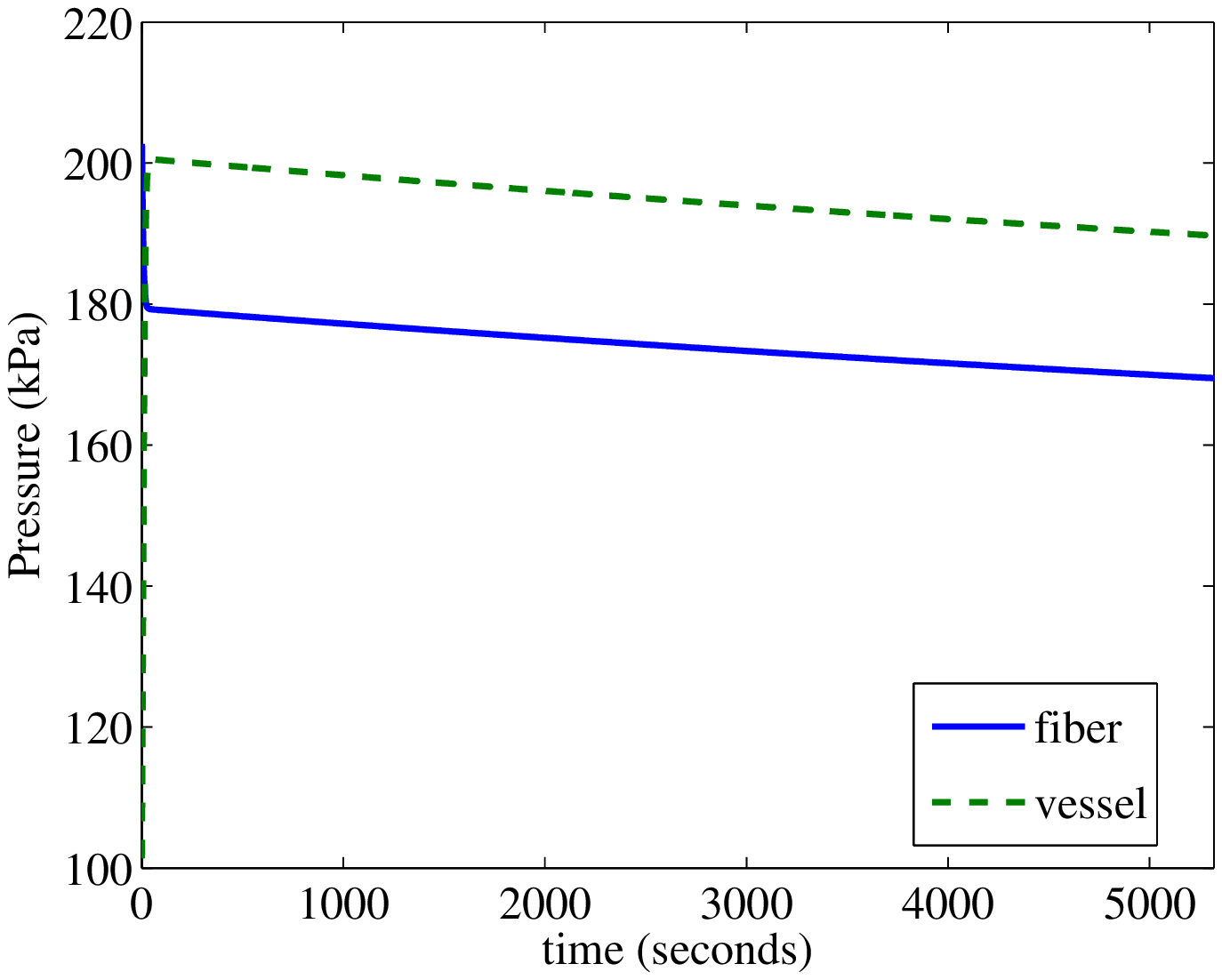}
      & 
      \myfig{0.31}{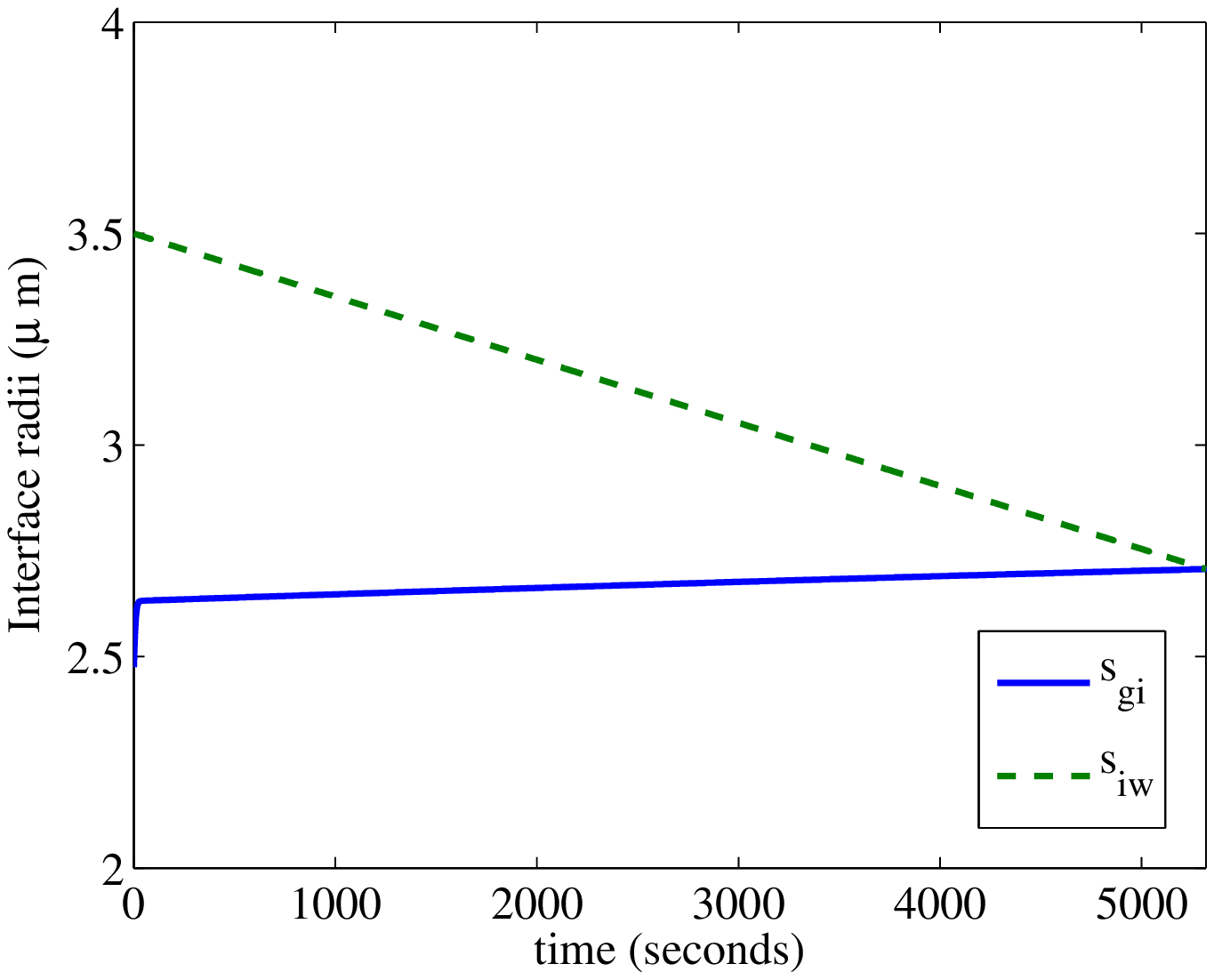}
    \end{tabular}
    \caption{Case~3 -- base case with ice: Milburn--O'Malley scenario,
      no osmosis.  Initial radii are $s_{gi}(0)=0.7\Rf$, $s_{iw}=\Rf$,
      $r(0)=0.3 \Rv$, and pressure $p^f_g(0)=200\;\units{kPa}$.}
    \label{fig:case3a}
  \end{center}
\end{figure}

In Figure~\ref{fig:case3a}, we present the simulation results using
initial conditions $s_{gi}=0.7\Rf$, $s_{wi}=\Rf$, $r(0)=0.3\Rv$,
$p^f_g(0)=200~\units{kPa}$ and $p^v_g(0)=100~\units{kPa}$, which can be
compared with the base case depicted in Figure~\ref{fig:case1a}.
Until a time of roughly 40~\units{s}, the qualitative features of the
two solutions are similar, with the water pressure decreasing in the
fiber and increasing in the vessel until they equilibrate at some
intermediate value.  Over this time period, the fiber bubble radius
increases smoothly from the initial $2.45\;\units{\mu m}$ to roughly
$2.6\;\units{\mu m}$, and the motion of the vessel bubble (not shown) is
also similar to that of the base case.  There is, however, a significant
difference in the gas and water pressures owing to the absence of
surface tension.

After the initial rapid equilibration phase, the two water pressures
remain equal and experience a gradual decrease over a much longer time
scale.  This slow adjustment in pressure is due to phase change and is
attributed to the roughly 10\%\ difference in density between water and
ice.  The ice/water interface $s_{iw}(t)$ has a roughly linear variation
in time according to Figure~\ref{fig:case3a}(c), which should be
contrasted with the $\sqrt{t}$ dependence that typically arises in
Stefan problems; this discrepancy is due to the fact that sap outflow
through the fiber wall dominates the interface motion.
The intersection of the $s_{gi}$ and $s_{iw}$ curves near $t\approx
2000\;\units{s}$ represents the time when the ice layer disappears.  Our
main conclusion from these results is that even when ice is present, the
vessel compartment is pressurized by the fiber, with the main difference
(comparing Figures~\ref{fig:case1a} and \ref{fig:case3a}) being that the
vessel sap pressure is higher because of the missing interfacial tension
force.

\begin{figure}[bthp]
  \centering
  \footnotesize
  \begin{tabular}{ccc}
    (a) Vessel bubble radius & & (b) Vessel water pressure \\
    \myfig{0.42}{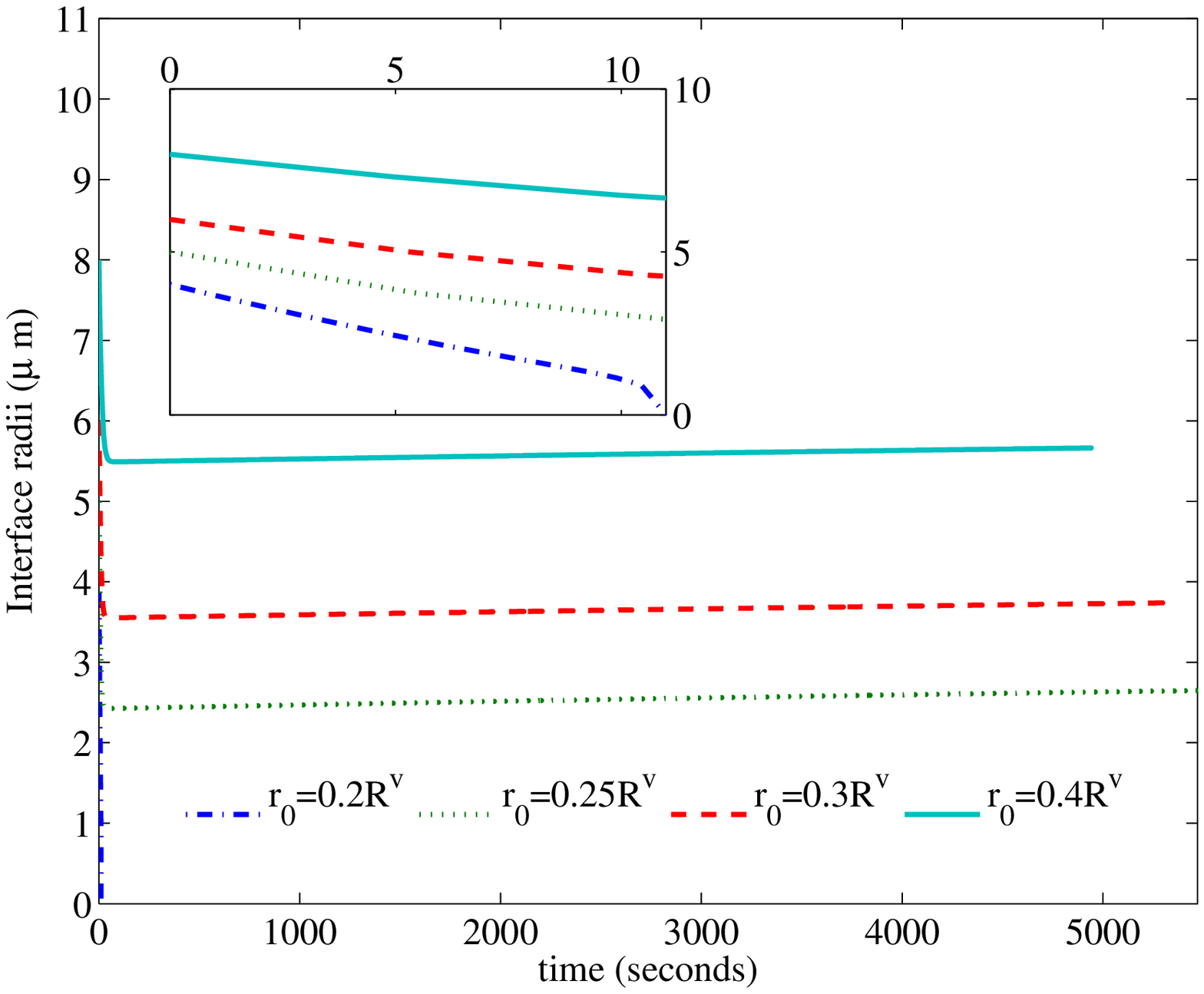}
    & & 
    \myfig{0.42}{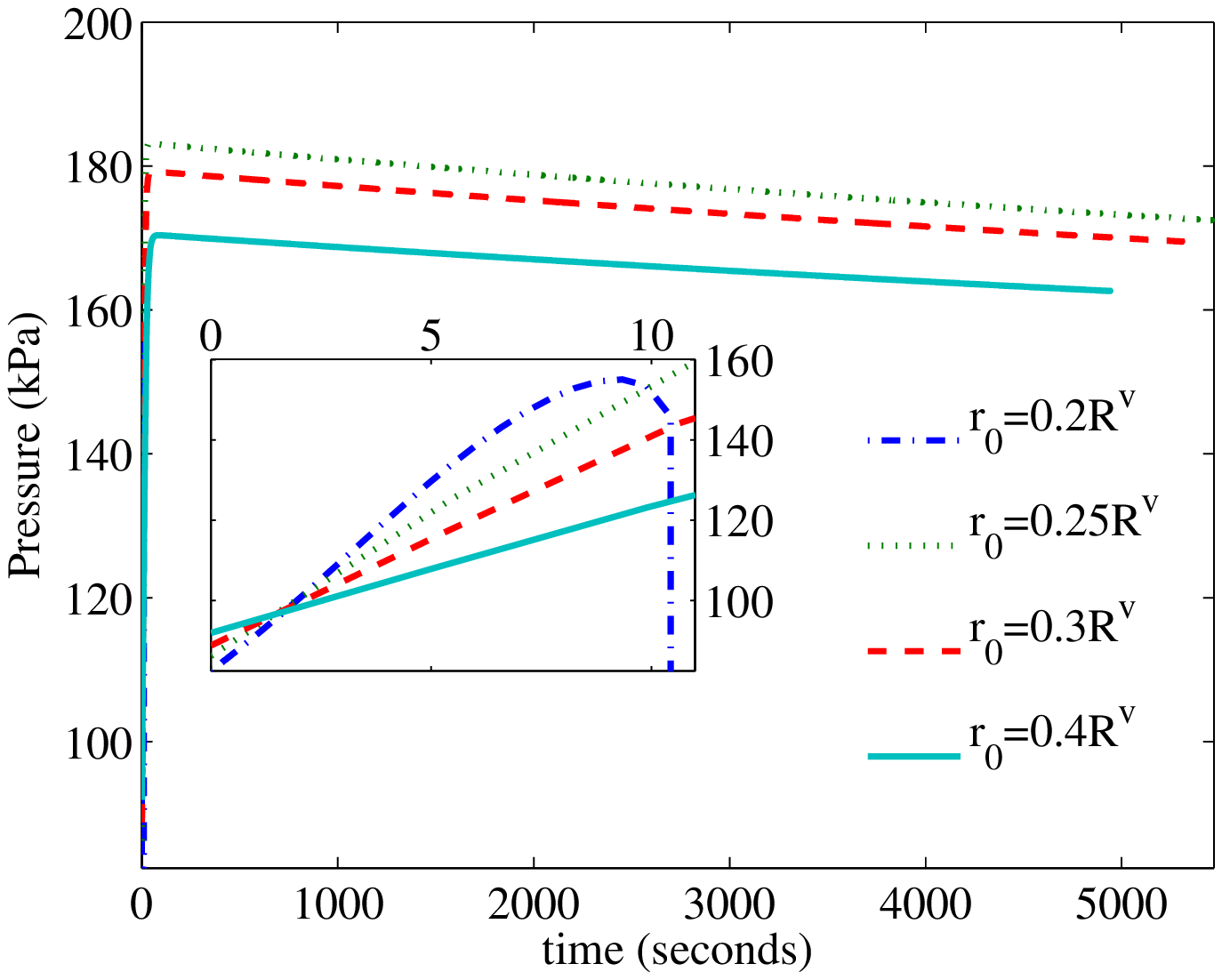}
  \end{tabular}
  \caption{Case~3 -- varying $r(0)$: Milburn--O'Malley scenario with ice
    and no osmosis, initial radius $s(0)=0.7\Rf$, and pressure
    $p^f_g(0)=200\;\units{kPa}$.  The ``zoomed-in'' sub-plot shows the
    curves near $t=0$.}
  \label{fig:case3-summary-r}
\end{figure}

A final series of simulations is then performed in which the initial
vessel bubble radius is varied over the range $r(0)=\{0.2, 0.25,
0.3, 0.4 \}\Rv$.  Figure~\ref{fig:case3-summary-r} shows that
the smallest bubble dissolves entirely within approximately 10 seconds,
after which the transfer of pressure from fiber to vessel stops.  In all
other cases, the fiber and vessel compartments equilibrate over a time
scale of roughly 2 hours.

\leavethisout{
  \begin{figure}[bthp]
    \centering
    \footnotesize
    \begin{tabular}{ccc}
      (a) Vessel bubble radius & & (b) Vessel water pressure \\
      \myfig{0.42}{Comp_thaw_25052012_Summary_vessel_radius_2.eps}
      & & 
      \myfig{0.42}{Comp_thaw_25052012_Summary_vessel_pressure_2.eps}
    \end{tabular}
    \caption{Case~3 -- varying $s_{gi}(0)$ and $p^f_g(0)$:
      Milburn--O'Malley scenario with ice and no osmosis, initial radius
      $r(0)=0.3\Rv$.}
    \label{fig:case3-summary-s-and-p}
  \end{figure}
}

\subsection{Case~4: Full model, with ice and osmosis}
\label{sec:sims-case4}

\begin{figure}[bthp]
  \begin{center}
    \footnotesize
    \begin{tabular}{ccc}
      (a) Water pressures & (b) Gas pressures & (c) Fiber phase interfaces\\
      \myfig{0.31}{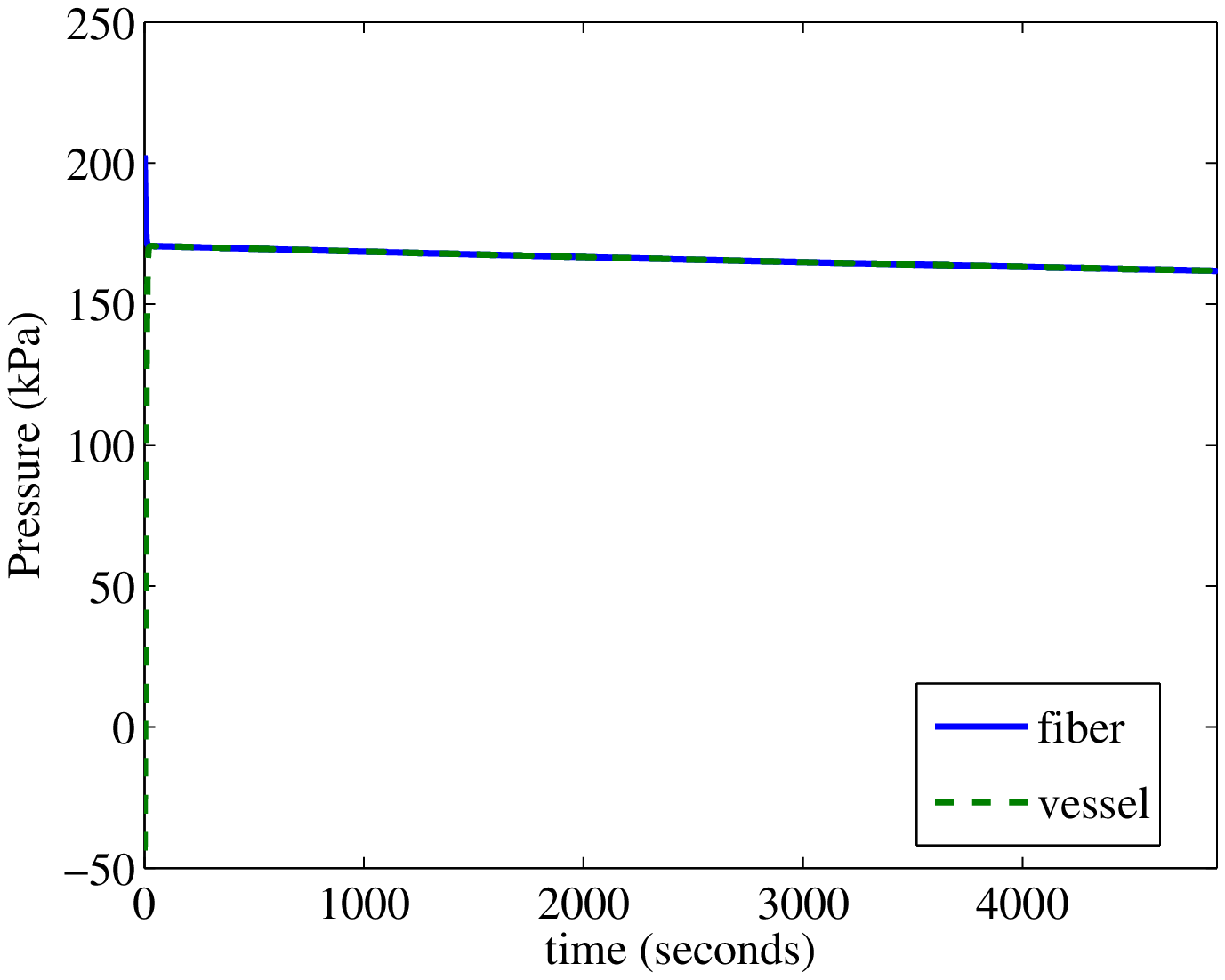}
      &
      \myfig{0.31}{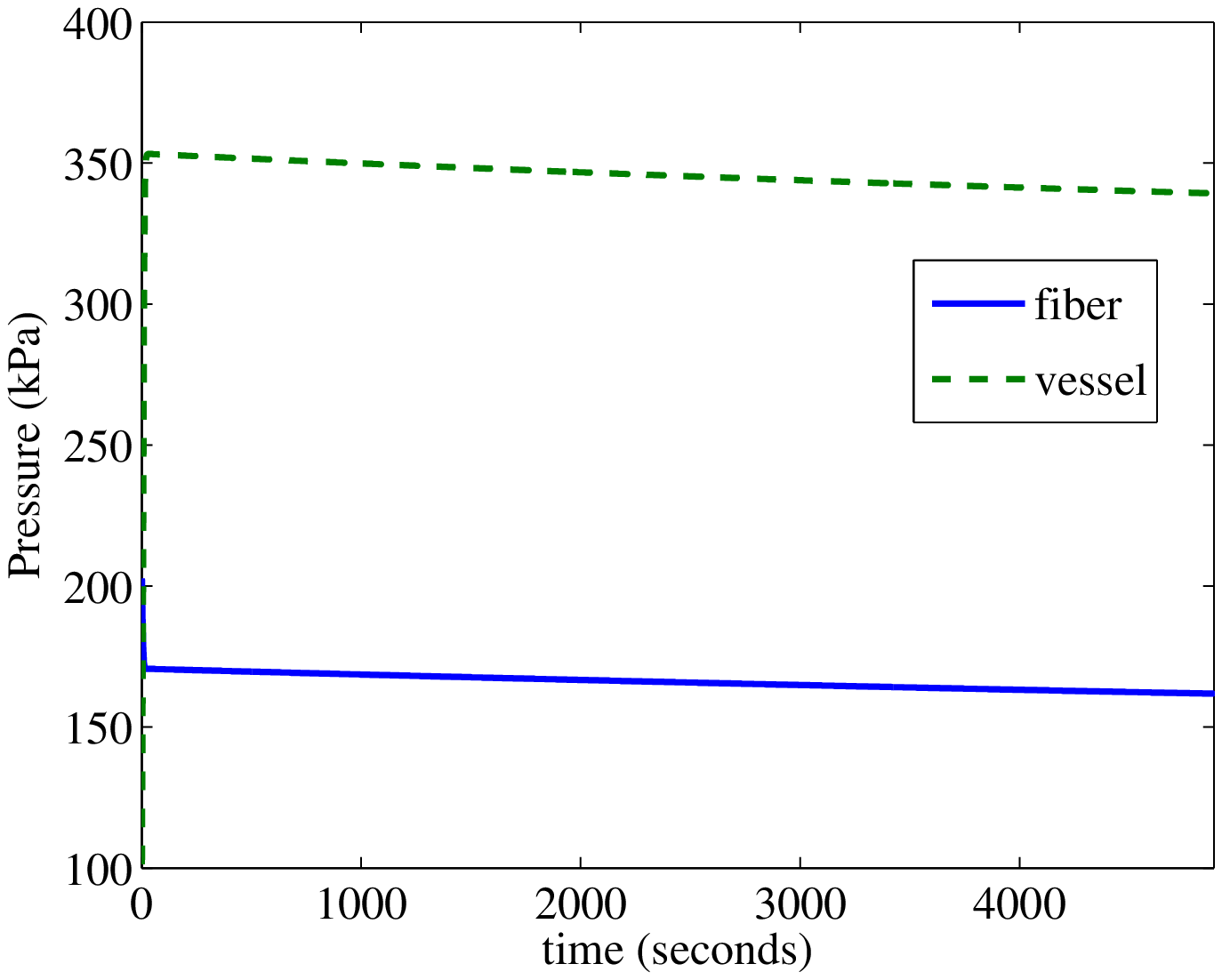}
      & 
      \myfig{0.31}{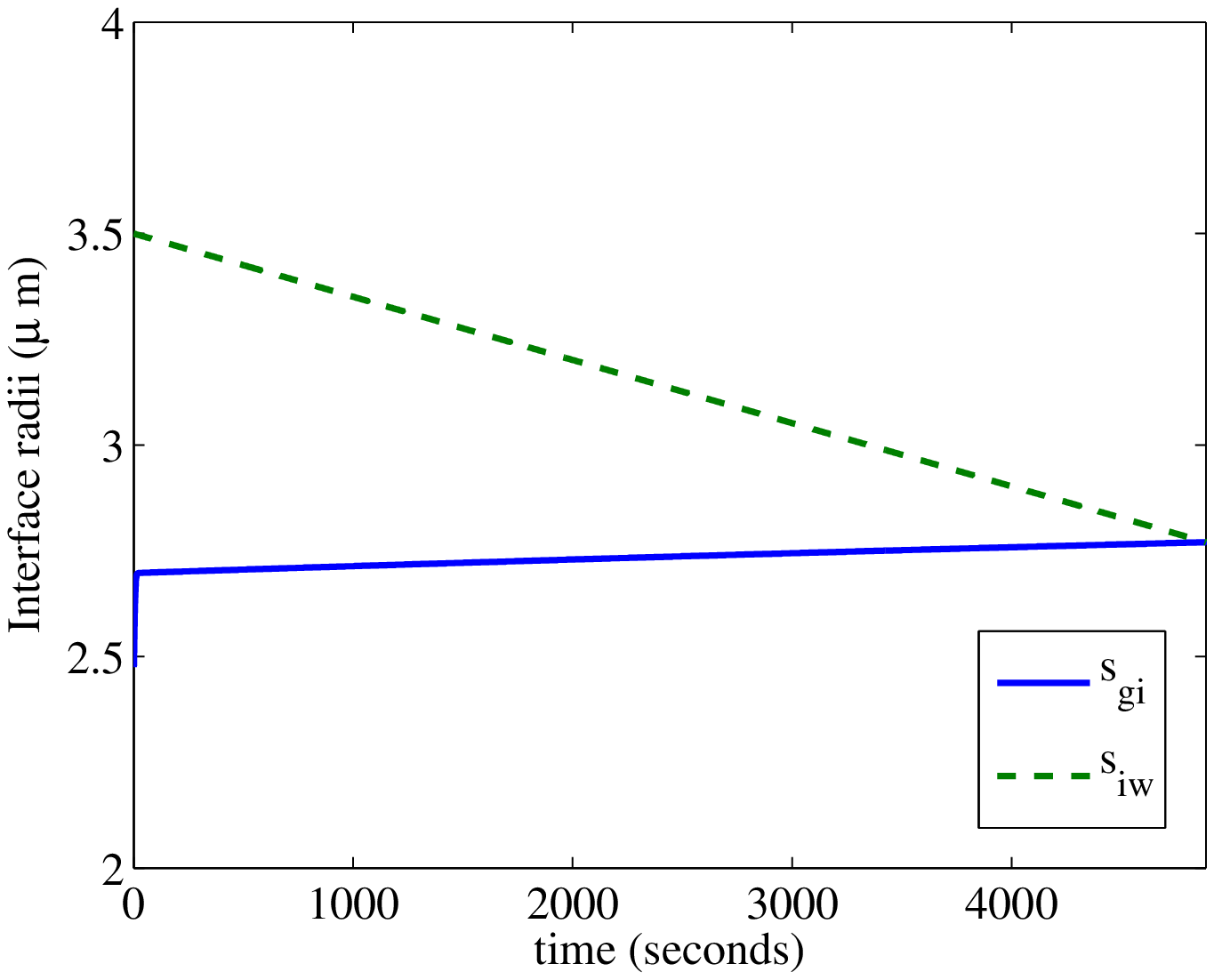}
    \end{tabular}
    \caption{Case~4 -- base case with ice and osmosis: Initial radii
      $s_{gi}(0) = 0.7\Rf$, $r(0)=0.3 \Rv$, and pressure
      $p^f_g(0)=200\;\units{kPa}$.}
    \label{fig:case4a}
  \end{center}
\end{figure}

In this section, we simulate the full model including ice and
osmosis, beginning again with the base case.  The solution plots
displayed in Figure~\ref{fig:case4a} should be compared with the
Milburn--O'Malley simulation in Figure~\ref{fig:case3a}, which differs
only in the lack of osmosis.  The two sets of results are qualitatively
similar, there being slight changes in the water pressure and
bubble radii owing to the inclusion of osmotic pressure.  Consistent
with the ice-free results in Cases~1 and~2, the vessel gas pressure is
increased by an amount roughly equal to the osmotic pressure
$\Pi=c_s{\cal R}T \approx 132\;\units{kPa}$.

\begin{figure}[bthp]
  \begin{center}
    \footnotesize
    \begin{tabular}{ccc}
      (a) Water pressures & (b) Gas pressures & (c) Gas bubble radii\\
      \myfig{0.31}{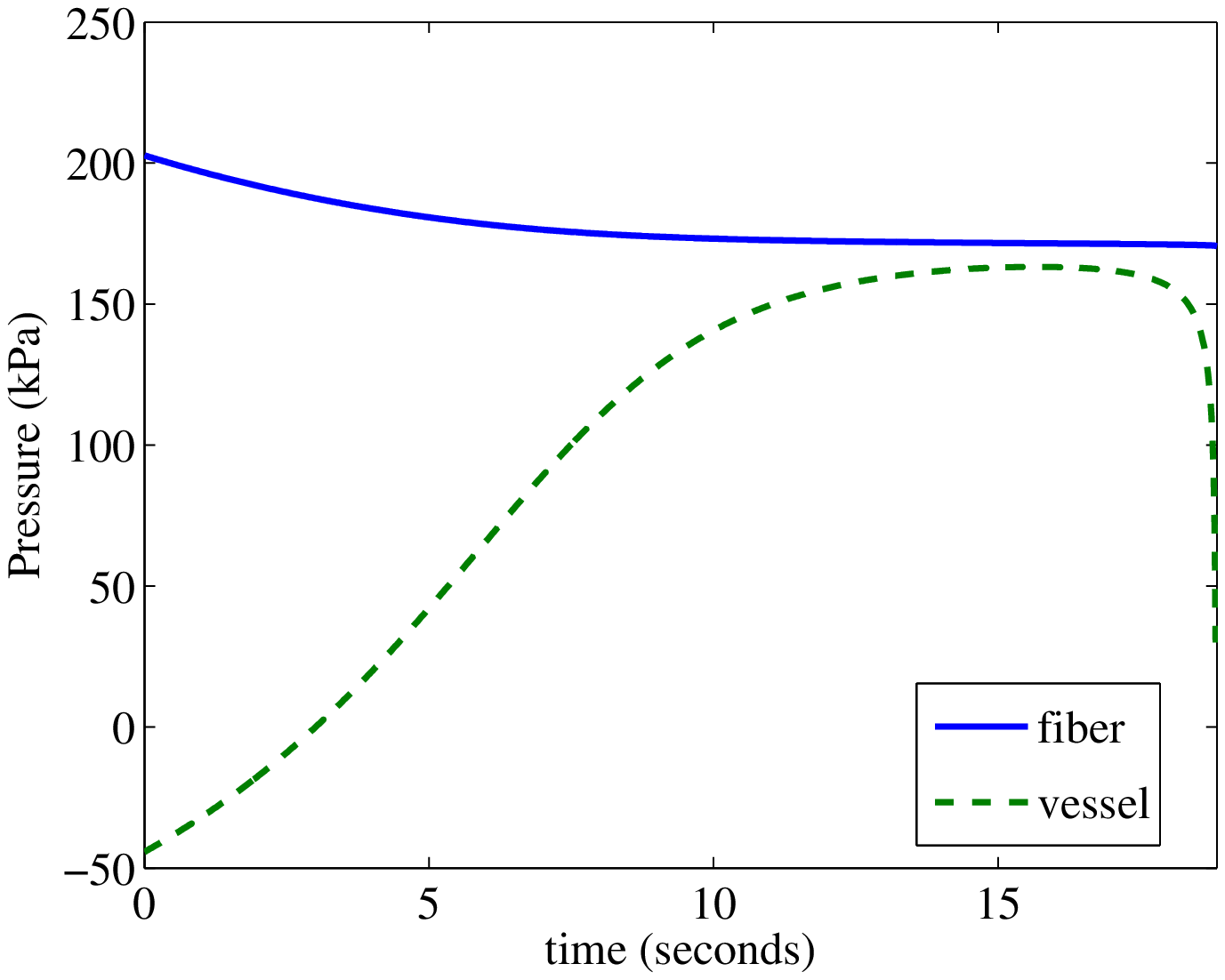}
      &
      \myfig{0.31}{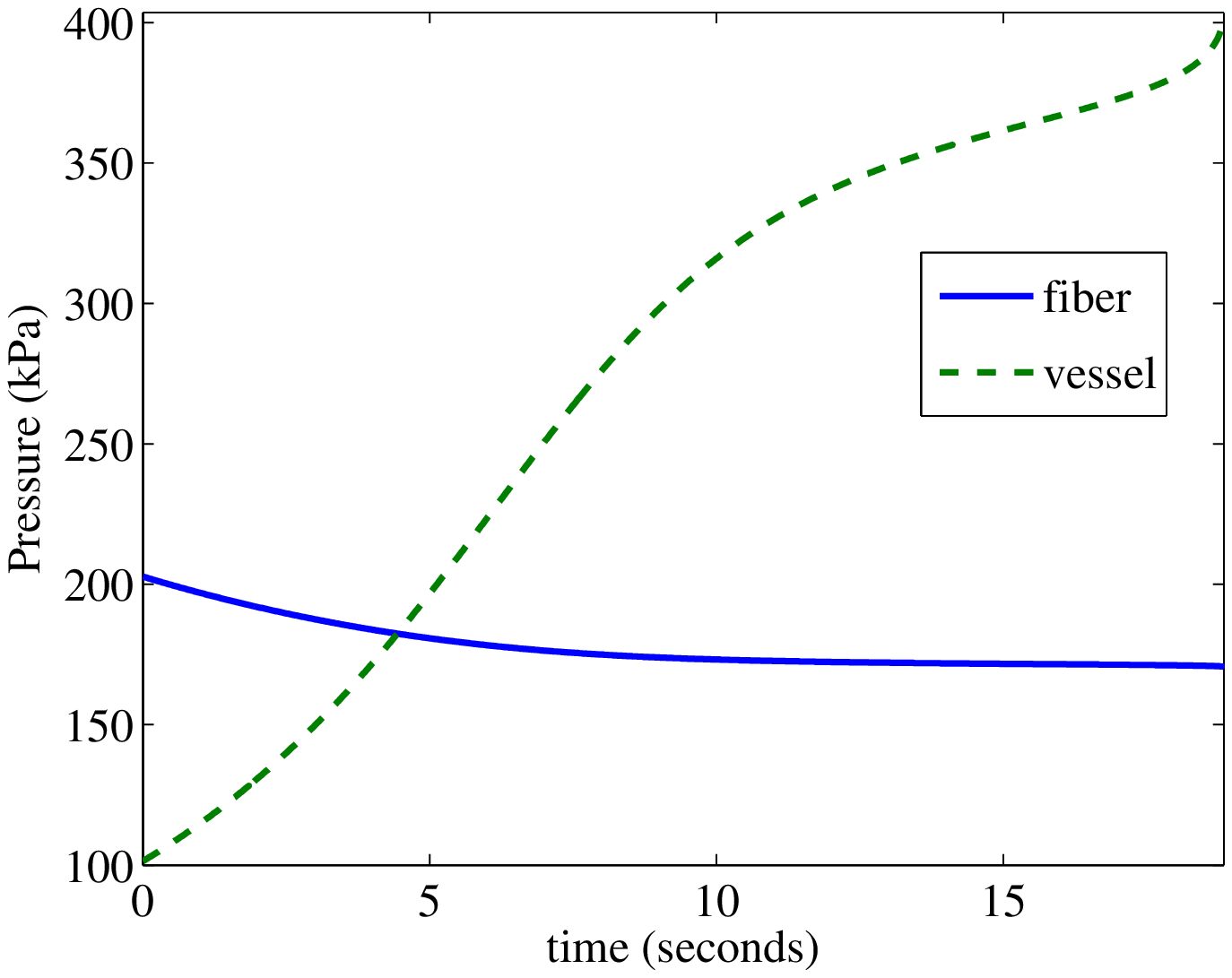}
      & 
      \myfig{0.31}{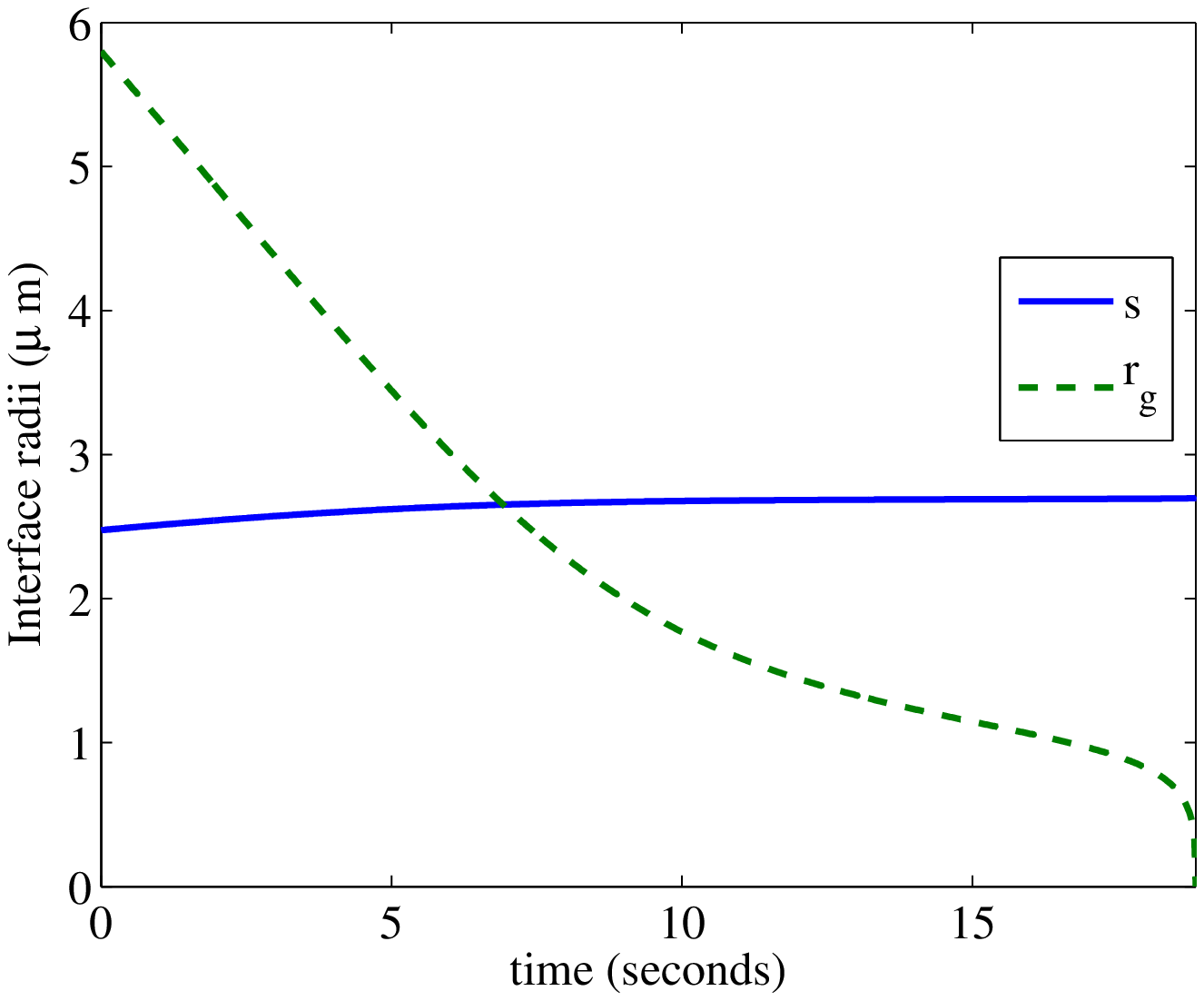}
    \end{tabular}
    \caption{Case~4 -- collapsing vessel bubble: With ice and osmosis,
      initial radii $s_{gi}(0) = 0.7\Rf$, $r(0)=0.29\Rv$, and pressure
      $p^f_g(0)=200\;\units{kPa}$.}
    \label{fig:case4i}
  \end{center}
\end{figure}

If the initial vessel bubble radius is decreased slightly from $0.30\Rv$
to $0.29\Rv$, then the vessel bubble is no longer unable to equilibrate
with the fiber and dissolves completely as shown in
Figure~\ref{fig:case4i}. This is consistent with the other results in
Figures~\ref{fig:case1i} and~\ref{fig:case2-summary}, except that the
extra osmotic pressure in the vessel causes the gas to disappear at
larger values of $r(0)$.  When viewed in the context of embolism
recovery, this is evidence that suggests osmosis can enhance the ability
of the tree to recover from embolism.  

\begin{figure}[bthp]
  \centering
  \footnotesize
  \begin{tabular}{ccc}
    (a) Vessel bubble radius & & (b) Vessel gas pressure\\
    \myfig{0.42}{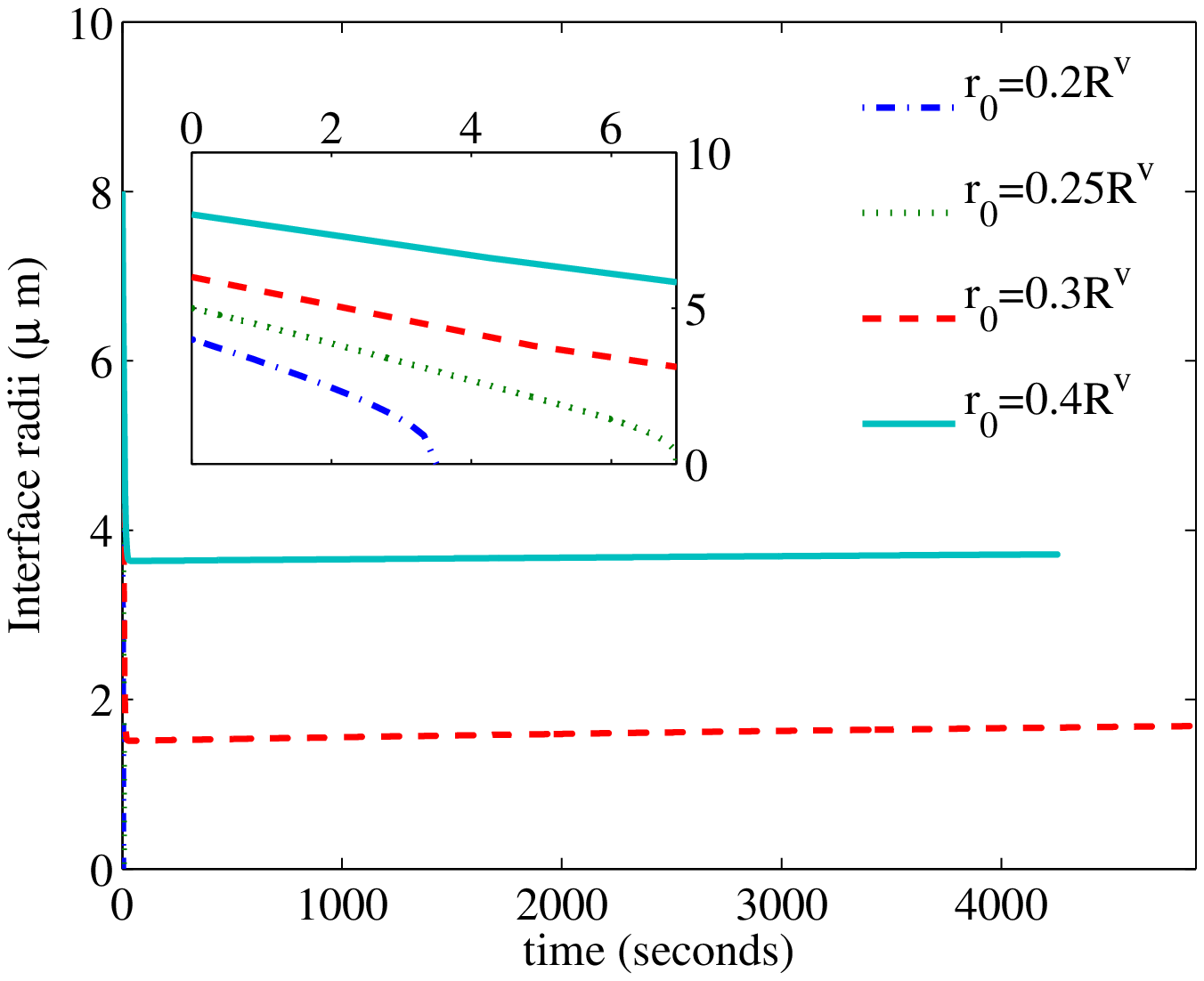}
    & & 
    \myfig{0.42}{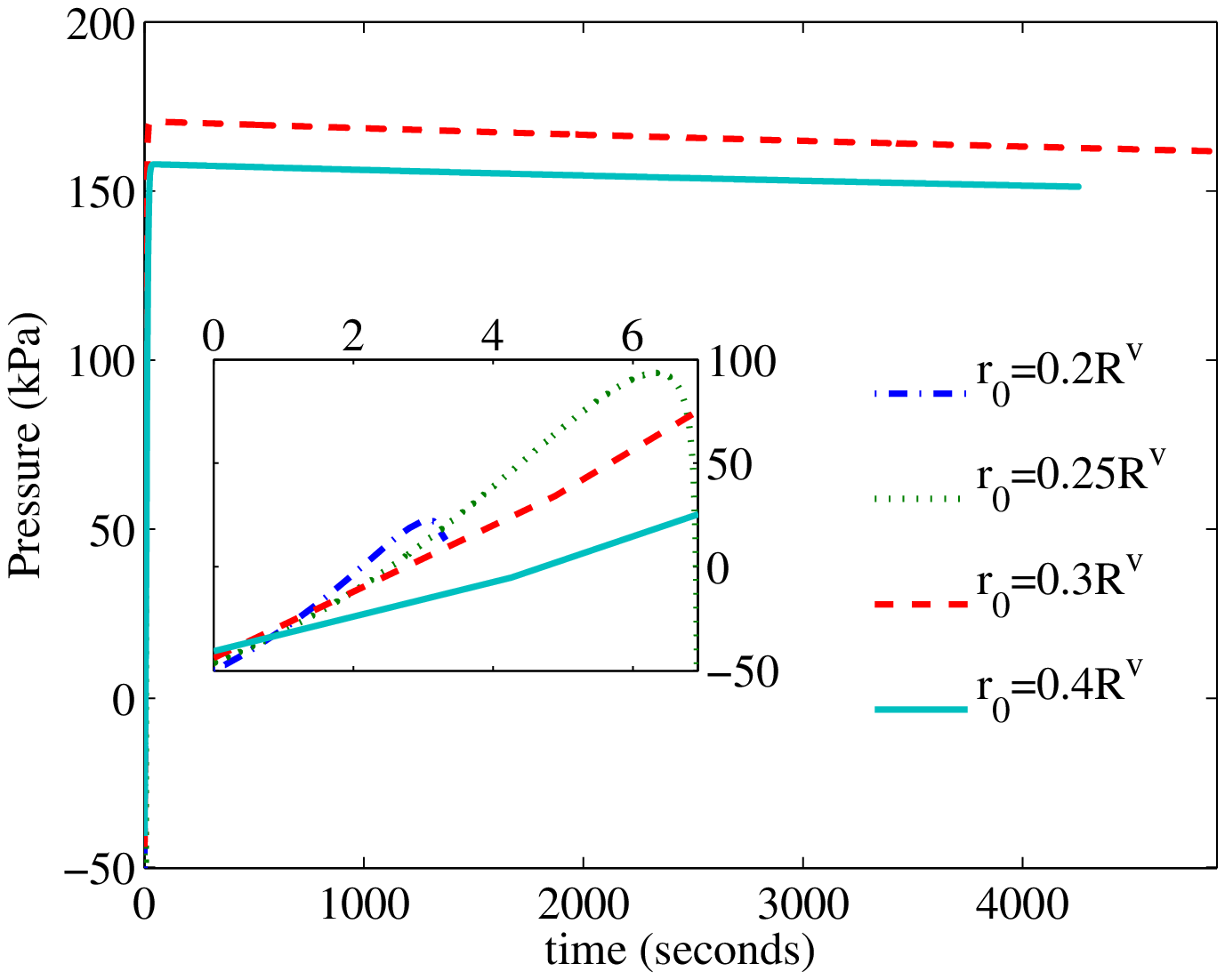}
  \end{tabular}
  \caption{Case~4 -- varying $r(0)$: With ice and osmosis, initial fiber
    radius $s_{gi}(0) = 0.7\Rf$, and pressure
    $p^f_g(0)=200\;\units{kPa}$.  The endpoint of each curve corresponds
    to the time that the ice layer completely melts. The ``zoomed-in''
    sub-plot shows the curves near $t=0$.}
  \label{fig:case4-summary-r}
\end{figure}

A series of simulations for different values of the initial vessel
bubble radius are reported in Figure~\ref{fig:case4-summary-r}, which
shows a similar outcome as in Case~3.  The main difference is that the
bubbles with radii $r(0)=0.2\Rv$ and $0.25\Rv$ both dissolve completely,
which is further evidence that osmosis enhances the effect of exudation
and makes the vessel bubbles more likely to collapse.

\leavethisout{
  \begin{figure}[bthp]
    \centering
    \footnotesize
    \begin{tabular}{ccc}
      (a) Vessel bubble radius & & (b) Vessel water pressure \\
      \myfig{0.42}{Comp_thaw_25052012_cs584_Summary_vessel_radius.eps}
      & & 
      \myfig{0.42}{Comp_thaw_25052012_cs584_Summary_vessel_Pressure.eps}
    \end{tabular}
    \caption{Case~4 -- varying $s_{gi}(0)$ and $p^f_g(0)$: With ice and
      osmosis, initial fiber radius $s_{gi}(0) = 0.7\Rf$, and pressure
      $p^f_g(0)=200\;\units{kPa}$.}
    \label{fig:case4-summary-s-and-p}
  \end{figure}
}

\subsection{Case~5: Long-time behavior after ice melts completely}
\label{sec:sims-case5}

The purpose of this final set of simulations is to study the behavior
over the longer term once the ice is totally melted, and to demonstrate
that our approach is capable of computing through the disappearance of
the ice layer and the corresponding change in the governing equations.
We present results for a few choices of parameters considered in the
previous section (Case 4) and emphasize that until the time that the ice
melts, the plots are identical.  Our main observation is that after the
ice disappears, the solution dynamics exhibit a rapid equilibration to a
constant equilibrium state that is consistent with what we saw earlier
in Case~2 (no ice, with osmosis).

\begin{figure}[bthp]
  \begin{center}
    \footnotesize
    \begin{tabular}{ccc}
      (a) Water pressures & (b) Gas pressures & (c) Gas bubble radii\\
      \myfig{0.31}{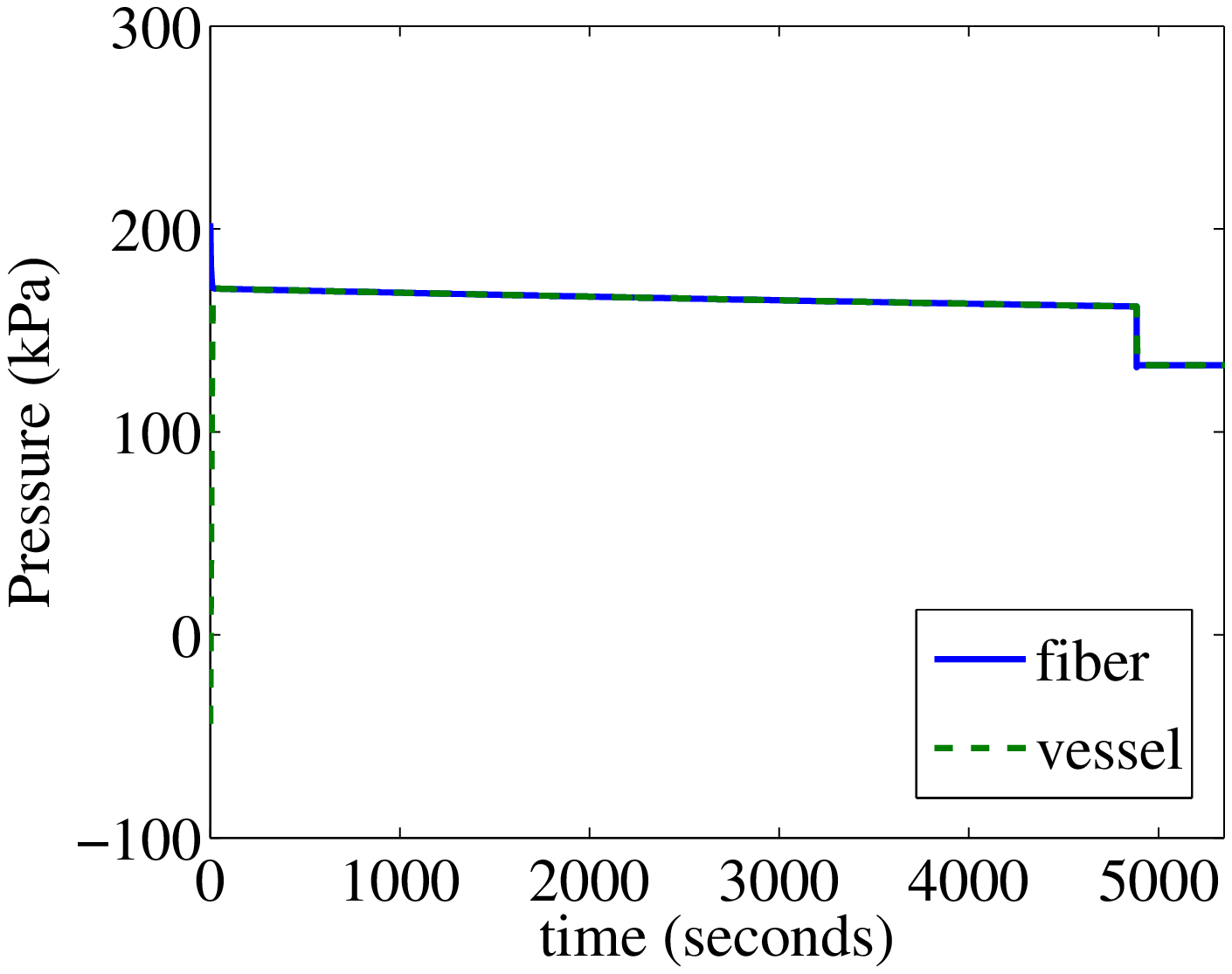}
      &
      \myfig{0.31}{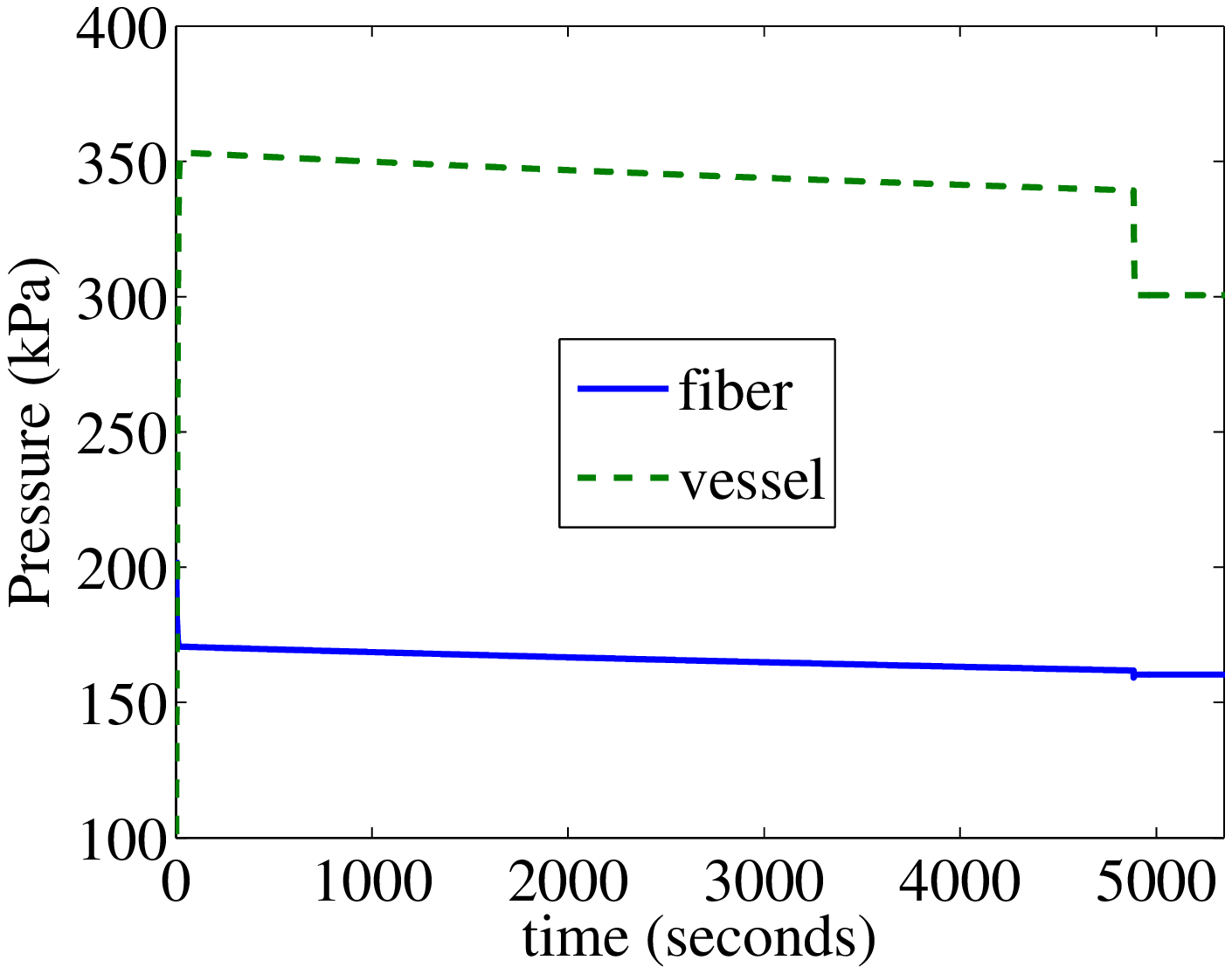}
      & 
      \myfig{0.31}{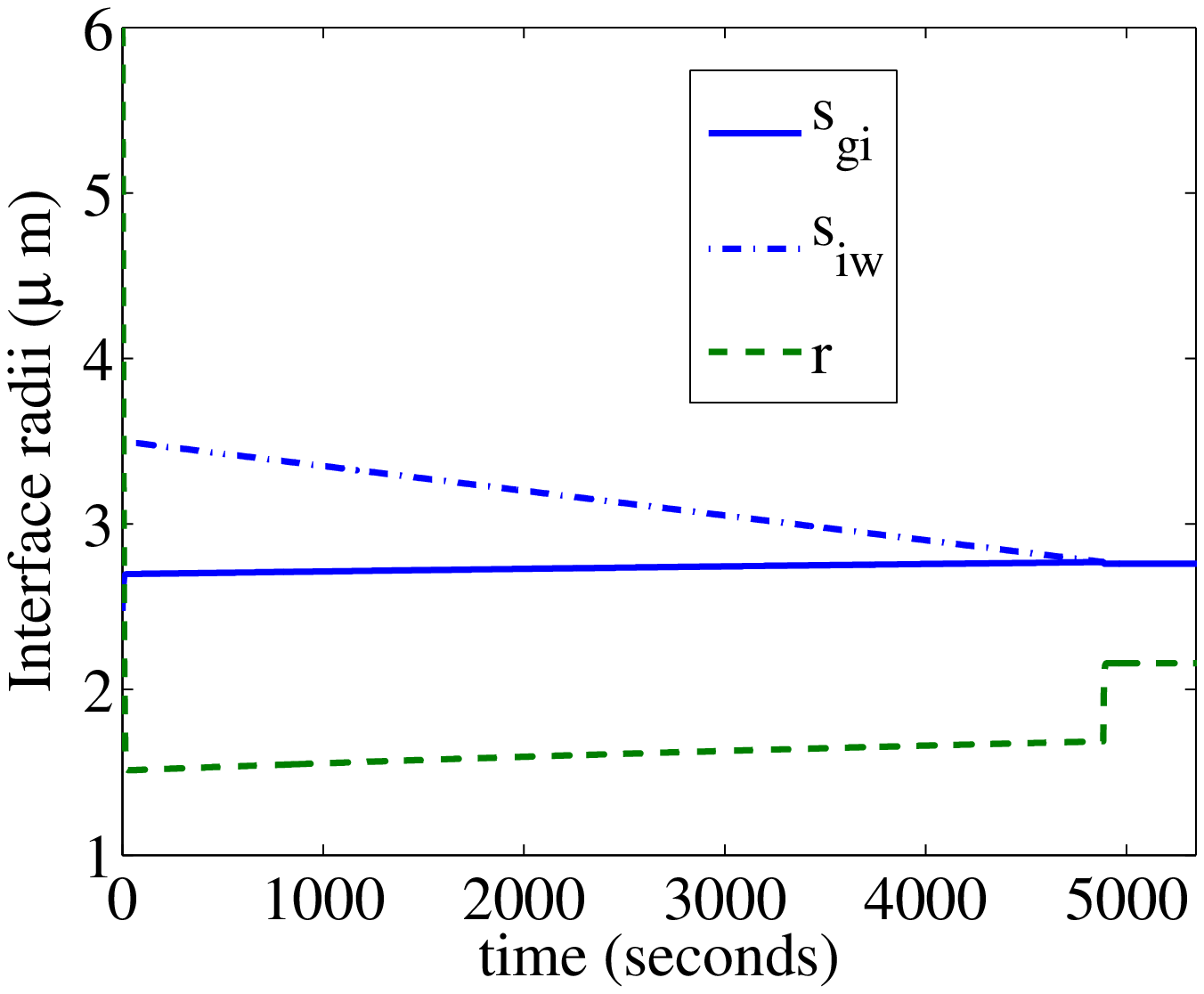}
    \end{tabular}
    \caption{Case~5 -- base case with ice and osmosis: Long-time
      behavior with initial radii $s_{gi}(0) = 0.7\Rf$, $r(0)=0.3 \Rv$,
      and pressure $p^f_g(0)=200\;\units{kPa}$.}
    \label{fig:case5a}
  \end{center}
\end{figure}

\begin{figure}[bthp]
  \begin{center}
    \footnotesize
    \begin{tabular}{ccc}
      (a) Water pressures & (b) Gas pressures & (c) Gas bubble radii\\
      \myfig{0.31}{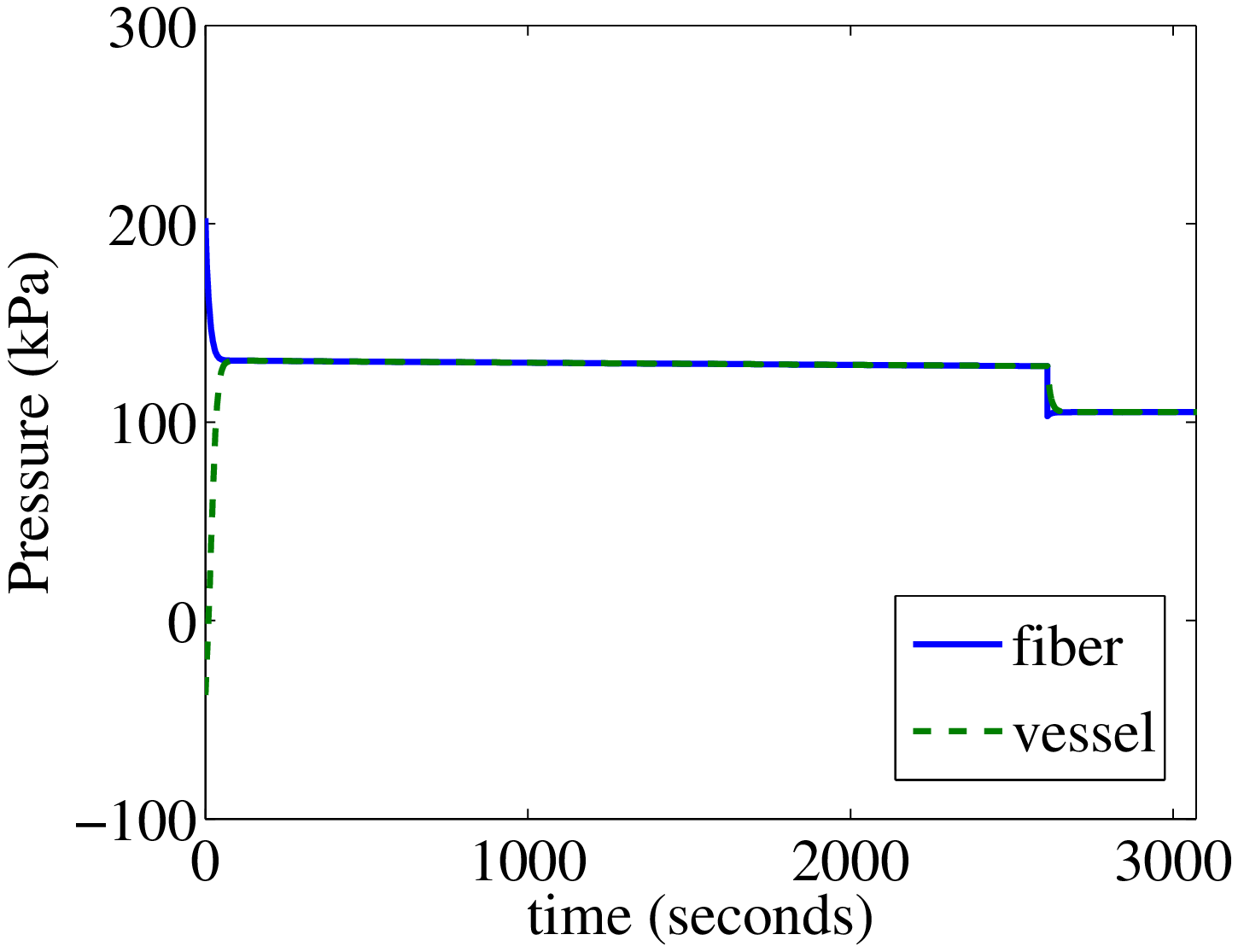}
      &
      \myfig{0.31}{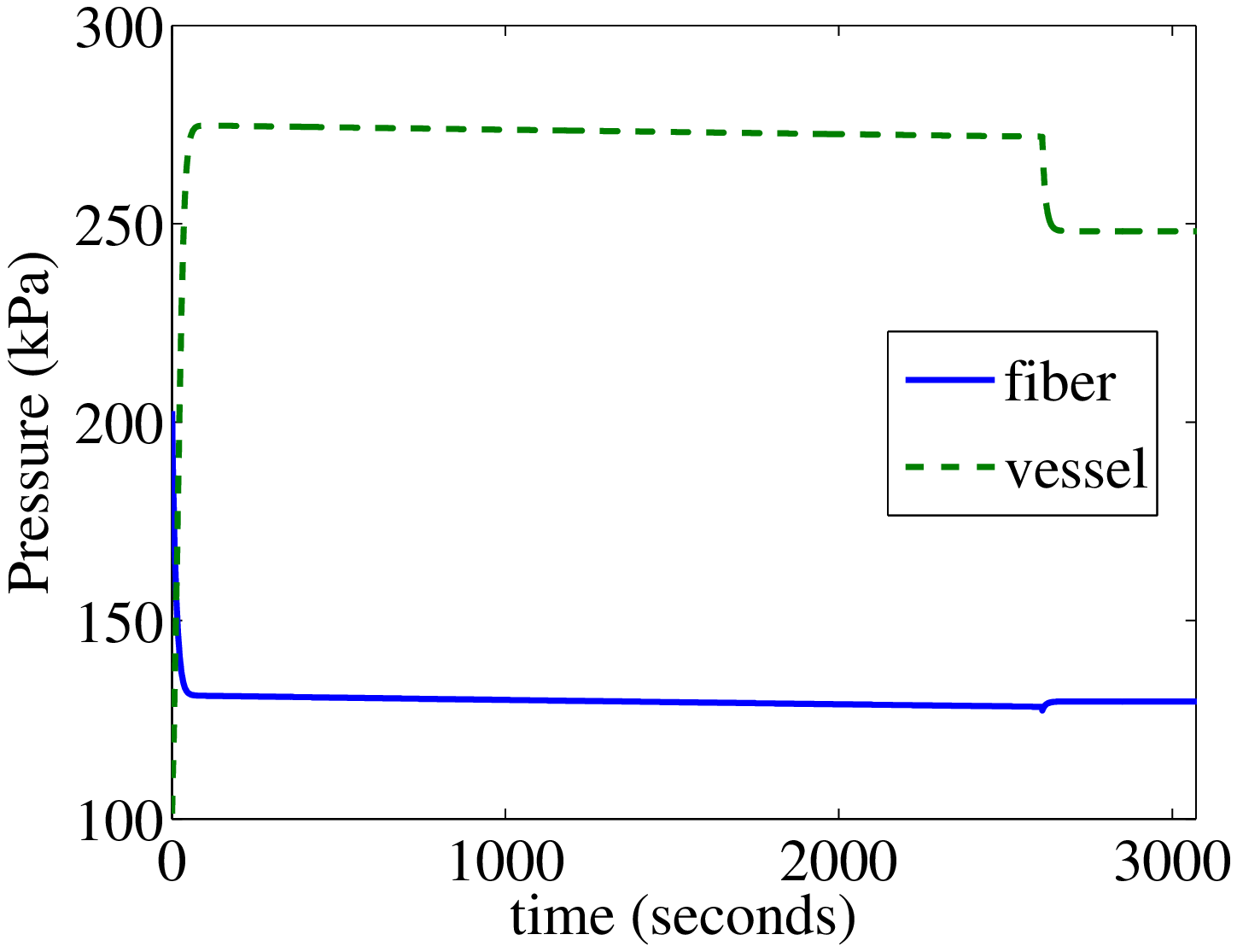}
      & 
      \myfig{0.31}{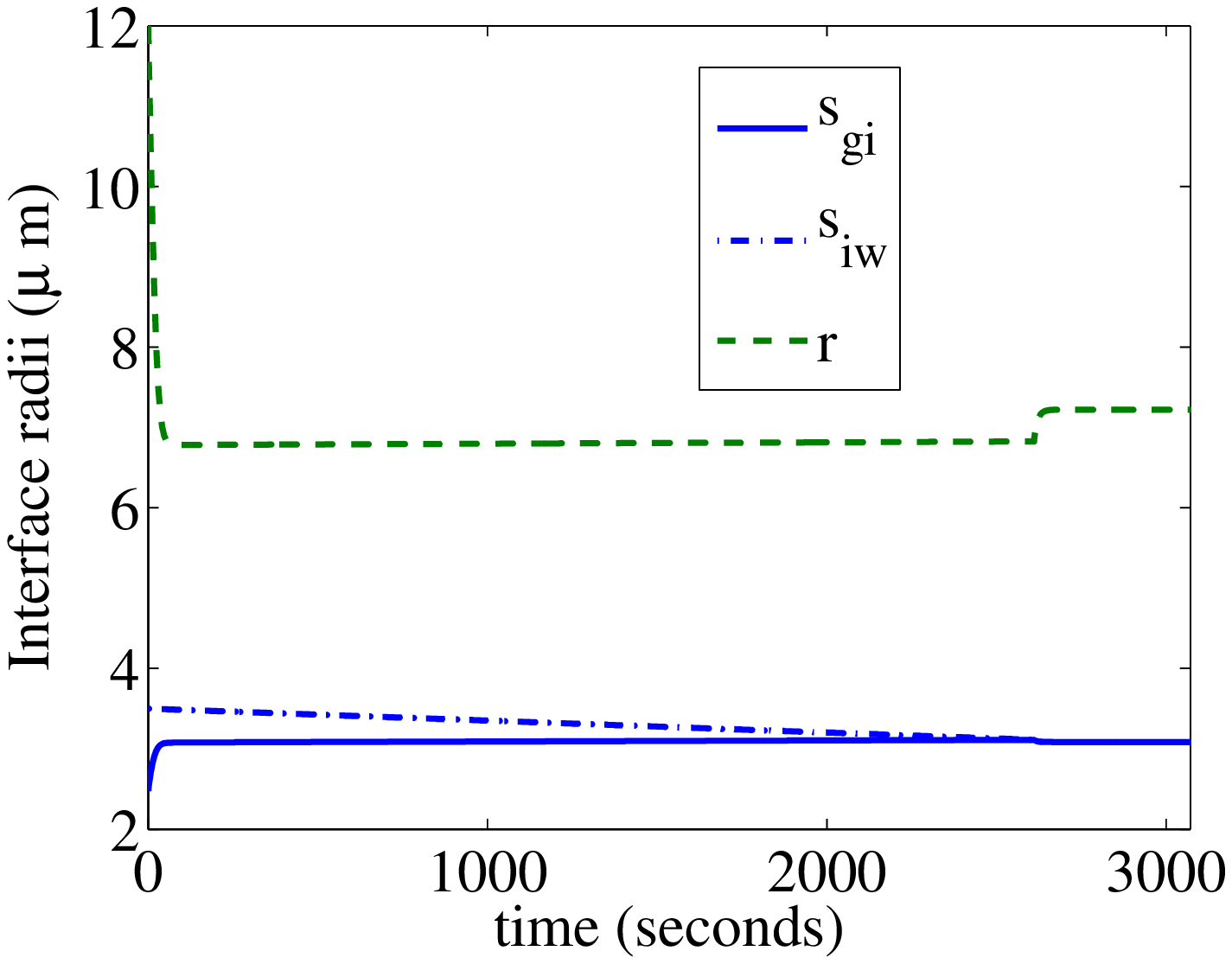}
    \end{tabular}
    \caption{Case~5 -- larger vessel bubble: With ice and osmosis,
      initial radii $s_{gi}(0) = 0.7\Rf$, $r(0)=0.6 \Rv$, and pressure
      $p^f_g(0)=200\;\units{kPa}$.}
    \label{fig:case5d}
  \end{center}
\end{figure}

Simulations are shown in Figures~\ref{fig:case5a} and \ref{fig:case5d}
for two values of vessel radius, $r(0)=0.3$ and $0.6$, that can be
compared with results in Figure~\ref{fig:case4a} and
\ref{fig:case4-summary-r}.  At the instant the ice layer melts away, the
water pressures undergo a sudden transition in which $p^f_w$ and $p^v_w$
drop sharply and then rapidly equilibrate to a new constant value.  This
pressure discontinuity arises from the sudden appearance of the surface
tension term in equation~\eqnref{surftens-nondim-noice}, when the ice
layer vanishes and the fiber sap comes into contact with the gas.  We
emphasize here that the numerical algorithm handles the discontinuity
and change in governing equations easily and without introducing any
spurious oscillations or other non-physical effects.

After the jump in water pressure, there is a rapid but smooth variation
in the radii and gas pressure as the system equilibrates (although in
the case when $r(0)=0.3\Rv$ the change in gas pressure is too small to
be seen on this scale).  Otherwise, the behavior after the loss of ice
is analogous to the earlier ice-free simulations.

In reality, we would expect to see a rapid but smooth variation in
liquid surface tension as the disappearing ice layer transitions through
a ``mushy region'' where both ice and water
coexist~\cite{lacey-tayler-1983}.  Including this additional level of
detail would not have a major effect on our solution, but it could form
the basis for an interesting future study of the detailed structure of
the transition layer.

\section{Conclusions and future work}
\label{sec:conclusion}

The aim of this work was to develop a detailed mathematical model that
captures the physical effects -- heat and mass transport, gas
dissolution, surface tension, osmosis and phase change -- that are
believed to play a significant role in sap exudation in maple trees.  In
particular, we were interested in investigating two competing hypotheses
for sap exudation based on a model of Milburn--O'Malley (without
osmosis)~\cite{milburn-omalley-1984} and another of Tyree (with
osmosis)~\cite{tyree-1995}.  The model focuses on the cellular scale and
the transfer of pressure between two main xylem components: fibers and
vessels.  We derived a coupled system of nonlinear, time-dependent,
differential-algebraic equations, which we solved using
\matlab. Extensive numerical simulations were performed that uncovered
the following:
\begin{itemize}
\item It is necessary to include the effect of gas bubbles in the vessel
  sap in order to allow a transfer of stored pressure in the fibers to
  the vessel sap.  This gas in the vessels was not explicitly mentioned
  by either Milburn--O'Malley or Tyree.
\item Taking parameter values consistent with data in the literature,
  our model captures positive vessel pressures provided that the gas in
  the fiber is sufficiently compressed during the previous year's
  freezing phase.  The observed increase in vessel sap pressure lies in
  the range of 30--60\;\units{kPa} that is consistent with experimental
  values reported in the literature for maple sap exudation.
\item For a wide range of parameters, and provided the vessel gas
  content is low enough, the fibers are able to completely dissolve the
  gas bubbles within the vessel sap.  This is consistent with the
  phenomenon of embolism repair that is known to occur in maple as
  well as other hardwood species.
\item Reasonable exudation pressures are obtained both with and without
  osmosis.  One impact of including osmotic effects is to increase
  slightly the threshold vessel bubble size below which gas in the
  vessel gas dissolves entirely; in this respect, our model suggests
  that osmosis may enhance embolism repair.  The second main impact of
  osmosis is to lessen the likelihood that the fiber gas bubble
  collapses, which enhances the ability to exude sap.
\end{itemize}

This paper can only be considered as a preliminary study of sap
exudation because the comparisons and conclusions we have drawn so far
are mainly qualitative.  More detailed comparisons with experimental
data in the literature are currently underway to further validate the
model, and will form the basis of a future publication.  There are a
number of other specific areas that we believe would provide fruitful
avenues for future research:
\begin{itemize}
\item The long-time behavior pictured in Figures~\ref{fig:case5a}
  and~\ref{fig:case5d} suggests that this problem is ripe for a
  multi-layer asymptotic analysis in which the solution is divided into
  two pieces: a linear solution, matched to initial conditions via a
  boundary layer on the left; and a constant equilibrium state that is
  matched across an interior layer to the linear solution.

\item The predictive power of this model would be strengthened with
  better estimates of two parameters: the gas content in the vessel, and
  the initial pressure in the fiber.  We will perform a more extensive
  search of the literature and focus in particular on data from studies
  of embolism.  In conjunction with this effort, we believe that our
  model (with only minor modifications) could also be applied to study
  the phenomenon of embolism repair coinciding
  with the spring thaw~\cite{yang-tyree-1992}.

\item The conditions under which the maple tree freezes at the end of
  the previous season are known to have a significant impact on sap
  exudation during the following spring.  To study this process, we
  could run our model ``in reverse'' to study the cooling sequence in
  the Milburn--O'Malley model (steps 1--4 in
  Figure~\ref{fig:milburn-omalley}).

\item We are currently developing a macroscopic tree-level model, based
  on the 1D model of Chuang et al.~\cite{chuang-etal-2006} that uses
  Richards' equation to capture sap flow in the porous xylem.  In order
  to apply this work to the study of exudation, we require
  solution-dependent transport coefficients that will be derived from
  the microscopic cell-level model derived in the current paper using an
  up-scaling procedure. 

  \leavethisout{
  \item Testable experimental hypotheses??: measure the size of the fiber
    gas bubble, the thickness of the ice layer, and/or the pressure of the
    compressed gas for various tree sections prepared under various
    conditions.  This can be compared directly to our model predictions. 
  }
  
  \leavethisout{
  \item Then perform pore network simulations of the same phenomenon
    (similar to Aumann and Ford) to compare.
    
  \item Experiments on plain why several days of temperatures
    oscillating above and below the freezing point are necessary to
    initiate exudation.
    
  \item Freezing point depression effect, in which sucrose concentration can
    reduce the freezing temperature by up to 0.2$\degC$.  This is
    considerably larger than the value $T_a=0.005\degC$, which suggests it
    could be important.  On the other hand, since the fiber contains no
    sucrose, this cannot play a role in our exudation model.  Perhaps it's
    important during the sap freezing process?
    
  \item Super-cooling effect~\cite{tyree-1984}. 
    
  \item Use the model to predict peak harvest times and to investigate the
    potential effects of climate change~\cite{tyree-1984}.
    
  \item Incorporate the effects of living cells as agents for conversion
    of starches stored in ``contact cells'' to
    sucrose~\cite{sauter-etal-1973}.  
  }
  
\end{itemize}



\providecommand{\noopsort}[1]{}

\end{document}